\documentclass[11pt,epsf]{article}
\DeclareMathAlphabet{\scr}{U}{rsfs}{m}{n}
\usepackage{latexsym,color}
\usepackage{epsfig}
\usepackage[mathscr]{eucal}
\usepackage{amsfonts}
\usepackage{amscd}
\usepackage{cite}
\usepackage{array}
\usepackage{amssymb}
\usepackage{colordvi}
\usepackage[centertags]{amsmath}
\usepackage{enumerate}
\usepackage{graphicx}
\usepackage{booktabs}
\usepackage{theorem}
\usepackage[footnotesize]{caption}
\usepackage{soul}

\newcommand{\lsim}{\raisebox{-0.13cm}{~\shortstack{$<$ \\[-0.07cm] $\sim$}}~}
\newcommand{\gsim}{\raisebox{-0.13cm}{~\shortstack{$>$ \\[-0.07cm] $\sim$}}~}
\newcommand{\newc}{\newcommand}
\newc{\be}{\begin{equation}}
\newc{\ee}{\end{equation}}
\newc{\bea}{\begin{eqnarray}}
\newc{\eea}{\end{eqnarray}}
\newc{\ol}{\overline}
\newc{\wt}{\widetilde}
\newc{\bs}{\boldsymbol}
\newc{\m}{\mathcal}
\newc{\la}{\langle}
\newc{\ra}{\rangle}

\newcommand{\gam}{\gamma \gamma}

\newcommand{\non}{\nonumber}
\newcommand{\beq}{\begin{eqnarray}}
\newcommand{\eeq}{\end{eqnarray}}
\newcommand{\s}{\smallskip}

\newcommand{\bc}{\begin{center}}
\newcommand{\ec}{\end{center}}

\newcommand{\ba}{\begin{array}}
\newcommand{\ea}{\end{array}}

\begin{document}

\title{
\vspace*{-3cm}
\phantom{h} \hfill\mbox{\small ADP-14-23-T881}\\[-1.1cm]
\phantom{h} \hfill\mbox{\small KA-TP-21-2014}\\[-1.1cm]
\phantom{h} \hfill\mbox{\small SFB/CPP-14-59}
\\[1cm]
\textbf{Discovery Prospects for NMSSM Higgs Bosons\\
at the High-Energy Large Hadron Collider}}

\date{}
\author{
S.~F.~King$^{1\,}$\footnote{E-mail: \texttt{king@soton.ac.uk}},
M. M\"{u}hlleitner$^{2\,}$\footnote{E-mail: \texttt{margarete.muehlleitner@kit.edu}},
R.~Nevzorov$^{3\,}$\footnote{E-mail: \texttt{roman.nevzorov@adelaide.edu.au}},
K. Walz$^{2\,}$\footnote{E-mail: \texttt{kathrin.walz@kit.edu}}
\\[9mm]
{\small\it
$^1$School of Physics and Astronomy,
University of Southampton,}\\
{\small\it Southampton, SO17 1BJ, U.K.}\\[3mm]
{\small\it
$^2$Institute for Theoretical Physics, Karlsruhe Institute of Technology,} \\
{\small\it 76128 Karlsruhe, Germany.}\\[3mm]
{\small\it
$^3$School of Chemistry and Physics, University of Adelaide,} \\
{\small \it SA 5005 Australia.}\\
}

\maketitle

\begin{abstract}
\noindent
We investigate the discovery prospects for NMSSM Higgs bosons during
the 13~TeV run of the LHC. While one of the neutral Higgs bosons
is demanded to have a mass around 125~GeV and
Standard Model (SM)-like properties, there can be substantially
lighter, nearby or heavier Higgs bosons, that have not been excluded
yet by LEP, Tevatron or the 8~TeV run of the LHC. The challenge consists in
discovering the whole NMSSM Higgs mass spectrum. We present the rates
for production and subsequent decay of the neutral NMSSM Higgs bosons
in the most promising final states and discuss their possible
discovery. The prospects for pinning down the Higgs sector of the
Natural NMSSM will be analysed taking into account alternative search
channels. We give a series of benchmark scenarios
compatible with the experimental constraints, that feature
Higgs-to-Higgs decays and entail (exotic) signatures with
multi-fermion and/or multi-photon final states. These decay chains
furthermore give access to the trilinear Higgs self-couplings.
We briefly discuss the possibility of exploiting coupling sum rules in
case not all the NMSSM Higgs bosons are discovered.
\end{abstract}
\thispagestyle{empty}
\vfill
\newpage
\setcounter{page}{1}

\section{Introduction}
The discovery of a new particle with mass around 125~GeV by the Large
Hadron Collider Experiments ATLAS and CMS~\cite{:2012gk,:2012gu} has
immediately triggered investigations of its properties such as spin and
CP quantum numbers and couplings to other SM particles. These
analyses conclude so far that the discovered particle is a Higgs boson
which behaves rather SM-like. Additionally no new resonances have been
discovered which could point to extensions beyond the SM (BSM). This
renders the detailed investigation of the Higgs properties
at highest possible accuracy accompanied by the search for new
particles even more important.
The Higgs sector of supersymmetric extensions of the SM
introduces an enlarged Higgs spectrum due to the requirement of at
least two complex Higgs doublets to ensure supersymmetry
\cite{susy}. These lead in the Minimal Supersymmetric Extension of the SM
(MSSM)~\cite{mssm} to five Higgs bosons. The
Next-to-Minimal Supersymmetric Extension of the SM
(NMSSM)~\cite{nmssm} is extended by an additional complex superfield
$\hat{S}$ and allows for a dynamical solution of the $\mu$ problem
\cite{muproblem} when the singlet field acquires a non-vanishing vacuum 
expectation value. After electroweak
symmetry breaking (EWSB) the NMSSM Higgs
sector features seven Higgs bosons, which are in the CP-conserving
case given by three neutral CP-even, two neutral CP-odd and two
charged Higgs bosons. A nice consequence of the singlet superfield, which
couples to the Higgs doublet superfields $\hat{H}_u$ and $\hat{H}_d$,
are new contributions to the quartic coupling proportional to the
singlet-doublet coupling $\lambda$ so that the tree-level
mass value of the lighter MSSM-like Higgs boson is
increased. Therefore less important radiative corrections are
required to shift the mass value to 125~GeV and in turn smaller stop
masses and/or mixing, allowing for less fine-tuning
\cite{finetune,ournmssmpapers}.
\s

The singlet admixture in the Higgs mass eigenstates entails reduced
couplings to the SM particles. Together with the enlarged Higgs sector
this leads to a plethora of interesting phenomenological scenarios and
signatures. Thus very light Higgs bosons are not yet excluded by the
LEP searches \cite{lep} if their SM couplings are small enough. In
turn heavier Higgs bosons can decay into a pair of lighter Higgs
bosons subsequently decaying into SM particles which leads to
interesting final state signatures
\cite{Nhung:2013lpa,Ellwanger:2013ova,Munir:2013dya}. Furthermore,
branching ratios into
LHC standard search channels such as $\gamma\gamma$ or vector boson
final states can be enhanced or suppressed
\cite{ournmssmpapers,brchanges}. An enhanced photonic rate of the
125~GeV Higgs boson can also be due to two almost degenerate Higgs
bosons with masses near 125~GeV \cite{ournmssmpapers,Gunion:2012gc}.
The investigation of double ratios of signal rates at the high-energy
option of the LHC would allow for the resolution of the double peak
\cite{Gunion:2012he}. It is obvious that on the theoretical side the
reliable interpretation of such BSM signatures and the disentanglement
of different SUSY scenarios as well as their distinction from the SM
situation requires precise predictions of the SUSY parameters such
as masses and Higgs couplings to other Higgs bosons
including higher order corrections. For the CP-conserving NMSSM the mass
corrections are available at one-loop accuracy \cite{effpot,leadlog,Degrassi:2009yq,full1loop,Ender:2011qh}, and two loop results of ${\cal O}
(\alpha_t \alpha_s + \alpha_b \alpha_s)$ in the
approximation of zero external momentum have been given in
Ref.~\cite{Degrassi:2009yq}. In the complex case the Higgs mass
corrections have been calculated at one-loop accuracy
\cite{effcorr1,effcorr2,effcorr3,Graf:2012hh} with the
logarithmically enhanced two-loop effects given in
\cite{complex2loop}. Higher-order corrections to the trilinear Higgs
self-coupling of the neutral NMSSM Higgs bosons have been provided in
\cite{Nhung:2013lpa}. But also constraints that arise from Dark
Matter, the low-energy observables, the direct Higgs boson searches at
LEP, Tevatron and the LHC as well as restrictions on the SUSY
parameter space due to the exclusion bounds on SUSY particle masses
have to be included when investigating viable NMSSM
scenarios. By performing a scan over wide ranges of the NMSSM
parameter space, incorporating in our analysis higher order corrections to
the NMSSM Higgs parameters and their production and decay rates
\cite{nmhdecay1,nmssmtools,Baglio:2013iia} and 
taking into account experimental constraints, we investigate the
features of the NMSSM Higgs spectrum, the Higgs mass values and
mixings.\footnote{For recent studies on the NMSSM phenomenology, see
  \cite{Badziak:2013bda,Munir:2013wka,Cerdeno:2013qta,Beskidt:2013gia,Choi:2013lda,Kozaczuk:2013spa,Cao:2013gba,Jia-Wei:2013eea,Huang:2013ima,Belanger:2014roa,Beskidt:2014oea,Buttazzo:2014nha,Ellwanger:2014dfa,Ellwanger:2014hia}.}
We present signal rates for the various SM 
final states and discuss the prospect of discovering the NMSSM Higgs
bosons during the next run of the LHC at a center-of-mass (c.m.) energy of
13~TeV. Subsequently, we restrict ourselves to a subspace of the NMSSM
with a slightly broken Peccei-Quinn symmetry, that is characterized
by a rather light Higgs spectrum with masses below about 530~GeV. This
subspace, which we call Natural NMSSM, turns out to allow for the
discovery of all Higgs bosons for a large fraction of its
scenarios. It can therefore be tested or ruled out at the next LHC
run. Besides their direct production, the Higgs bosons can also be
produced in the decays of heavier Higgs bosons into lighter Higgs pairs
or into a Higgs and gauge boson pair. The Higgs-to-Higgs decay
processes give access to the trilinear Higgs self-couplings, which are
a necessary ingredient for the reconstruction and test of the Higgs potential. We
present benchmark scenarios that cover various aspects of
Higgs-to-Higgs decays. We find that they lead to, in part unique, in part exotic,
multi-photon and/or multi-lepton final states. These
scenarios can therefore be viewed as benchmark scenarios and
should be considered in the experimental searches to make sure that no
BSM states will be missed. We also briefly discuss what can be learned 
from the couplings of the discovered Higgs bosons in case not all of
the NMSSM Higgs bosons are accessible.
\s

The layout of the remainder of this paper is as follows. In
section~\ref{sec:nmssm} we briefly introduce the NMSSM
Lagrangian. In section~\ref{sec:scan} the details of the
scan in the NMSSM parameter space are given. Our results are shown
in section~\ref{sec:analysis}. After presenting the parameter and mass
distributions in subsections~\ref{subsec:pardistr} and
\ref{subsec:massdistr}, the NMSSM Higgs signal rates in various final
states will be discussed in subsection~\ref{subsec:signalrates}. 
Section~\ref{sec:naturalnmssm} is devoted to the analysis of
the natural NMSSM. In Section~\ref{sec:scenarios} we present
several benchmark scenarios that feature Higgs-to-Higgs decays. 
Section~\ref{sec:concl} summarises and concludes the
paper. 

\section{The NMSSM Lagrangian \label{sec:nmssm}}
The NMSSM differs from the MSSM in the superpotential and the
soft SUSY breaking Lagrangian. In terms of the (hatted) superfields and
including only the third generation fermions, the scale invariant
NMSSM superpotential reads
\beq
{\cal W} = \lambda \widehat{S} \widehat{H}_u \widehat{H}_d +
\frac{\kappa}{3} \, \widehat{S}^3 + h_t
\widehat{Q}_3\widehat{H}_u\widehat{t}_R^c - h_b \widehat{Q}_3
\widehat{H}_d\widehat{b}_R^c  - h_\tau \widehat{L}_3 \widehat{H}_d
\widehat{\tau}_R^c \; .
\label{eq:superpot}
\eeq
The first term replaces the $\mu$-term $\mu \hat{H}_u \hat{H}_d$ of
the MSSM superpotential. The second term, which is cubic in the
singlet superfield, breaks the Peccei-Quinn symmetry \cite{pqsymm}, so that
no massless axion can appear. The last three terms are the Yukawa
interactions. The soft SUSY breaking Lagrangian receives contributions
from the scalar mass parameters for the Higgs and sfermion fields,
which in terms of the fields corresponding to the complex scalar
components of the superfields read
\beq
\label{eq:Lmass}
 -{\cal L}_{\mathrm{mass}} &=&
 m_{H_u}^2 | H_u |^2 + m_{H_d}^2 | H_d|^2 + m_{S}^2| S |^2 \non \\
  &+& m_{{\tilde Q}_3}^2|{\tilde Q}_3^2| + m_{\tilde t_R}^2 |{\tilde t}_R^2|
 +  m_{\tilde b_R}^2|{\tilde b}_R^2| +m_{{\tilde L}_3}^2|{\tilde L}_3^2| +
 m_{\tilde  \tau_R}^2|{\tilde \tau}_R^2|\; .
\eeq
The Lagrangian including the trilinear soft SUSY breaking interactions
between the sfermions and Higgs fields is given by
\beq
\label{eq:Trimass}
-{\cal L}_{\mathrm{tril}}=  \lambda A_\lambda H_u H_d S + \frac{1}{3}
\kappa  A_\kappa S^3 + h_t A_t \tilde Q_3 H_u \tilde t_R^c - h_b A_b
\tilde Q_3 H_d \tilde b_R^c - h_\tau A_\tau \tilde L_3 H_d \tilde \tau_R^c
+ \mathrm{h.c.}
\eeq
The soft SUSY breaking Lagrangian with the gaugino mass parameters
finally reads
\beq
-{\cal L}_\mathrm{gauginos}= \frac{1}{2} \bigg[ M_1 \tilde{B}
\tilde{B}+M_2 \sum_{a=1}^3 \tilde{W}^a \tilde{W}_a +
M_3 \sum_{a=1}^8 \tilde{G}^a \tilde{G}_a  \ + \ {\rm h.c.}
\bigg].
\eeq
We will work in the unconstrained NMSSM with non-universal soft terms
at the GUT scale. After EWSB the Higgs doublet and singlet fields
acquire non-vanishing vacuum expectation values (VEVs). The SUSY
breaking masses squared for $H_u$, $H_d$
and $S$ in ${\cal L}_{\mathrm{mass}}$ are traded for their tadpole parameters by
exploiting the three minimisation conditions of the scalar potential. \s

The NMSSM Higgs potential is obtained from the superpotential, the
soft SUSY breaking terms and the $D$-term contributions.
Expanding the Higgs fields about their VEVs $v_u$, $v_d$ and
$v_s$, which we choose to be real and positive,
\beq
H_d = \left( \begin{array}{c} (v_d + h_d + i a_d)/\sqrt{2} \\
   h_d^- \end{array} \right) \;, \quad
H_u = \left( \begin{array}{c} h_u^+ \\ (v_u + h_u + i a_u)/\sqrt{2}
 \end{array} \right) \;, \quad
S= \frac{v_s+h_s+ia_s}{\sqrt{2}} \;,
\eeq
the Higgs mass matrices for the three scalar, two pseudoscalar and  the
charged Higgs bosons are derived from the tree-level scalar
potential. The squared $3 \times 3$ mass matrix $M_S^2$ for the
CP-even Higgs fields is diagonalised through a rotation matrix ${\cal
  R}^S$ yielding the CP-even mass eigenstates $H_i$ ($i=1,2,3$),
\beq
(H_1, H_2, H_3)^T = {\cal R}^S (h_d,h_u,h_s)^T \;,
\label{eq:scalarrot}
\eeq
with the $H_i$ ordered by ascending mass, $M_{H_1} \le M_{H_2} \le
M_{H_3}$. The CP-odd mass eigenstates $A_1$, $A_2$ and the massless
Goldstone boson $G$ are obtained by consecutively applying a rotation
${\cal R}^G$ to separate $G$, followed by a rotation ${\cal R}^P$ to
obtain the mass eigenstates
\beq
(A_1,A_2,G)^T = {\cal R}^P {\cal R}^G (a_d,a_u,a_s)^T \;,
\label{eq:pseudorot}
\eeq
which are ordered such that $M_{A_1} \le M_{A_2}$.  \s

At tree-level the NMSSM Higgs
sector can be parameterised by the six parameters
\beq
\lambda\ , \ \kappa\ , \ A_{\lambda} \ , \ A_{\kappa}, \
\tan \beta =\langle H_u^0 \rangle / \langle H_d^0 \rangle \quad \mathrm{and}
\quad \mu_\mathrm{eff} = \lambda \langle S \rangle\; .
\eeq
The brackets around the fields denote the corresponding VEVs of the
neutral components of the Higgs fields. The sign
conventions for $\lambda$ and $\tan\beta$ are chosen such that they
are positive, while $\kappa$, $A_\lambda$, $A_\kappa$ and
$\mu_{\mathrm{eff}}$ can have both signs. Including the important
higher order corrections, also the soft SUSY breaking mass terms for
the scalars and the gauginos as well as the trilinear soft SUSY
breaking couplings have to be taken into account.

\section{The NMSSM Parameter Scan \label{sec:scan}}
We perform a scan in the NMSSM parameter space with the aim of finding
scenarios that are compatible with the LHC Higgs search results and
which lead to Higgs spectra that can be tested at the LHC with high
c.m.~energy, $\sqrt{s} = 13$~TeV. We demand them to
contain at least one scalar Higgs boson with mass value around 125~GeV and
rates that are compatible with those reported by the LHC experiments
ATLAS and CMS. \s

We used the program package {\tt NMSSMTools}
\cite{nmhdecay1,nmssmtools} for the calculation of the SUSY particle
and NMSSM Higgs boson spectrum and branching ratios. The higher
order corrections to the NMSSM Higgs boson masses
\cite{effpot,leadlog,Degrassi:2009yq,full1loop,Ender:2011qh} have been
incorporated in {\tt NMSSMTools} up to ${\cal O}(\alpha_t \alpha_s +
\alpha_b \alpha_s)$ for vanishing external momentum. The Higgs decays
widths and branching ratios are
obtained from {\tt NMHDECAY} \cite{nmhdecay1}, an NMSSM extension of
the Fortran code {\tt HDECAY} \cite{hdecay,susyhit}. The SUSY particle
branching ratios are obtained from the Fortran code {\tt NMSDECAY}
\cite{nmsdecay} which is a generalisation of the Fortran code {\tt
  SDECAY} \cite{susyhit,sdecay} to the NMSSM particle spectrum. The NMSSM
particle spectrum, mixing angles, decay widths and branching ratios
are given out in the SUSY Les Houches Accord (SLHA) format
\cite{slha}. For various parameter sets we have cross-checked the
NMSSM Higgs branching ratios against the ones obtained with the
recently released Fortran package {\tt NMSSMCALC}
\cite{Baglio:2013iia}. Differences arise in the treatment of the radiative
corrections to the Higgs boson masses and in the more sophisticated
and up-to-date inclusion of the dominant higher order corrections to
the decay widths as well as the consideration of off-shell effects in
{\tt NMSSMCALC}. The overall picture, however, remains unchanged. \s

We have performed the scan over a large fraction of the NMSSM
parameter space in order to get a general view of the NMSSM Higgs
boson phenomenology. For the mixing angle $\tan\beta$
and the NMSSM couplings $\lambda$ and $\kappa$ the following ranges
have been considered,
\beq
1 \le \tan\beta \le 30 \; , \qquad 0 \le \lambda \le 0.7 \; , \qquad
-0.7 \le \kappa \le 0.7 \;.
\label{eq:cond1}
\eeq
We have taken care of not violating perturbativity by applying
the rough constraint
\beq
\sqrt{\lambda^2 + \kappa^2} \le 0.7 \;. \label{eq:unitary}
\eeq
The soft SUSY breaking trilinear NMSSM couplings $A_\lambda$ and
$A_\kappa$ and the effective $\mu_{\text{eff}}$ parameter have been
varied in the ranges
\beq
-2 \mbox{ TeV} \le A_\lambda \le 2 \mbox{ TeV} \; , \;
-2 \mbox{ TeV} \le A_\kappa \le 2 \mbox{ TeV} \; , \;
-1 \mbox{ TeV} \le \mu_{\text{eff}} \le 1 \mbox{ TeV} \;.
\label{eq:cond2}
\eeq
Note, that too large positive values for $A_\kappa$ for negative
$\kappa$ lead to non self-consistent solutions, which have been discarded.
The parameter $A_\lambda$ is related to the charged
Higgs boson mass. The compatibility with the lower bound on the
charged Higgs mass has been checked for \cite{chargedhiggs}. 
In the choice of the remaining soft SUSY breaking
trilinear couplings and masses care has been taken to respect the
exclusion limits on the SUSY particle masses
\cite{susyexclusion1,susyexclusion1a,susyexclusion2}. The trilinear soft SUSY 
breaking couplings of the up- and down-type quarks and the charged
leptons, $A_U, A_D$ and $A_L$ with $U\equiv u,c,t, D\equiv d,s,b$ and
$L\equiv e,\mu,\tau$, are varied independently in the range
\beq
-2 \mbox{ TeV} \le A_U, A_D, A_L \le 2 \mbox{ TeV} \,.
\label{eq:cond3}
\eeq
The soft SUSY breaking right- and left-handed masses of
the third generation are
\beq
600 \mbox{ GeV} \le M_{\tilde{t}_R} = M_{\tilde{Q}_3} \le 3 \mbox{
 TeV} \; , \;
600 \mbox{ GeV} \le M_{\tilde{\tau}_R} = M_{\tilde{L}_3} \le 3 \mbox{
 TeV} \; , \;
M_{\tilde{b}_R} = 3 \mbox{ TeV} \;.
\label{eq:cond4}
\eeq
And for the first two generations we chose
\beq
M_{\tilde{u}_R,\tilde{c}_R} =
M_{\tilde{d}_R,\tilde{s}_R}=M_{\tilde{Q}_{1,2}}=M_{\tilde{e}_R,\tilde{\mu}_R}=
M_{\tilde{L}_{1,2}} = 3 \mbox{ TeV} \;.
\label{eq:cond5}
\eeq
The gaugino soft SUSY breaking masses finally are varied in the ranges
\beq
100 \mbox{ GeV} \le M_1 \le 1 \mbox{ TeV} \;, \;
200 \mbox{ GeV} \le M_2 \le 1 \mbox{ TeV} \;, \;
1.3 \mbox{ TeV} \le M_3 \le 3 \mbox{ TeV} \;.
\label{eq:cond6}
\eeq
Note, that in {\tt NMSSMTools} the
NMSSM-specific input parameters $\lambda,\kappa,A_\lambda$ and
$A_\kappa$ as well as all other soft SUSY breaking masses and
trilinear couplings, according to the SLHA format, are understood as
$\overline{\mbox{DR}}$ parameters taken at the SUSY scale
$\tilde{M}=1$~TeV, while $\tan\beta$ is taken at the mass of the $Z$
boson, $M_Z$. \s

The program package {\tt NMSSMTools} is interfaced
with {\tt micrOMEGAS} \cite{micromegas} so that the compatibility of
the relic abundance of the lightest neutralino as the NMSSM Dark
Matter candidate with the latest PLANCK results \cite{Ade:2013zuv} can
be checked for. The package furthermore tests for the constraints from
low-energy observables as well as from Tevatron and LEP. Details can be
found on the webpage of the program \cite{nmssmtools}.\footnote{Note
  that in our analysis we do not take into account the constraint from
  $g-2$, as it is non-trivial to find parameter combinations which can
  explain the 2$\sigma$ deviation from the SM value.} In addition we
have included in {\tt 
  NMSSMTools} the latest LHC Higgs exclusion limits, given in
Refs.~\cite{exclgamgam,exclzz,exclww,exclbb,excltautau,exclzgam,exclmumu}.
Among the parameter points, that survive the constraints incorporated
in {\tt NMSSMTools}, only those are kept that feature an NMSSM Higgs
spectrum with at least one CP-even Higgs boson in the range of 124~GeV
to 127~GeV, which is fulfilled by either $H_1$ or $H_2$, and which will
be denoted by $h$ in the following. In some cases
$H_1$ and $H_2$ are almost degenerate with a mass value near
125~GeV. In this case the signal rates observed at the LHC are built
up by two Higgs bosons. Applying the narrow width approximation, the
reduced signal rate $\mu_{XX}$ into a final state
particle pair $XX$ is given by the production cross section
$\sigma_{\text{prod}}$ of the NMSSM Higgs boson $H_i$ times its
branching ratio $BR$ into the final state $XX$, normalised to the
corresponding SM values for a SM Higgs boson $H^{\text{SM}}$, {\it
  i.e.}
\beq
\mu_{XX} (H_i) = \frac{\sigma_{\text{prod}}
  (H_i) BR(H_i \to XX)}{\sigma_{\text{prod}} (H^{\text{SM}}) BR
  (H^{\text{SM}} \to XX)} \equiv R_{\sigma_{\text{prod}}} (H_i) \,
R_{XX}^{BR} (H_i) \; , \label{eq:ratedef}
\eeq
with
\beq
M_{H^{\text{SM}}} = M_{H_i} \equiv M_h =
124-127\mbox{ GeV} \;.
\eeq
In Eq.~(\ref{eq:ratedef}) we have introduced the ratio  of production
cross sections
\beq
R_{\sigma_{\text{prod}}} (H_i) \equiv \frac{\sigma_{\text{prod}}
  (H_i) }{\sigma_{\text{prod}} (H^{\text{SM}})}
\eeq
and the ratio of branching ratios
\beq
R_{XX}^{BR} (H_i) \equiv \frac{BR(H_i \to XX)}{BR
  (H^{\text{SM}} \to XX)} \;.
\eeq
If not stated otherwise we approximate the inclusive cross section by the
dominant gluon fusion production cross section. In the SM the loop
induced process is mediated by top and bottom quark loops and known at
next-to-leading order (NLO) QCD including the full mass dependence
\cite{Spira:1995rr}. The next-to-next-to-leading order (NNLO) QCD
corrections have been calculated in the heavy top mass
approximation \cite{nnlo}. The latter is valid to better than 1\% for
Higgs masses below 300~GeV \cite{masseffects}. The soft gluon
resummation \cite{softresum,softresum1} and partial, respectively
approximate, next-to-next-to-next-to-leading order (N$^3$LO) results
have been presented in \cite{softresum1,n3lo}. In the NMSSM, gluon
fusion production additionally contains stop and sbottom quark
loops, with the squark loops becoming particularly important for
squark masses
below $\sim 400$~GeV \cite{squarkeffects}. The cross section can be
adapted from the corresponding result
in the MSSM \cite{Spira:1995rr,fortschr}, by replacing the respective
MSSM Higgs-quark-quark and Higgs-squark-squark couplings with the
NMSSM couplings. As we use in our scan the program package {\tt
  NMSSMTools}, which provides the ratio of the NMSSM Higgs boson decay
width into a gluon pair at NLO QCD with respect to the 
NLO QCD decay width into $gg$ of a SM Higgs boson with same
mass, we approximate the gluon fusion production cross section for a CP-even
NMSSM Higgs boson $H_i$ by
\beq
\sigma^{\text{NMSSM}}_{gg \to H_i} =
\frac{\Gamma^{\text{NMSSM}}_{\text{NLO}} (H_i \to
  gg)}{\Gamma^{\text{SM}}_{\text{NLO}} (H^{\text{SM}} \to gg)} \times
\sigma^{\text{SM}}_{gg \to H^{\text{SM}}} \;, \quad \mbox{with } \; M_{H_i} = M_{H^{\text{SM}}} \;. \label{eq:ggfuscxn}
\eeq
The SM gluon fusion cross section has been calculated at NNLO QCD for
$M_{H^{\text{SM}}}=126$~GeV with the Fortran program {\tt HIGLU}
\cite{higlu}.\footnote{For comparison we also calculated for some
  parameter points the NMSSM
  Higgs production cross sections at NNLO with {\tt HIGLU}. 
  For the SM-like Higgs boson the approximation
  Eq.~(\ref{eq:ggfuscxn}) works better than 1\%. For the heavy
  MSSM-like Higgs bosons there can be deviations of up to ${\cal
    O}(10\%)$, and for the light singlet-like Higgs bosons they can reach
  the level of ${\cal O}(20\%)$ for some scenarios. For our purposes
  the approximation is good enough.} The cross section changes by
less than 5\% in the 
range $124\mbox{ GeV } \lsim M_{H^{\text{SM}}} \lsim 127$~GeV.
In {\tt NMSSMTools} the gluon decay width does not include
mass effects at NLO QCD. The higher order electroweak (EW) corrections
\cite{ggew} are of ${\cal O}$(5\%) in the SM. As they are not
available for the NMSSM and cannot be obtained by simple coupling
modifications, we consistently neglect them. \s

In case the signal is built up by the superposition of the rates from
the 125~GeV $h$ boson and another Higgs boson $\Phi=H_i,A_j$, which is
almost degenerate, we define the reduced signal strength as
\beq
\mu_{XX} (h) \equiv R_{\sigma_{\text{prod}}} (h) \, R_{XX}^{BR} (h)
\;\; + \hspace*{-0.4cm}
\sum_{\scriptsize \begin{array}{c} \Phi\ne h \\ |M_{\Phi}\!-\!M_h| \le
  \delta \end{array}} \hspace*{-0.4cm} R_{\sigma_{\text{prod}}} (\Phi)
\, R_{XX}^{BR} (\Phi) \, F(M_h, M_\Phi, d_{XX}) \;. \label{eq:mudef}
\eeq
Here $\delta$ denotes the mass resolution in the $XX$ final state and
$F(M_h, M_\Phi, d_{XX})$ the Gaussian weighting function as
implemented in {\tt NMSSMTools}. The experimental resolution $d_{XX}$ of the
channel $XX$ influences the width of the weighting function. In the
following only those parameter values are retained that lead to
reduced rates according to Eq.~(\ref{eq:mudef}) into the
$bb,\tau\tau,\gamma\gamma,WW$ and $ZZ$ final states, that are within 2
times the 1$\sigma$ interval around the respective best fit value, as
reported by ATLAS and CMS. Here we have combined the signal rates and errors,
given in Refs.~\cite{atlasnotemu} and \cite{cmsnotemu},
of the two experiments according to Eq.~(5) in
\cite{Espinosa:2012vu}. The combined signal rates and errors are given
in Table~\ref{tab:comb}.\footnote{In order to check the compatibility
  in the $b\bar{b}$ final state we have replaced gluon fusion
  production with production in association with $W,Z$.} Note, that in
the following we will use the shorthand notation $2\times 1\sigma$ to
indicate the interval around the measured rate, that we allow for. 
\begin{table}[!h]
  \centering
  \begin{tabular}{|c||c|c|}
    \hline
channel & best fit value & $2\times 1\sigma$ error \\ \hline
$VH \to Vbb$ & 0.97 & $\pm 1.06$ \\ \hline
$H\to \tau\tau$ & 1.02 & $\pm 0.7$
\\ \hline
$H\to \gamma\gamma$ & 1.14  & $\pm 0.4$
\\ \hline
$H \to WW$ & 0.78 & $\pm 0.34$
\\ \hline
$H\to ZZ$ & 1.11 & $\pm 0.46$
\\ \hline
\end{tabular}
\caption{The ATLAS and CMS combined signal rates and errors for the
  $bb,\tau\tau,\gamma\gamma,WW$ and $ZZ$ final states. Apart from the $bb$
  final state, where Higgs-strahlung $VH$ ($V=W,Z$) is the production
  channel, they are based on the inclusive production cross
  section. Details can be found in Refs.~\cite{atlasnotemu} and
  \cite{cmsnotemu}.}
\label{tab:comb}
\end{table}
In summary, in addition to the implemented restrictions in
{\tt NMSSMTools}, the conditions on our parameter scan are
\beq
&&\hspace*{-1.6cm}
\mbox{\underline{Conditions on the parameter scan:} } \nonumber \\[0.2cm]
&&\hspace*{-1.5cm}\begin{array}{ll}
\mbox{At least one CP-even Higgs boson $H_i\equiv h$ with: } & 124 \mbox{ GeV }
\lsim M_{h} \lsim 127 \mbox{ GeV } \\[0.2cm]
\mbox{Compatibility with $\mu_{XX}^{\text{exp}}$
  ($X=b,\tau,\gamma,W,Z$): } &
|\mu_{XX}^{\text{scan}} (h) - \mu_{XX}^{\text{exp}}| \le 2 \times 1\sigma
\end{array}
\label{eq:cond}
\eeq
In our case we have $H_i =H_1$ or $H_2$, and $\mu_{XX}^{\text{exp}}$
and $2 \times 1\sigma$ as given in Table~\ref{tab:comb}. \s

For all scenarios we require that they lead to relic densities
$\Omega_c h^2$ that are not larger than the result given by PLANCK
\cite{Ade:2013zuv},
\beq
\Omega_c h^2 = 0.1187 \pm 0.0017 \;.
\eeq
While this can be achieved for a large fraction of parameter points,
there are only a few points that reproduce the relic density within
the given errors. This is to be expected, however, in
view of the remarkably small error on the relic density reported by
PLANCK.

\section{Results \label{sec:analysis}}
In this section we first present the general features of the scenarios
as the outcome of our scan that survive all imposed
constraints. Subsequently the parameter sets shall be
investigated in more detail with respect to their prospects of discovering NMSSM
Higgs bosons or testing the coupling structure of the Higgs sector.

\subsection{NMSSM and Stop Sector Parameter Distributions \label{subsec:pardistr}}
\begin{figure}[ht]
\begin{center}
\includegraphics[width=7.9cm]{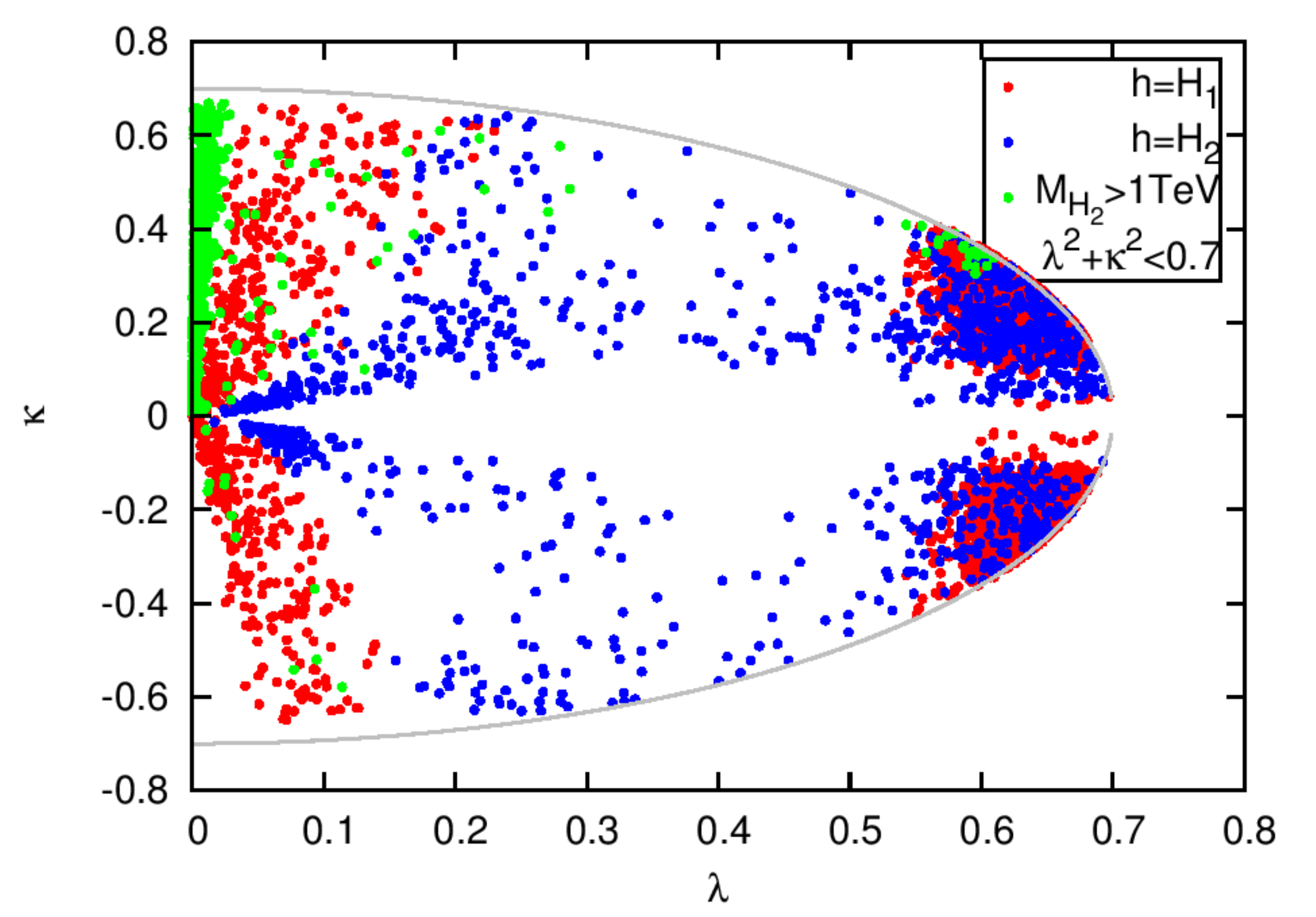}
\includegraphics[width=7.9cm]{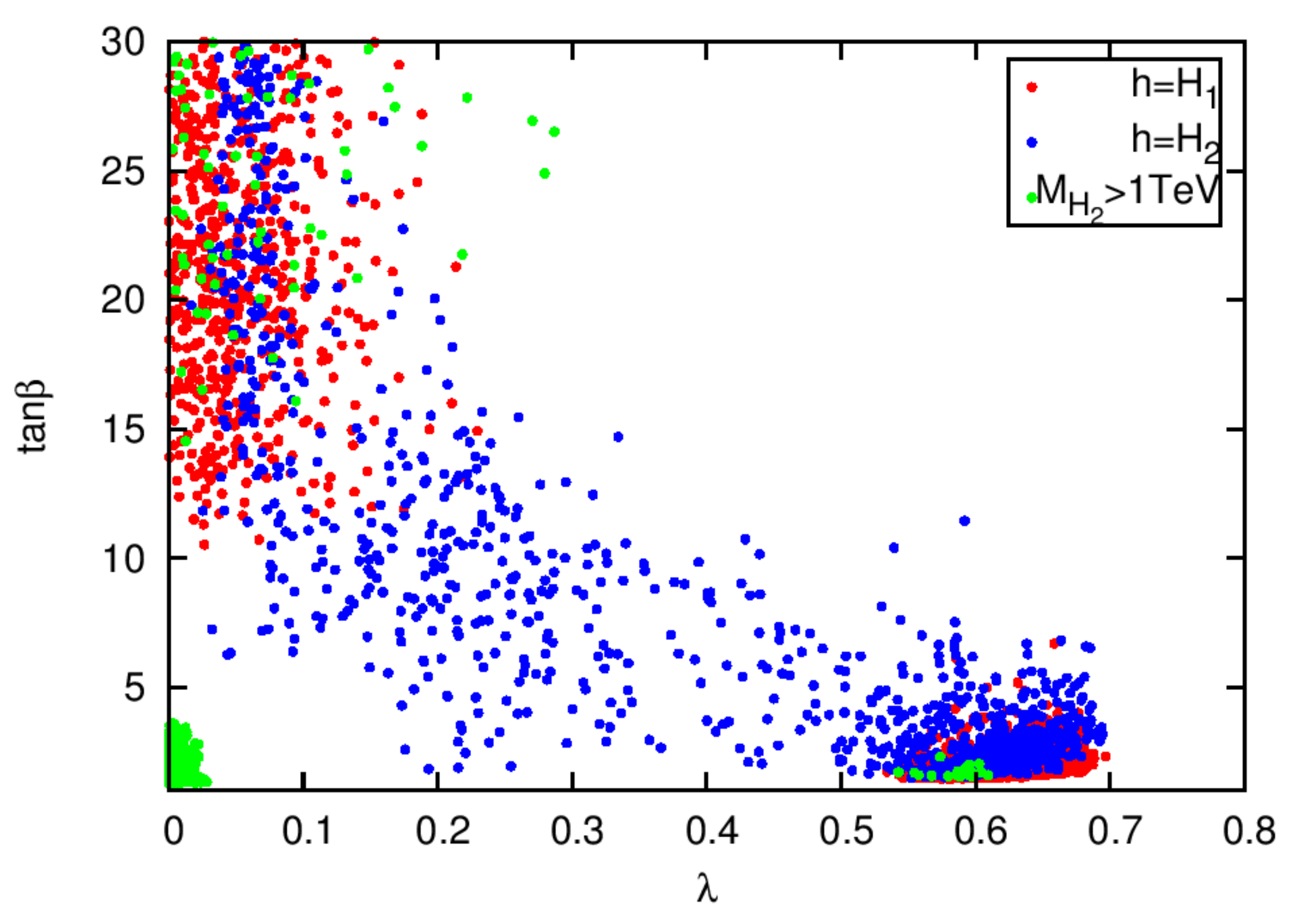}
\caption{Parameter distributions in the $\kappa-\lambda$ plane (left)
  and in the $\tan\beta-\lambda$ plane (right). The SM-like Higgs boson
  $h$ corresponds to either $H_1$ (red points) or $H_2$ (blue
  points). Green points correspond to $h \equiv H_1$ and $H_2$ mass
  values above 1 TeV. \label{fig:lamkaptanb}}
\end{center}
\end{figure}

Figure~\ref{fig:lamkaptanb} (left) shows the distribution of the
$\lambda$ and $\kappa$ values for the scenarios resulting from our
scan. The right figure shows the viable values in the
$\tan\beta-\lambda$ plane. The particular shape of the
$\kappa-\lambda$ distribution is the result of the requirement
Eq.~(\ref{eq:unitary}). As can be inferred from the figures, in the
scenarios passing the constraints either the lightest Higgs boson
$H_1$ (red points) or the second lightest $H_2$ (blue points) is the
SM-like Higgs boson $h$ with mass around 125~GeV. While the $\lambda$
and $\kappa$ values cover the whole allowed region, most scenarios are
found for either low ($\lsim 0.1$) or high ($\gsim 0.55$) $\lambda$
values. In the low-$\lambda$ region in particular $\kappa$ values
close to 0 lead to the $H_2$ being the SM-like Higgs boson, higher $\kappa$
values imply the lightest scalar boson to have SM properties. For
small $\lambda$ values and $\kappa$ non-zero the singlet-like Higgs
boson is at tree-level already comparatively heavy, so that it
corresponds to the second lightest Higgs boson and $H_1$ is SM-like.  
If, however, $\kappa$ is also small the singlet mass becomes smaller than
125~GeV so that $H_2$ can be SM-like. Large $\lambda$ values allow
independently of $\kappa$ for either $H_1$ or $H_2$ being
SM-like. Note, however, that in this region, $\kappa$ values close to
zero are precluded\footnote{In the limit when $\kappa$ goes to zero
the mass of the lightest neutralino $m_{\tilde{\chi}^0_1}$, which is predominantly 
singlino, becomes much smaller than $M_Z$ and the couplings of this state to the 
SM particles and their superpartners tend to be negligibly small
leading to rather small  annihilation cross section for
$\tilde{\chi}^0_1\tilde{\chi}^0_1\to \mbox{SM particles}$. 
Since the dark matter density is inversely proportional to the annihilation cross 
section at the freeze-out temperature such a light neutralino state gives rise to 
a relic density that is typically substantially larger than its measured value. 
As a result $\kappa$ values close to zero are basically ruled out, unless there 
also exists a very light CP--even or CP--odd Higgs state with mass
$\approx 2 m_{\tilde{\chi}^0_1}$ that can facilitate lightest
neutralino annihilation (for a recent discussion see \cite{Ellwanger:2014dfa}).}. \s

Figure~\ref{fig:lamkaptanb} (right) shows the well known fact that in
the NMSSM large $\lambda$ values ($\sim 0.5-0.7$) in
conjunction with small $\tan\beta$ values below $\sim 5$ allow for the
lightest or next-to-lightest scalar Higgs to be SM-like around
125~GeV. The other parameter combination leading to scenarios
compatible with the Higgs data, though less in number, is the
combination of $\lambda \lsim 0.1$ and $\tan\beta \gsim 10$. This is
in accordance with the behaviour of the upper mass bound on the
MSSM-like light Higgs boson given by
\beq
m_h^2 \approx M_Z^2 \cos^2 2\beta + \frac{\lambda^2 v^2}{2} \sin^2
2 \beta + \Delta m_h^2 \;, \label{eq:upperbound}
\eeq
\begin{figure}[!b]
\hspace*{-0.1cm}
\includegraphics[width=7.9cm]{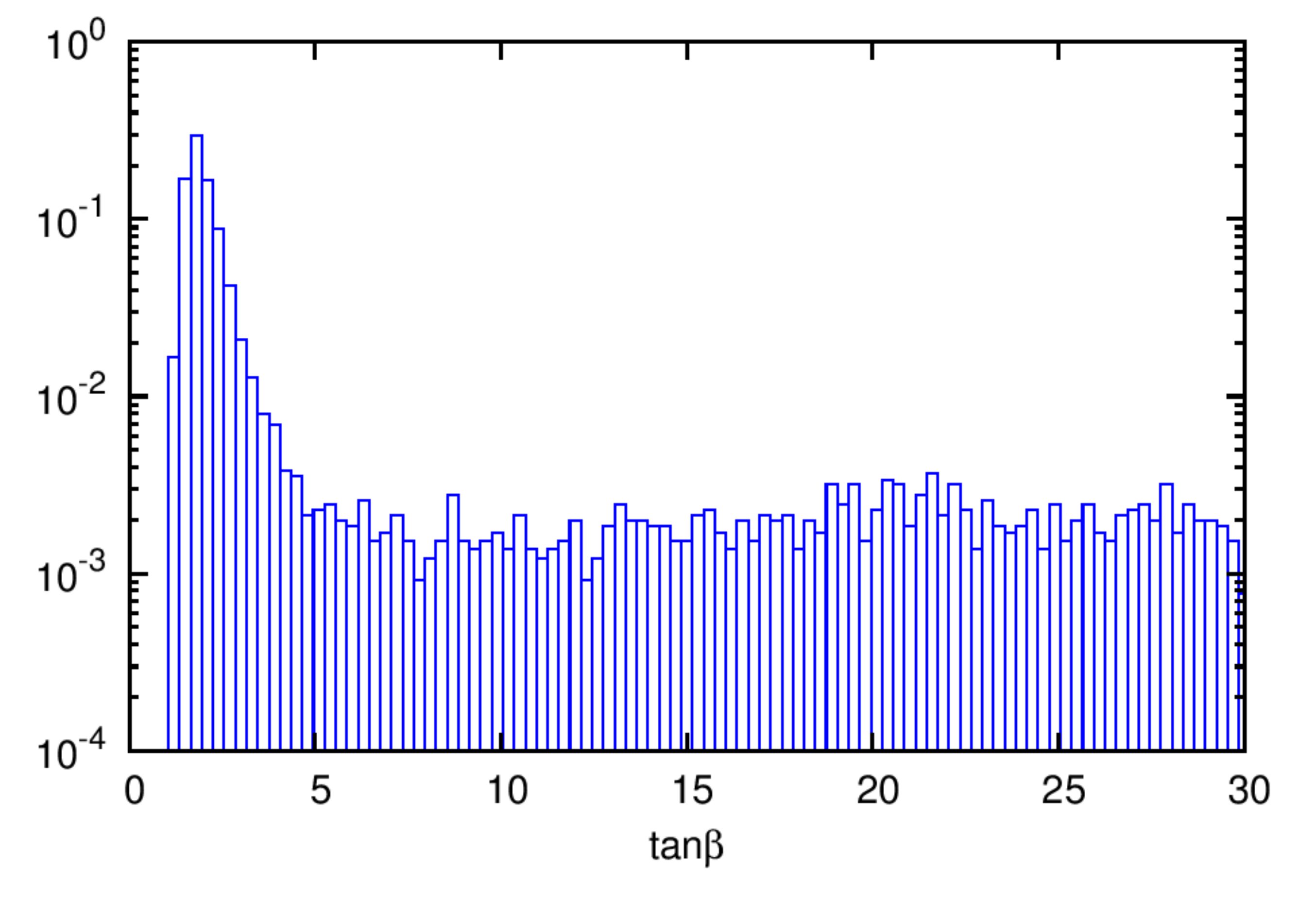} 
\includegraphics[width=7.9cm]{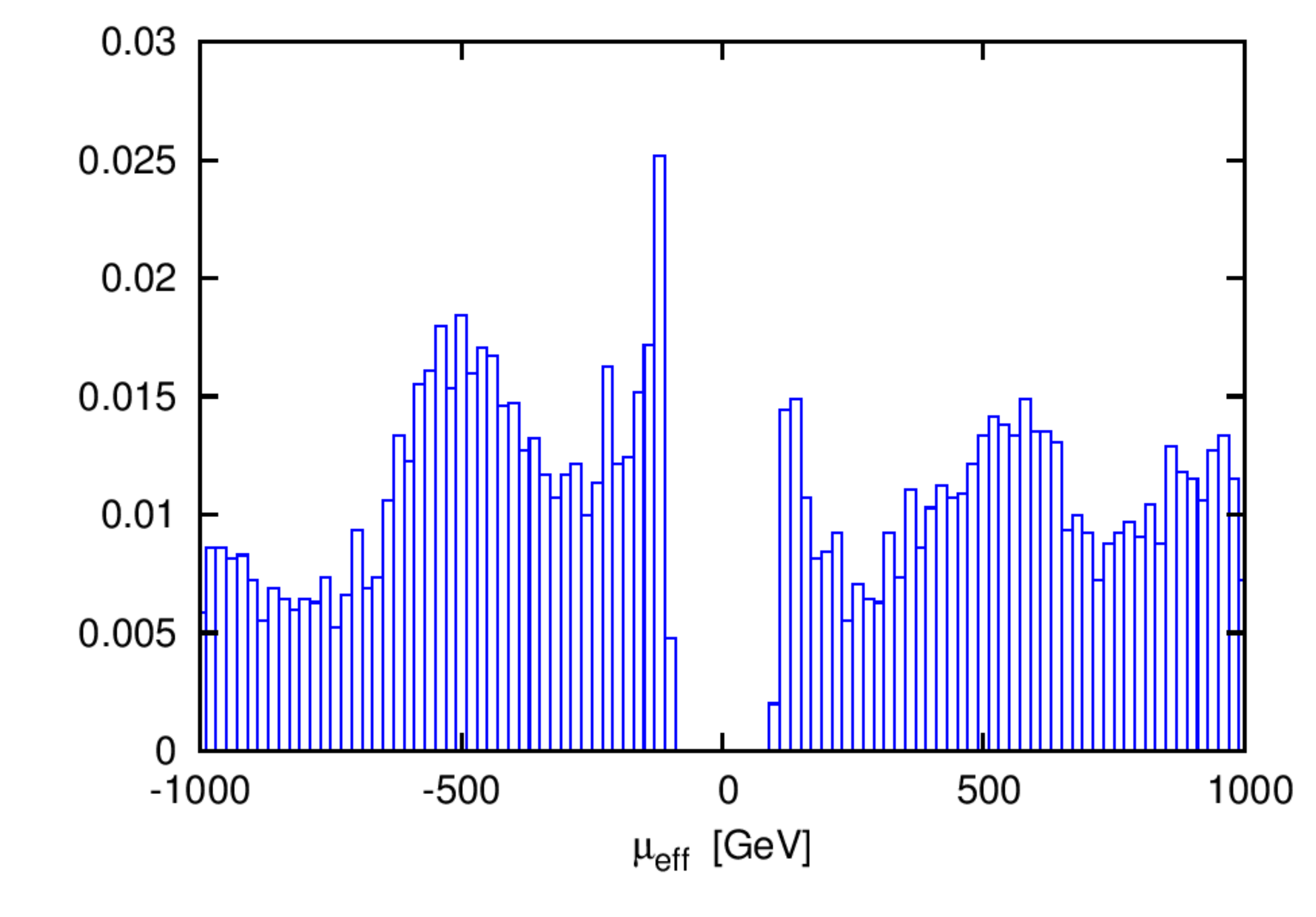}\\[-0.8cm]
\begin{center}
\caption{Distribution of $\tan\beta$ (left) and $\mu_{\text{eff}}$ (right). Normalised
  to the total number of parameter points of $\sim 8000$. 
\label{fig:tbdistr}}
\end{center}
\end{figure}
with $v\approx 246$~GeV, and
which is to be identified with the SM-like Higgs $h$ at 125~GeV after
the inclusion of the radiative corrections $\Delta m_h^2$. Also 
for intermediate $\lambda$ values viable scenarios can be found,
although, being away from the maximum of the function
Eq.~(\ref{eq:upperbound}), only for $h\equiv H_2$. For the lightest
CP-even Higgs in this case the tree-level mass bound is too low to be
shifted through radiative corrections to large enough mass values
compatible with all constraints imposed. \s

Note that in scenarios, where $M_{H_1} < 125$~GeV, this Higgs boson is
mostly singlet-like thus escaping the constraints from LEP, Tevatron
and LHC searches in this mass region. The strategies to search for
$H_1$ in this case shall be discussed in the next section. We also
found scenarios where the mass of the second lightest scalar $H_2$ is
larger than 1~TeV and can indeed become very large. This large
increase is caused by either small $\lambda$ values or large values
for $A_\lambda$ or $\mu_{\text{eff}}$. As these scenarios are
extremely fine-tuned we will discard them in the
following investigations from Fig.~\ref{fig:mhima1mass} on. \s

Figure~\ref{fig:tbdistr} shows the $\tan\beta$ and $\mu_{\text{eff}}$
distributions.  Note that for the $\tan\beta$ distribution we use a
logarithmic scale. Most
$\tan\beta$ values are clustered around $\tan\beta \approx 2$. 
The range of the effective $\mu_{\text{eff}}$ values is
$100\;\mbox{GeV} \lsim |\mu_{\text{eff}}| \lsim 1$~TeV.
Absolute $\mu_{\text{eff}}$ values of less than $\sim 100$~GeV are excluded
due to the lower bounds on the chargino masses. \s

\begin{figure}[h]
\begin{center}
\includegraphics[width=7.9cm]{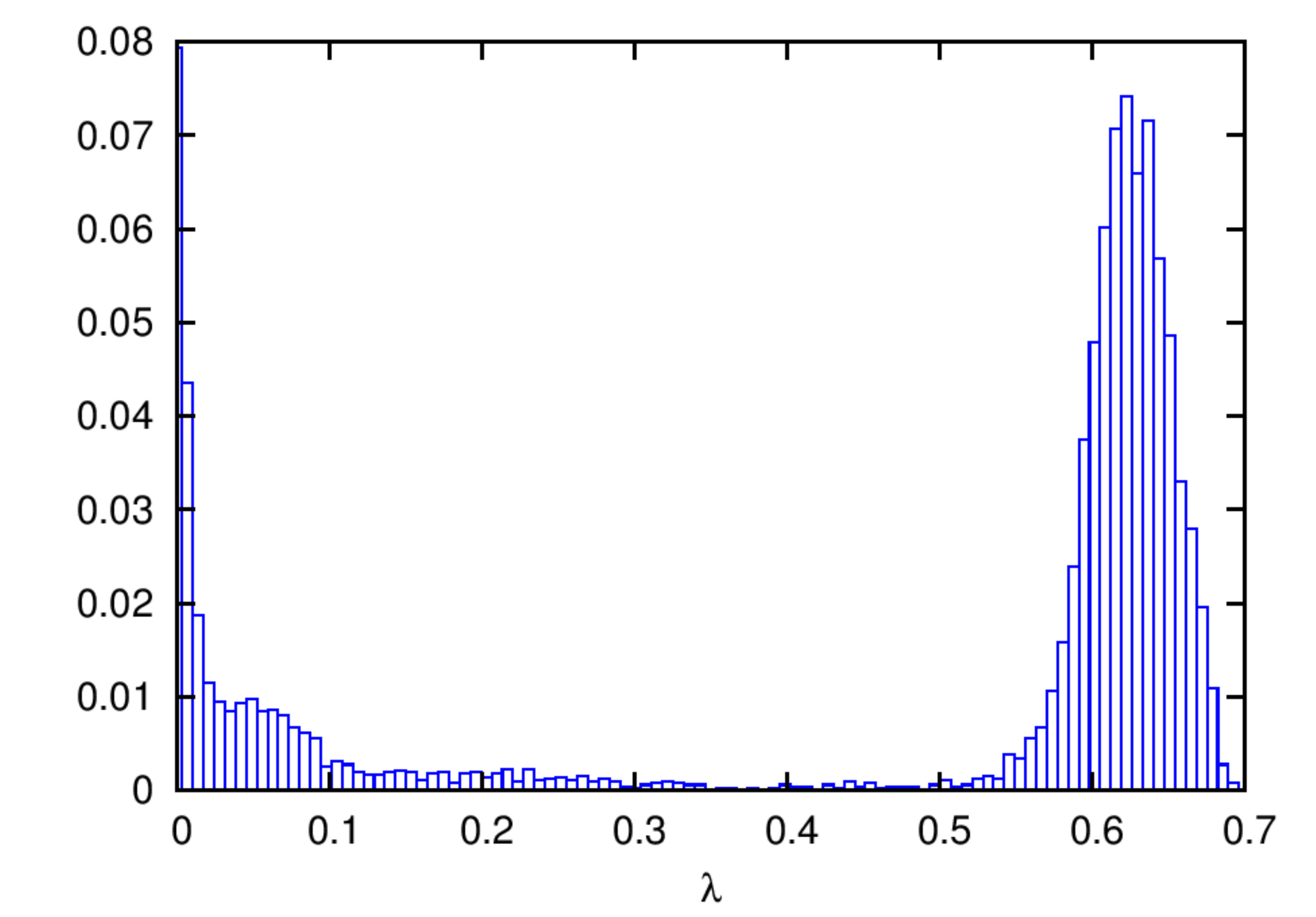}
\includegraphics[width=7.9cm]{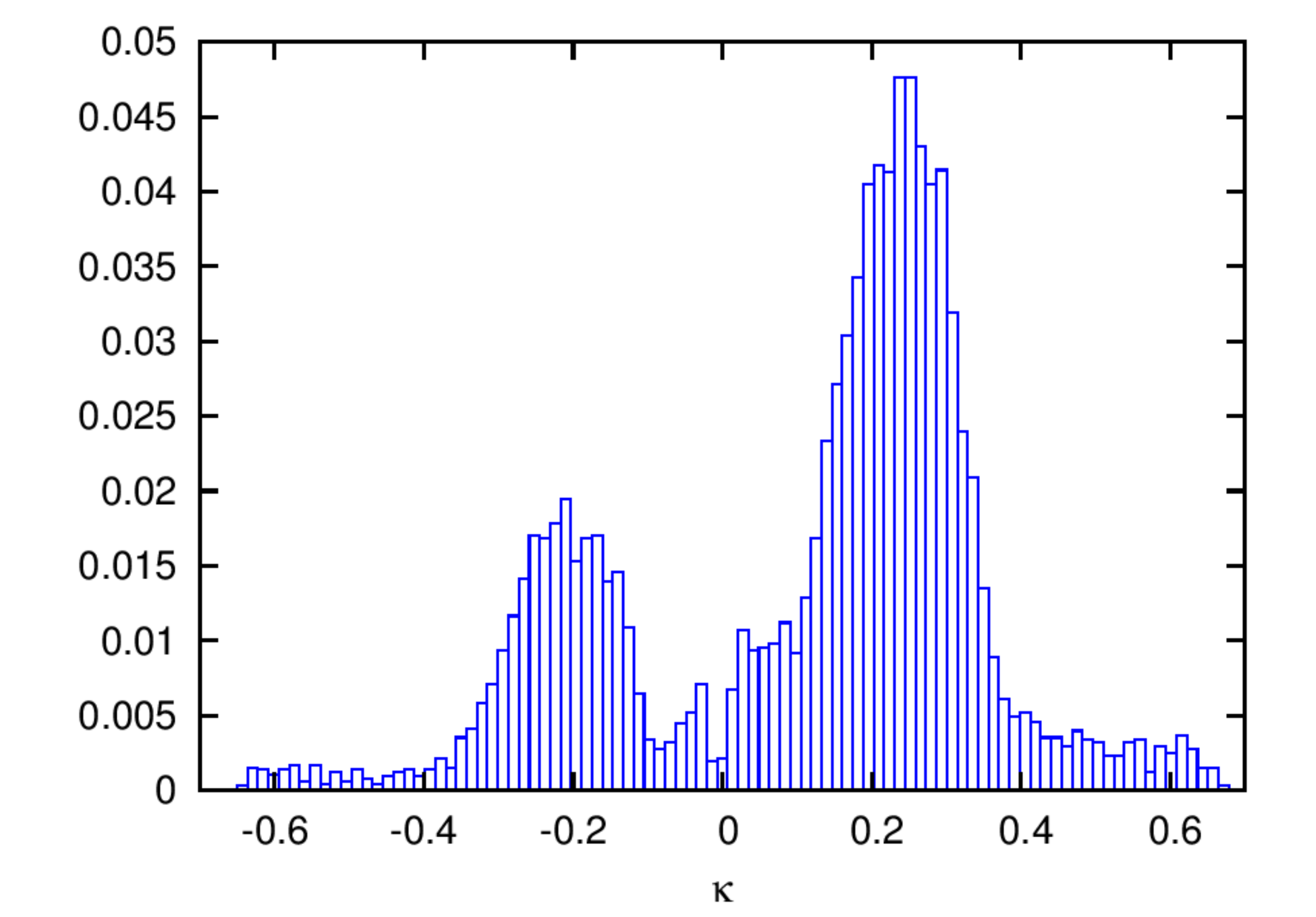}\\[0.2cm]
\includegraphics[width=7.9cm]{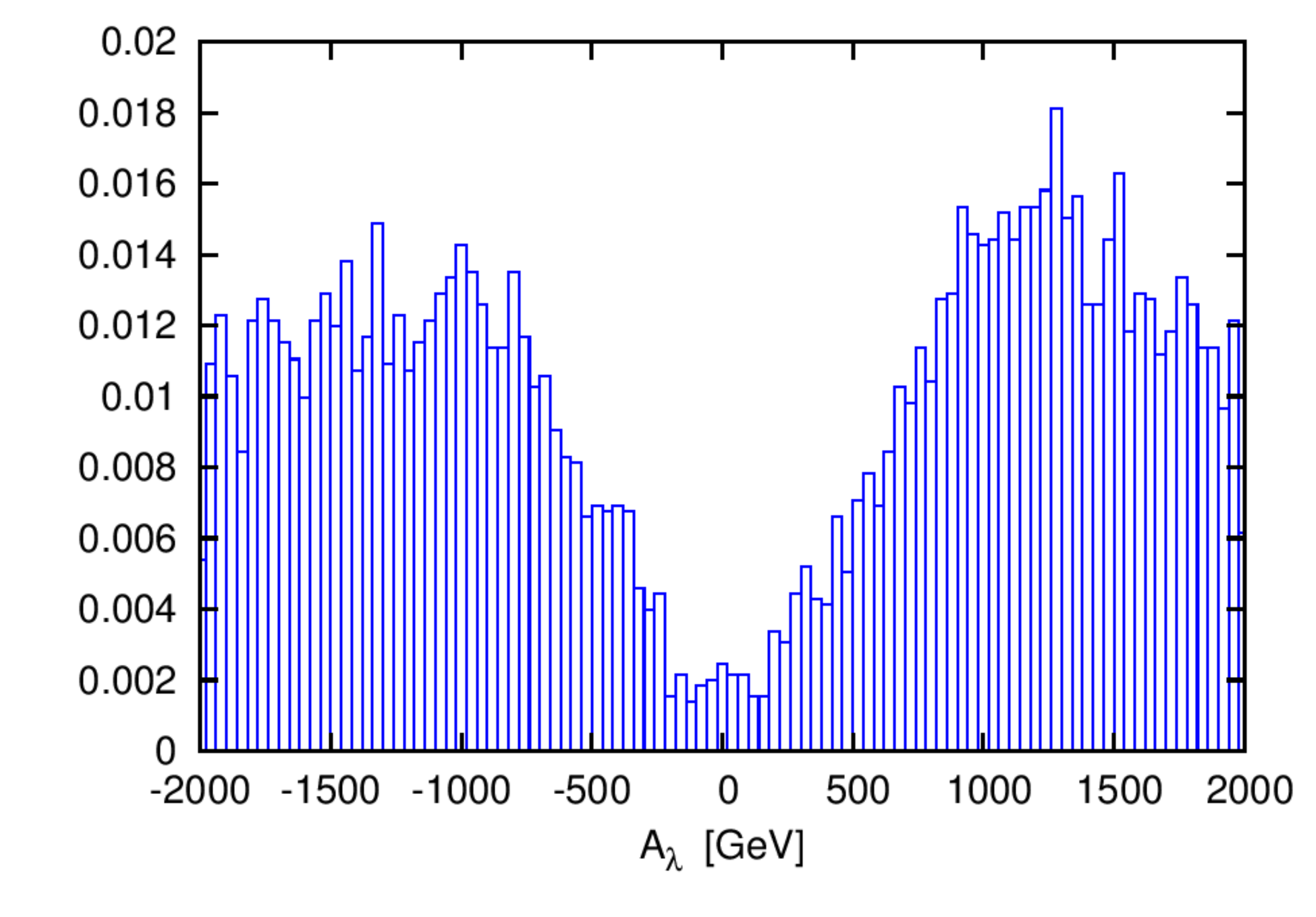}
\includegraphics[width=7.9cm]{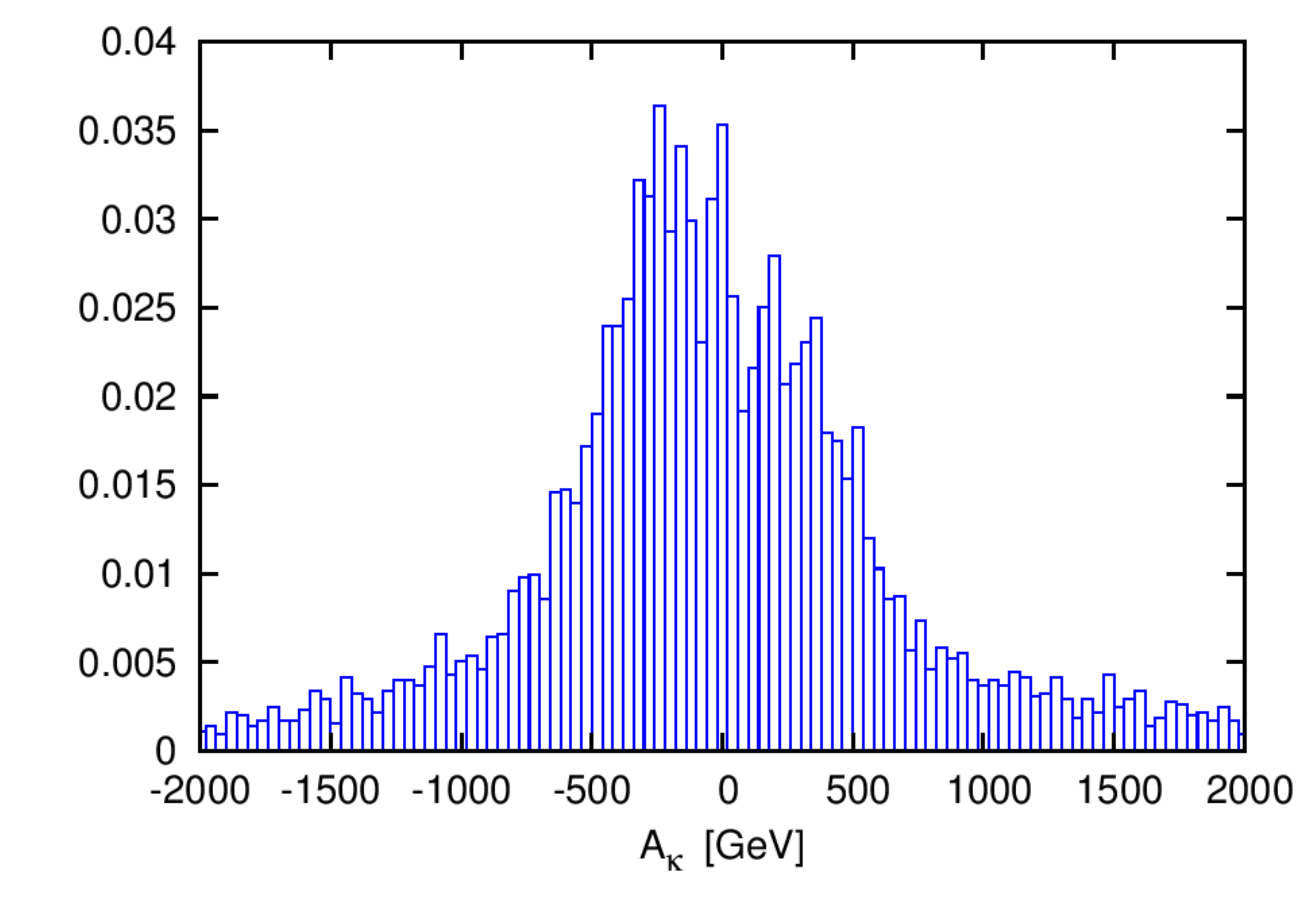}\\
\caption{Upper: Distribution of $\lambda$ (left) and $\kappa$ values
  (right). Lower: Distribution of $A_\lambda$ (left) and $A_\kappa$
  values (right). Normalised to the total number of parameter points
  of $\sim 8000$.
\label{fig:nmssmpardistr}}
\end{center}
\end{figure}
\begin{figure}[h]
\begin{center}
\includegraphics[width=7.9cm]{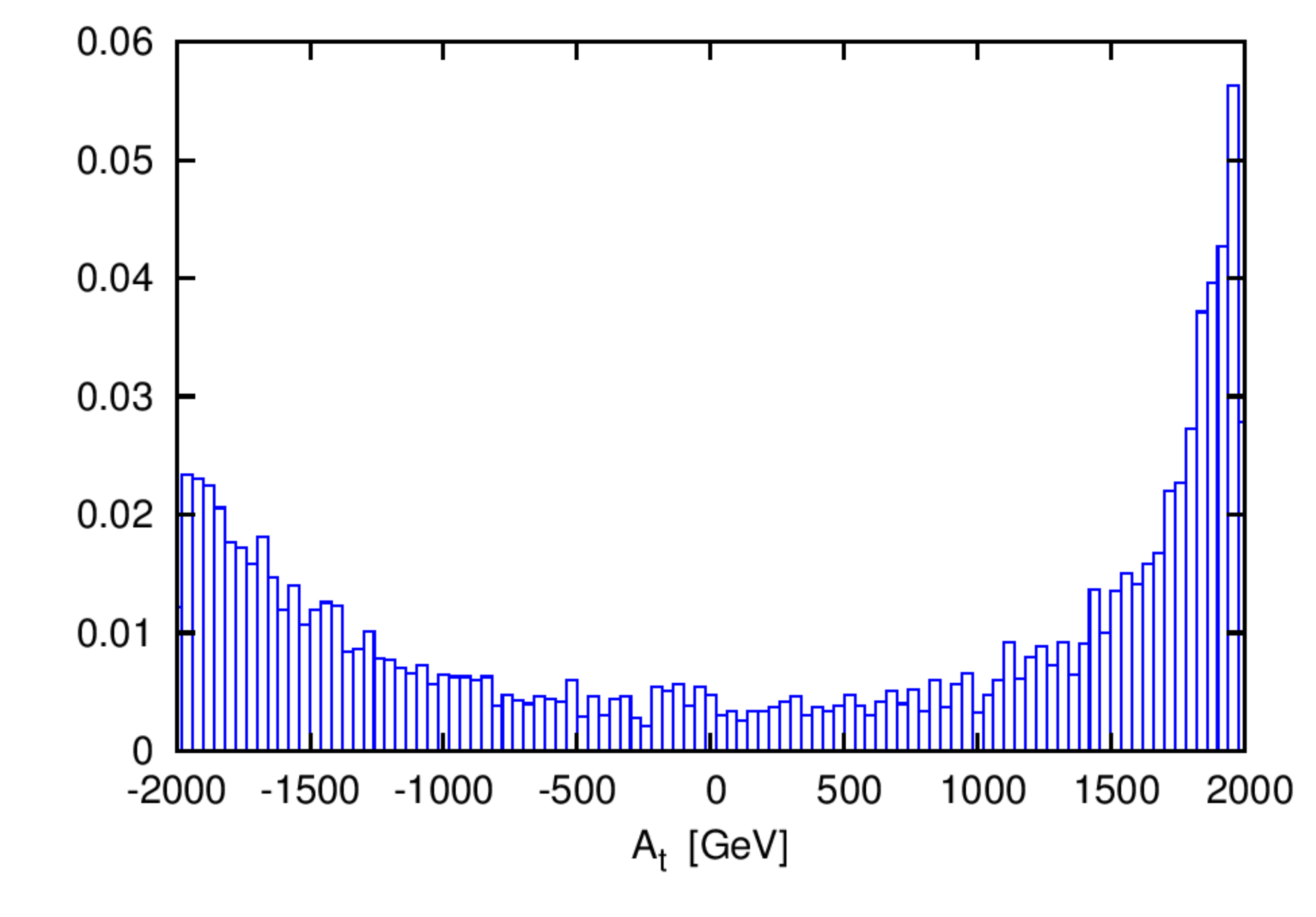}
\includegraphics[width=7.9cm]{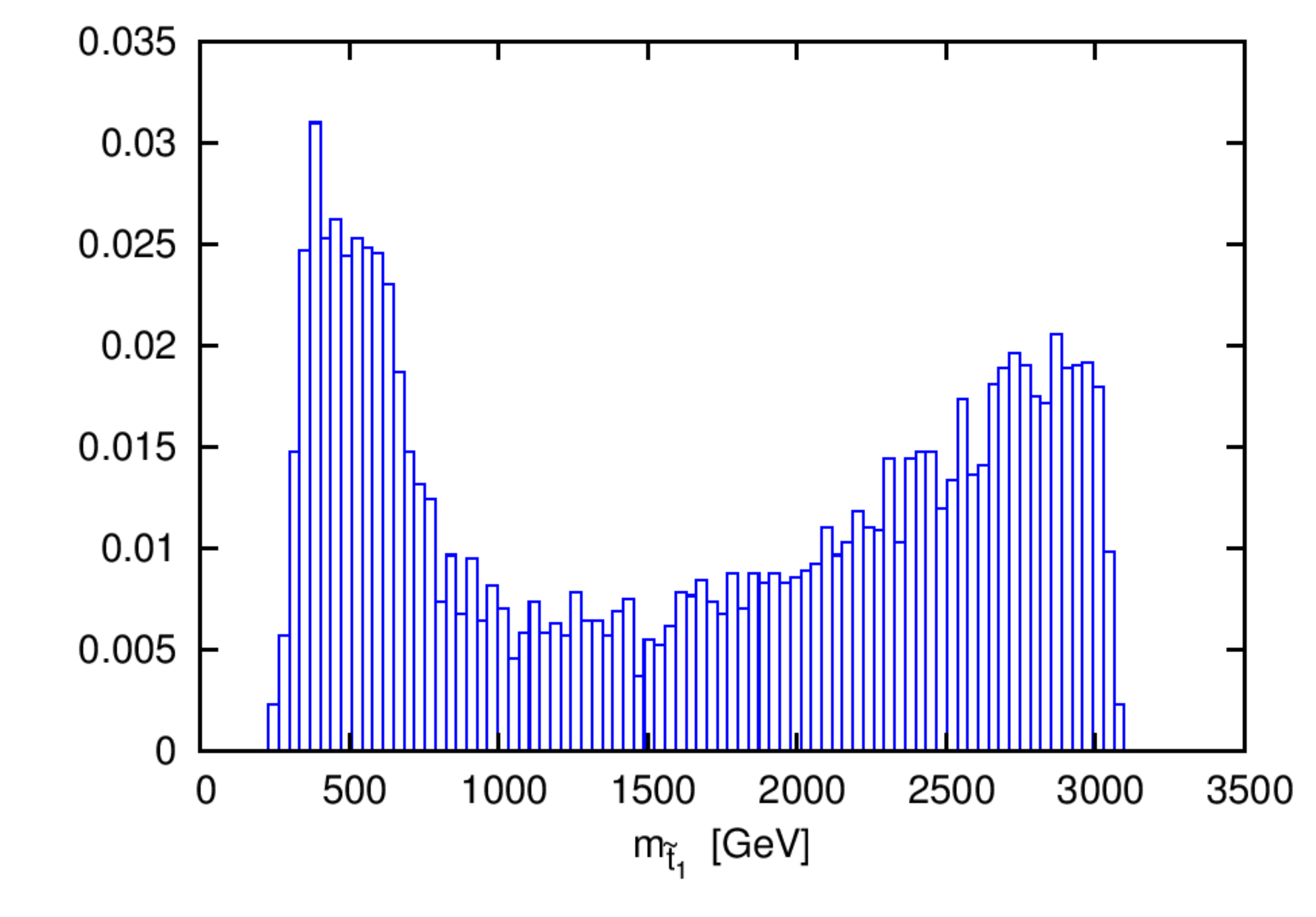}\\
\caption{Distribution of $A_t$ (left) and $m_{\tilde{t}_1}$
 (right), normalised to the total number of parameter points
 of $\sim 8000$.}
\label{fig:stopsecdistr}
\end{center}
\end{figure}
The distributions of the NMSSM parameters $\lambda$, $\kappa$,
$A_\lambda$ and $A_\kappa$ normalised to the total number of parameter
points of about 8000 are shown in Fig.~\ref{fig:nmssmpardistr}. The
$\lambda$ values cluster close to 0 and around 0.65, the preferred $\kappa$
values are close to $-0.25$ and $+0.25$. The preferred values of $A_\kappa$
around 0 can be understood by the fact that this is the range of
$A_\kappa$ allowing for the lightest scalar and pseudoscalar
masses squared to be positive.\footnote{This range has been derived in
  \cite{Nevzorov:2004ge} applying approximate tree-level mass formulae.}
For $A_\lambda$ larger values
are preferred, as these lead to sufficiently heavy Higgs bosons
\cite{Nevzorov:2004ge}, that are not excluded by the experiment and
entail a heavy enough SM-like Higgs boson. All these
considerations of course apply modulo the higher order corrections which are
indispensable to shift the SM-like Higgs boson mass to 125~GeV. These
corrections are dominated by the (s)top loop contributions. \s

The distributions of
the trilinear soft SUSY breaking stop-sector coupling $A_t$ and of the
lightest stop mass $m_{\tilde{t}_1}$ are shown in
Fig.~\ref{fig:stopsecdistr}.
Due to the additional contribution proportional to $\lambda$ to the
tree-level mass, {\it cf.}~Eq.~(\ref{eq:upperbound}), less important
radiative corrections $\Delta m_h^2$ are necessary to shift $m_h$ to
125~GeV. Contrary to the MSSM, therefore zero mixing in the stop
sector leads to allowed scenarios in the NMSSM. For the same reason,
the lightest stop mass can still be rather light, down to about
270~GeV. The upper bound is limited by our scan range. 
Experiments set a low mass bound in searches for stop quarks
decaying into a charm quark and the lightest neutralino,
$\tilde{\chi}_1^0$ 
\cite{susyexclusion1a,atlasstopnote,cmsstopsearch,tevsearch,lepsearch}.\footnote{Investigating
  the stop four-body decays into a neutralino, bottom quark and fermion pair
  \cite{boehm,Delgado:2012eu}, 
similar bounds can be derived in the searches for monojets
\cite{susyexclusion1a,monoj}.}
This is the dominating decay for mass differences
$m_{\tilde{t}_1} - m_{\tilde{\chi}_1^0} < M_W$, where $M_W$ denotes
the charged $W$ boson mass,
\cite{Hikasa:1987db,Muhlleitner:2011ww}. The latest results by ATLAS
\cite{susyexclusion1a} exclude top squark masses up to about 240~GeV at
95\%  C.L.~for arbitrary neutralino masses, within the kinematic
boundaries. \s

The distribution for $A_t$ peaks at large values of $\sim\pm
2$~TeV. Large $A_t$ values entail large radiative corrections to the
SM-like Higgs boson mass, so that the tree-level mass value
can be shifted to the required 125~GeV. At the same time this results
in a large splitting of the stop mass values, so that the lightest
stop mass $m_{\tilde{t}_1}$ distribution reaches its maximum around
500~GeV. Another accumulation is at the upper $m_{\tilde{t}_1}$ range
as a result of large radiative mass corrections induced by heavy stop
masses. 

\subsection{Mass Distributions \label{subsec:massdistr}}
\begin{figure}[t]
\begin{center}
\includegraphics[width=10cm]{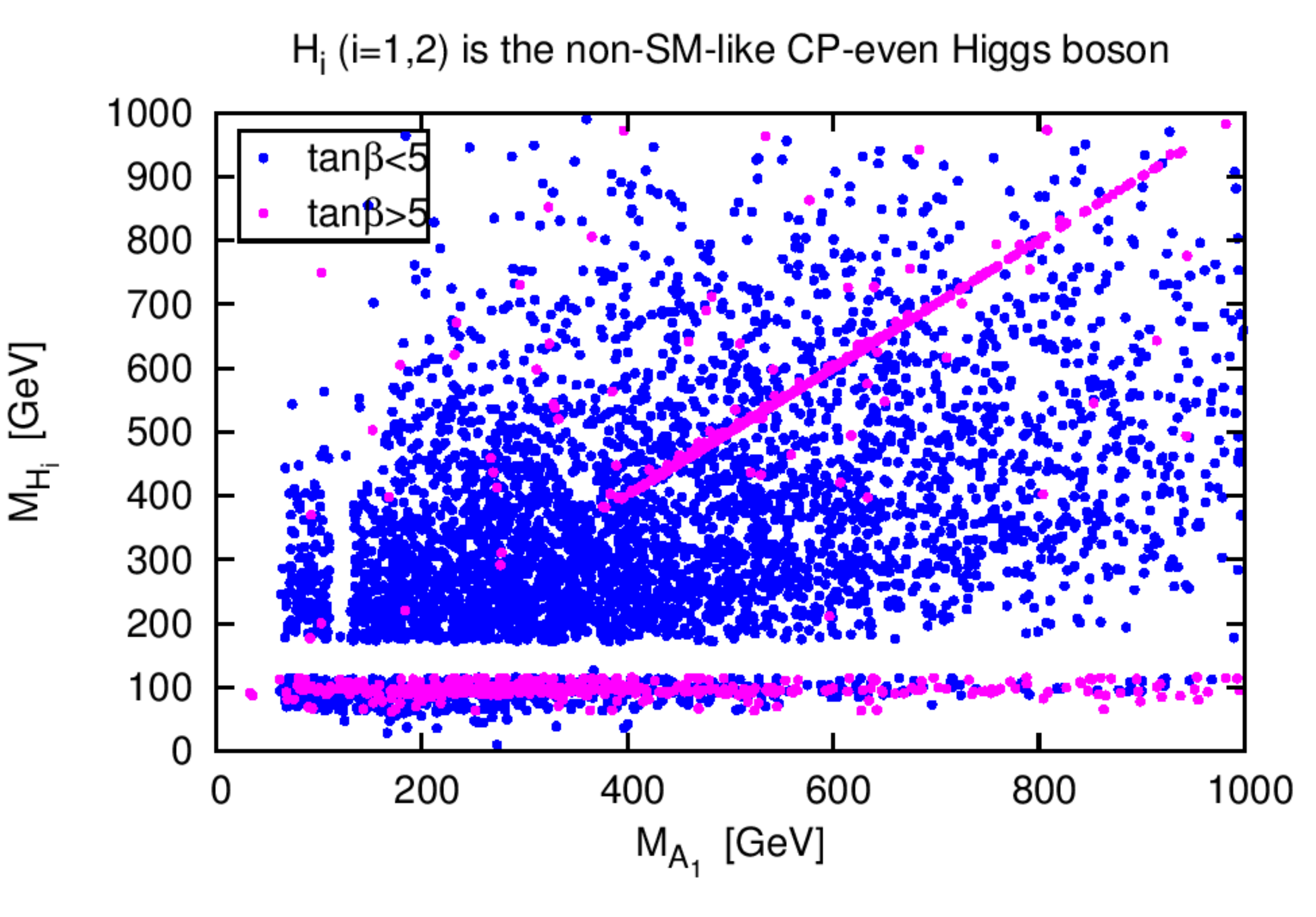}
\caption{Mass values of the non-SM-like CP-even Higgs boson $H_i$
  ($i=1$ or 2) and the lightest pseudoscalar $A_1$ for all points
  passing our constraints. Blue (pink) points are for scenarios with
  $\tan\beta <(>) 5$.}
\label{fig:mhima1mass}
\end{center}
\end{figure}

\begin{figure}[!t]
\begin{center}
\includegraphics[width=7.9cm]{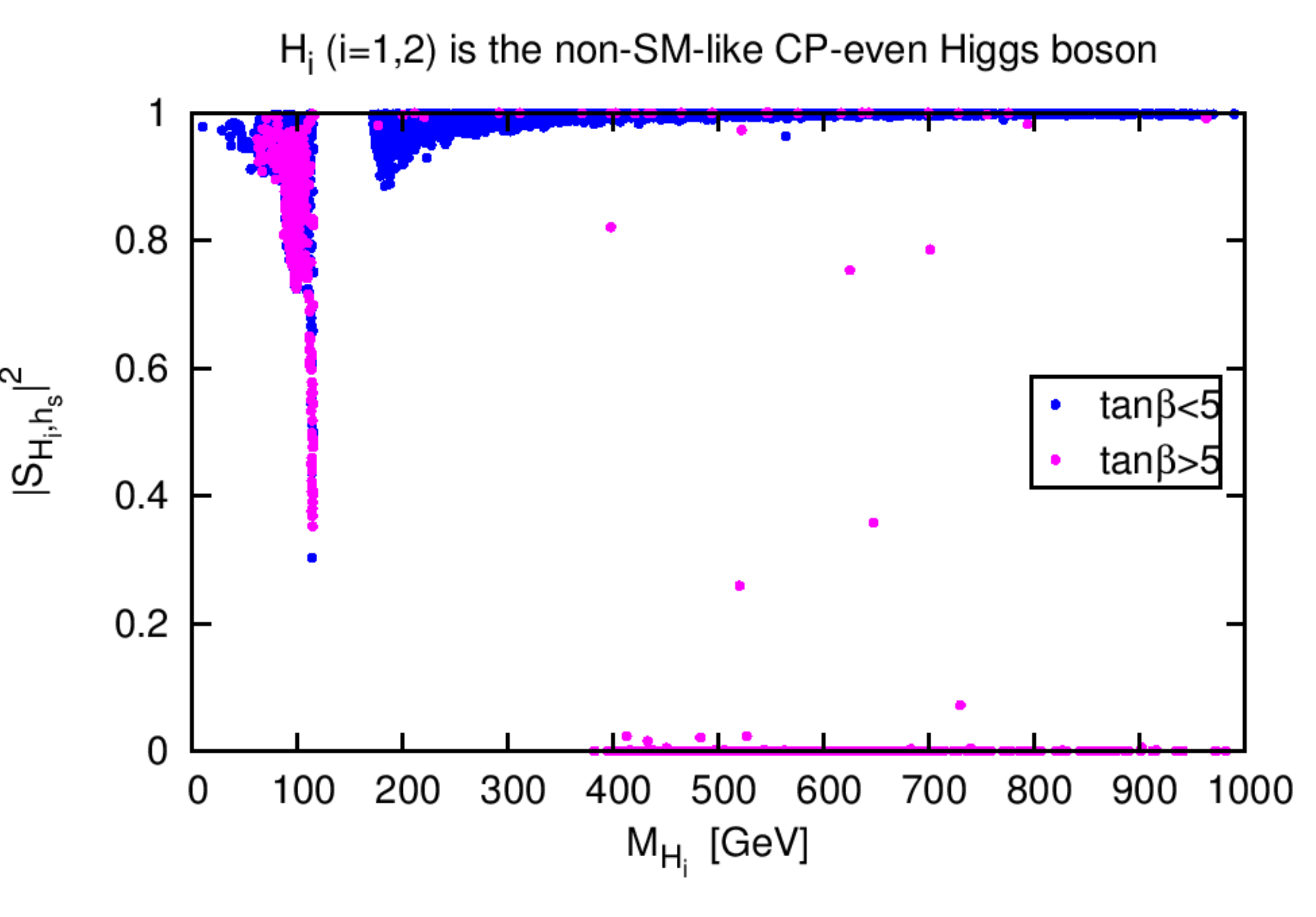}
\includegraphics[width=7.9cm]{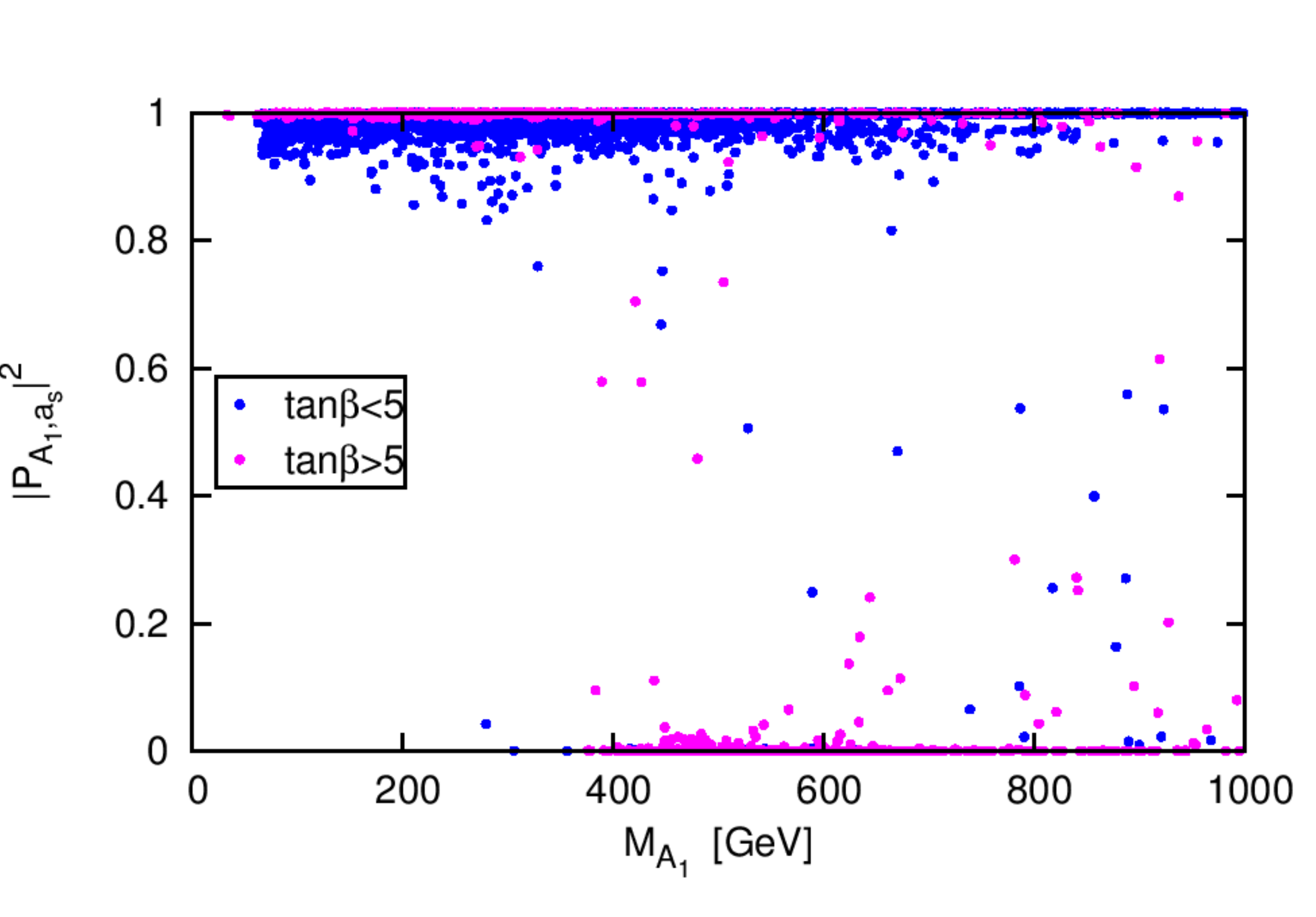} \\[0.2cm]
\includegraphics[width=7.9cm]{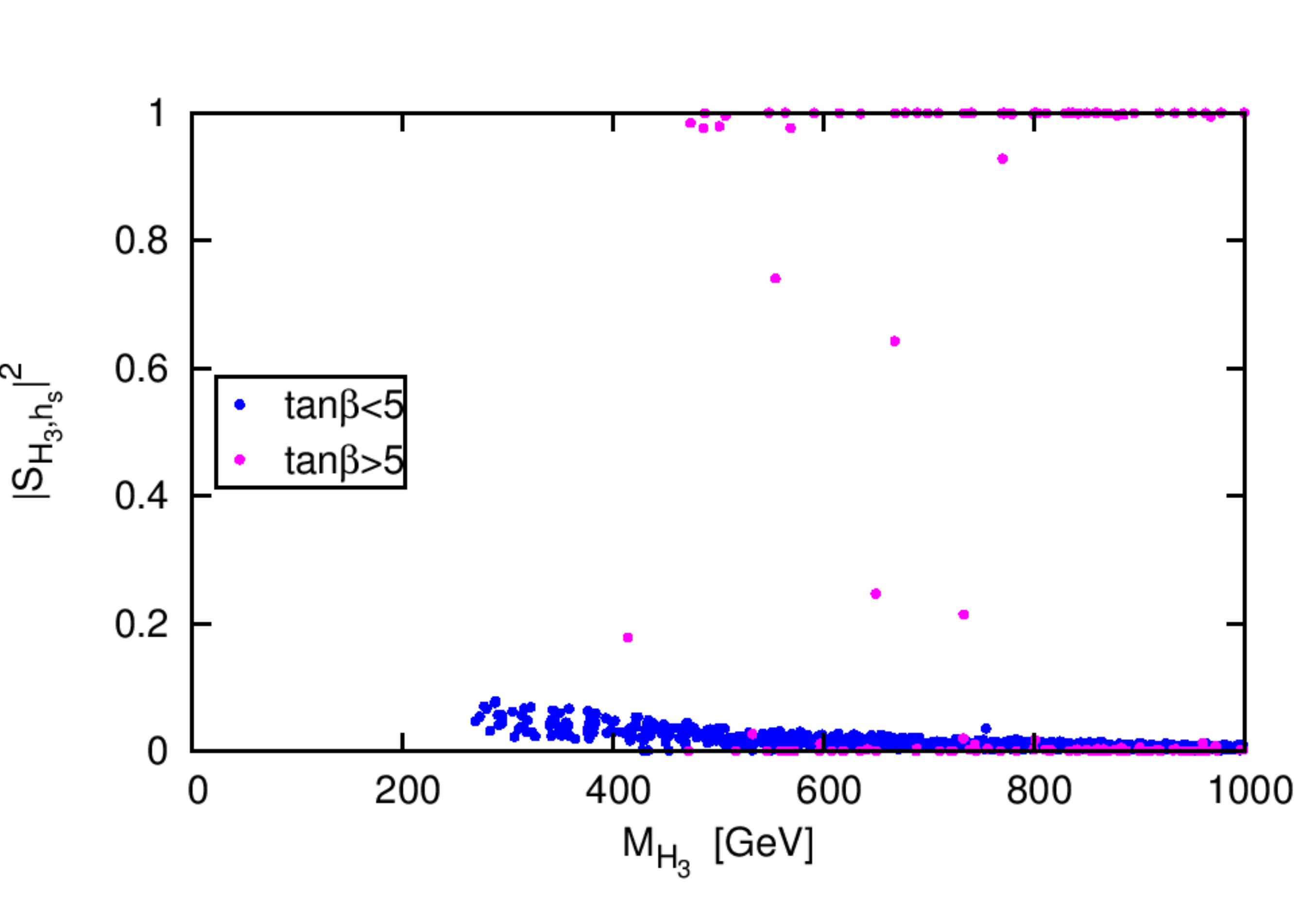}
\includegraphics[width=7.9cm]{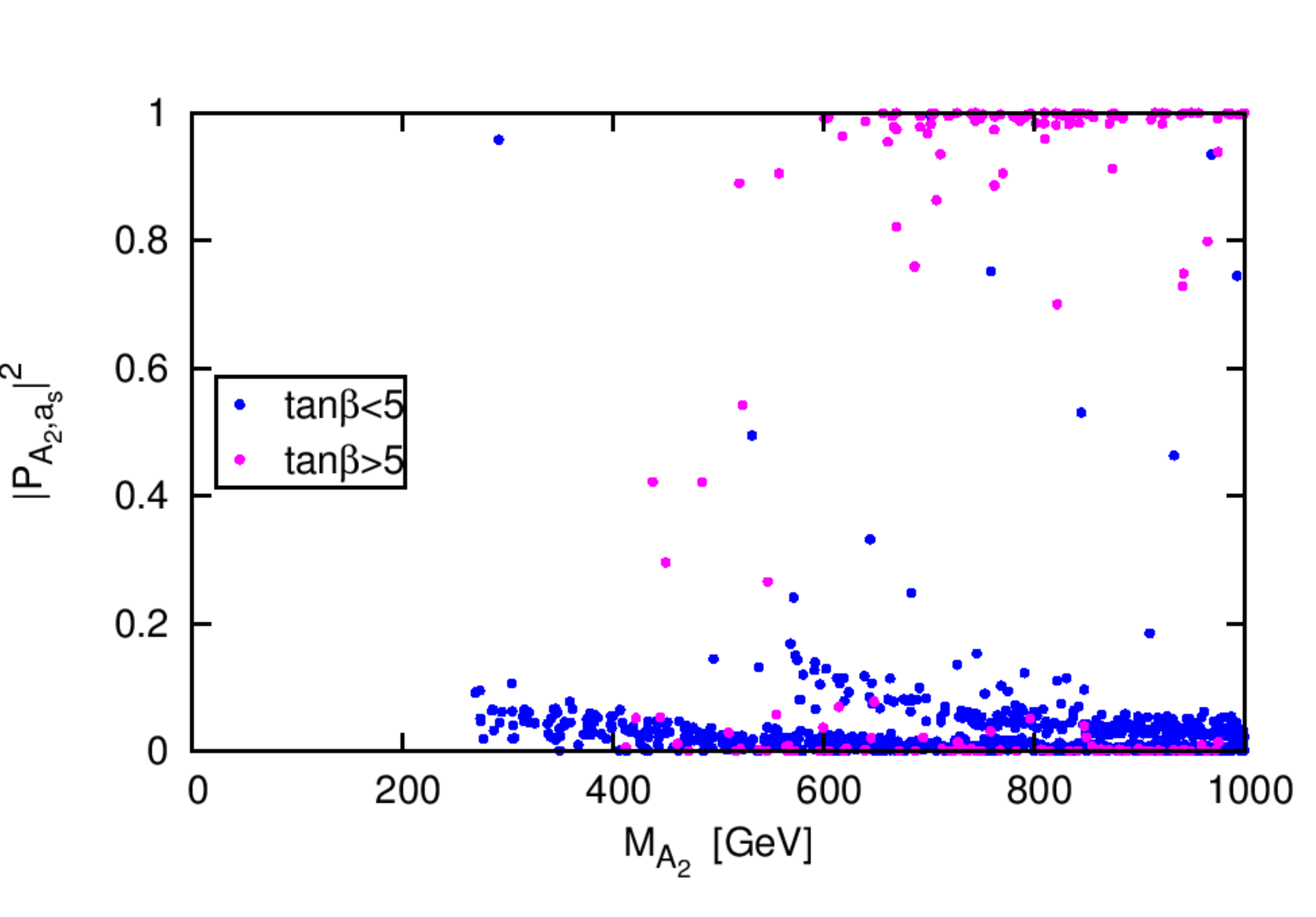}
\caption{The singlet component $|S_{H_i,h_s}|^2$ of the non-SM-like
  CP-even Higgs boson $H_i$ (upper left), $|P_{A_1,a_s}|^2$ of the
  lightest pseudoscalar $A_1$ (upper right), $|S_{H_3,h_s}|^2$ of the heavy 
  CP-even Higgs boson $H_3$ (lower left) and $|P_{A_2,a_s}|^2$ of the
  heavy pseudoscalar $A_2$ (lower right). Blue (pink) points refer to scenarios with
  $\tan\beta<(>) 5$.}
\label{fig:mhima1singl}
\end{center}
\end{figure}

In Fig.~\ref{fig:mhima1mass} we show the mass distribution for the
non-SM-like $H_i$, which is either $H_1$ or $H_2$, versus the lightest
pseudoscalar mass values $M_{A_1}$. For mass values below about
115~GeV the non-SM-like Higgs boson is $H_1$, while for $M_{H_i} \gsim
170$~GeV it is $H_2$. There is a gap for $115 \mbox{ GeV } \lsim M_{H_i}
\lsim 170$~GeV and another one for $115 \mbox{ GeV } \lsim M_{A_1}
\lsim 130$~GeV. Both are due to the LHC exclusion limits. 
There are only a few points for $M_{H_i}, M_{A_1} \lsim 62$~GeV. Here
the decay of the SM-like 125 GeV Higgs boson into very light scalars
or pseudoscalars would be kinematically possible, inducing reduced
branching ratios in the SM decays and hence reduced signal rates not
compatible with the experiment any more.
The blue (pink) points denote scenarios with $\tan\beta<(>) \,5$. Both
low and high $\tan\beta$ values yield scenarios with $H_2 \equiv
h$ (here $M_{H_i}\equiv M_{H_1} \lsim 115$~GeV). Scenarios with the
lightest scalar being SM-like arise
mostly for low $\tan\beta$ values, since then the tree-level mass of the
lightest MSSM-like Higgs boson is maximized. As can be inferred from the plot,
for large $\tan\beta$ values and mass values above $\sim 400$~GeV,
the mass values of $H_i (\equiv H_2)$ are (almost) equal to those of
$A_1$. Here, both $H_i$ and $A_1$ are MSSM-like with almost the same
mass \cite{Nevzorov:2004ge}. Figures~\ref{fig:mhima1singl} show the
amount of singlet component of $H_i$, $|S_{H_i,h_s}|^2$ (upper left), and $A_1$,
$|P_{A_1,a_s}|^2$ (upper right), as a function of $M_{H_i}$ and
$M_{A_1}$, respectively.\footnote{The matrix $S$ corresponds to the
  rotation matrix ${\cal R}^S$ Eq.~(\ref{eq:scalarrot}), however taking into account
  loop corrections to the scalar Higgs mass eigenstates. Accordingly $P$
  corresponds to the matrix $({\cal R}^P {\cal R}^G)$
  Eq.~(\ref{eq:pseudorot}), performing the rotation 
  from the interaction to the loop corrected pseudoscalar mass eigenstates.} Values of 1 (0) correspond to pure singlet-like
(MSSM-like) states. For mass values above 400~GeV and $\tan\beta >5$,
$H_i$ and $A_1$ can be MSSM-like. \s

In general the mass values of $H_1$ are $35 \mbox{ GeV} \,\lsim M_{H_1}
\lsim 115 \mbox{ GeV}$, the lowest possible mass values for $A_1$ are 
$M_{A_1} \gsim 30$~GeV, ranging up to ${\cal O} (\mbox{TeV})$, and for
the next-to-lightest scalar we have $170 \mbox{ GeV}\,\lsim M_{H_2}
\lsim {\cal O} (\mbox{TeV})$. The
masses of the heavier Higgs bosons $H_3$ and $A_2$ lie between about
300~GeV up to ${\cal O} (\mbox{TeV})$. Their singlet components
$|S_{H_3,h_s}|^2$ and $|P_{A_2 a_s}|^2$ are shown in
Fig.~\ref{fig:mhima1singl} (lower left) and (lower right), respectively. \s

The singlet-/doublet-composition of the various Higgs bosons
determines their production and decay rates, which depend on the Higgs  
coupling strengths to the SM-particles, and hence their discovery
prospects.  
Inspection of the singlet components for the two $\tan\beta$ ranges
below and above 5 and the scenarios with either $H_1$ or $H_2$ being
SM-like reveals the approximate pattern for the compositions of the
NMSSM Higgs bosons, given in Table~\ref{tab:pattern}. Note that for
an SM-like $H_2$ the lightest $H_{i=1}$ can become doublet-like in the
regions with strong singlet-doublet mixing, {\it i.e.} in mass regions
close to 125~GeV. The unitarity of the mixing matrix does not allow
for all Higgs bosons being simultaneously doublet-like, so that
alternative search techniques for the Higgs bosons with significant
singlet component, like Higgs-to-Higgs decays or SUSY particle
decays into Higgs bosons, have to be exploited. Another class of scenarios are
those with Higgs bosons that mix strongly and that are not exclusively singlet- or
doublet-like. They may challenge the Higgs searches by too small cross
sections and/or Higgs signals built up by more than one Higgs
boson.\footnote{The resolution of degenerate Higgs signals will require
  particularly high luminosity accumulated at the end of the LHC
  operation \cite{Gunion:2012he,Grossman:2013pt}, and this discussion
  is not subject of  this paper.} 

\begin{table}[!h]
 \centering
 \begin{tabular}{|c||l||l|}
   \hline
$\tan\beta <5$ & $H_{i=1}$ SM-like & $H_{i=2}$ SM-like \\ \hline \hline
$H_{j=1,2 \ne i}$ & singlet & singlet- up to almost doublet \\ \hline
$H_3$ & doublet & doublet\\ \hline
$A_1$ & mostly singlet (few doublet) & mostly singlet (few
doublet) \\ \hline
$A_2$ & mostly doublet (few singlet) & mostly doublet
(few singlet) 
\\ \hline \hline \hline 
$\tan\beta \ge 5$ & $H_{i=1}$ SM-like & $H_{i=2}$ SM-like \\ \hline \hline
$H_{j=1,2\ne i}$ & mostly doublet & singlet- up to almost doublet \\ \hline
$H_3$ & singlet (few doublet) & doublet \\ \hline
$A_1$ & doublet or singlet (for small $M_{A_1}$)  & doublet or
singlet (for small $M_{A_1}$) \\ \hline
$A_2$ & singlet or doublet & singlet or doublet
\\ \hline
\end{tabular}
\caption{The approximate singlet-/doublet-composition of the NMSSM
  Higgs bosons for small and large $\tan\beta$ values and 
  scenarios with either $H_1$ (left) or $H_2$ (right) SM-like. \label{tab:pattern}}
\end{table}

\subsection{Signal Rates \label{subsec:signalrates}}
In order to investigate the discovery prospects of the non-SM-like
NMSSM Higgs bosons, we analyse in the following their signal rates in
various SM particle final states. \s

\begin{figure}[t]
\begin{center}
\includegraphics[width=7.9cm]{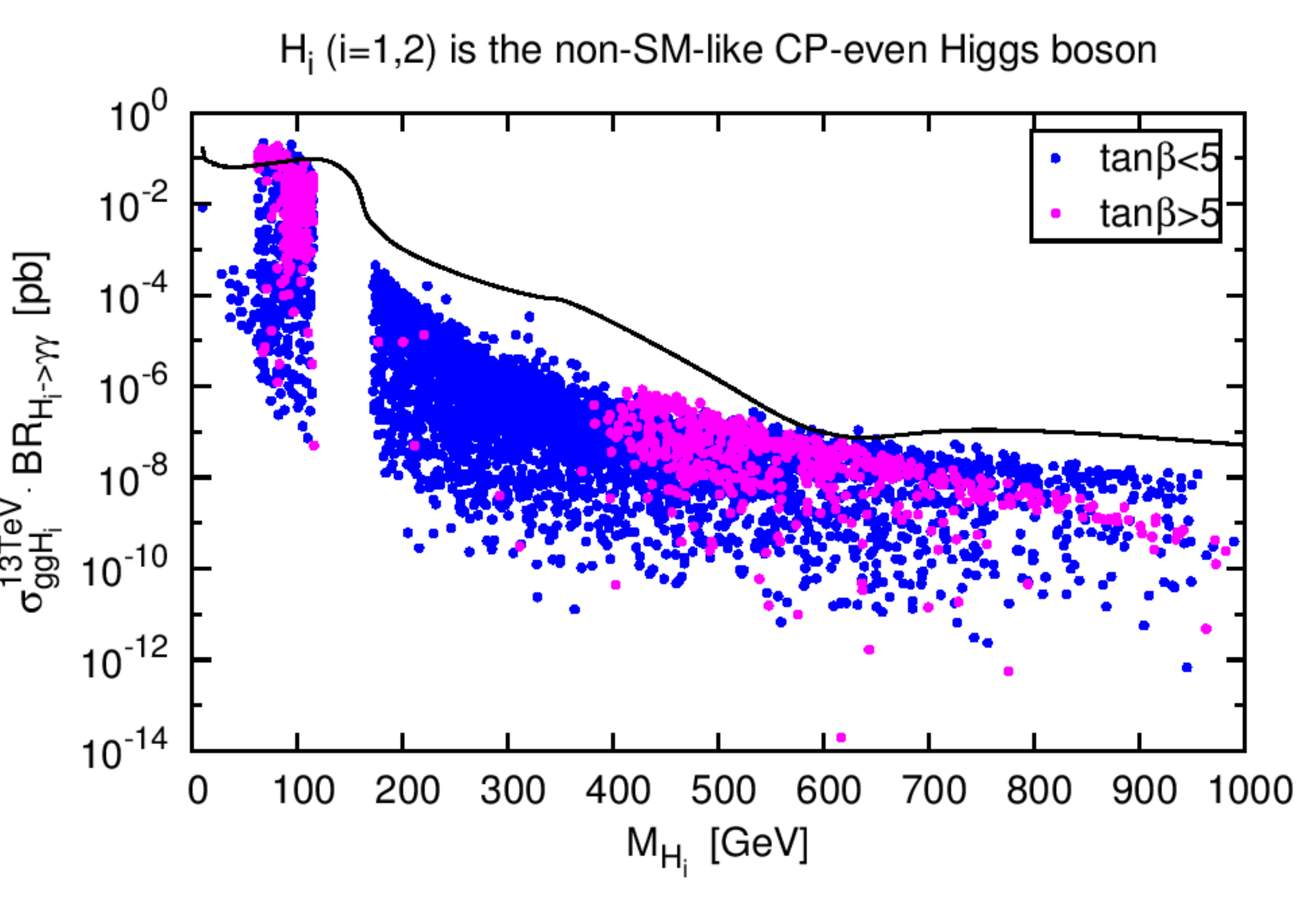}
\includegraphics[width=7.9cm]{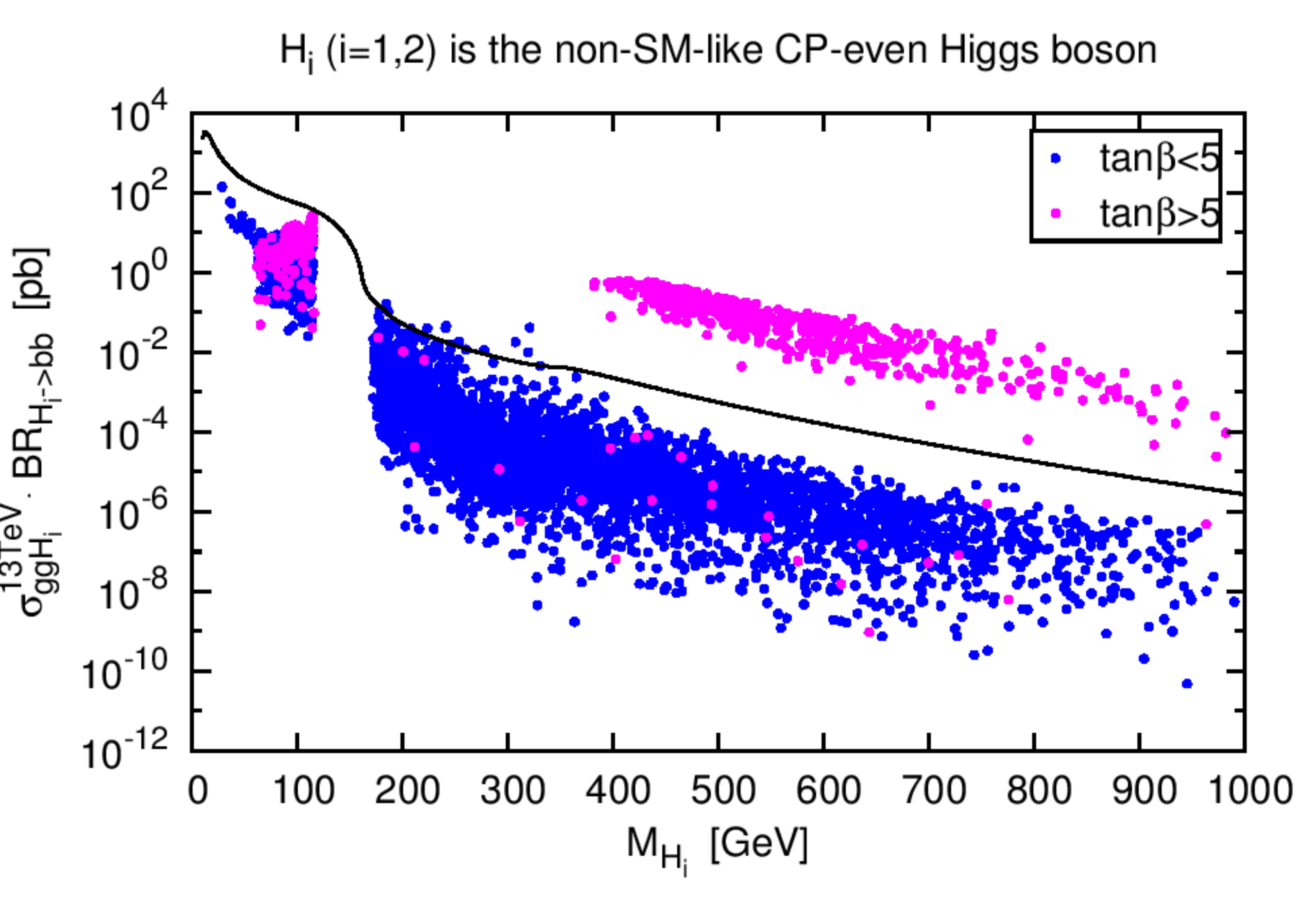} \\[0.2cm]
\includegraphics[width=7.9cm]{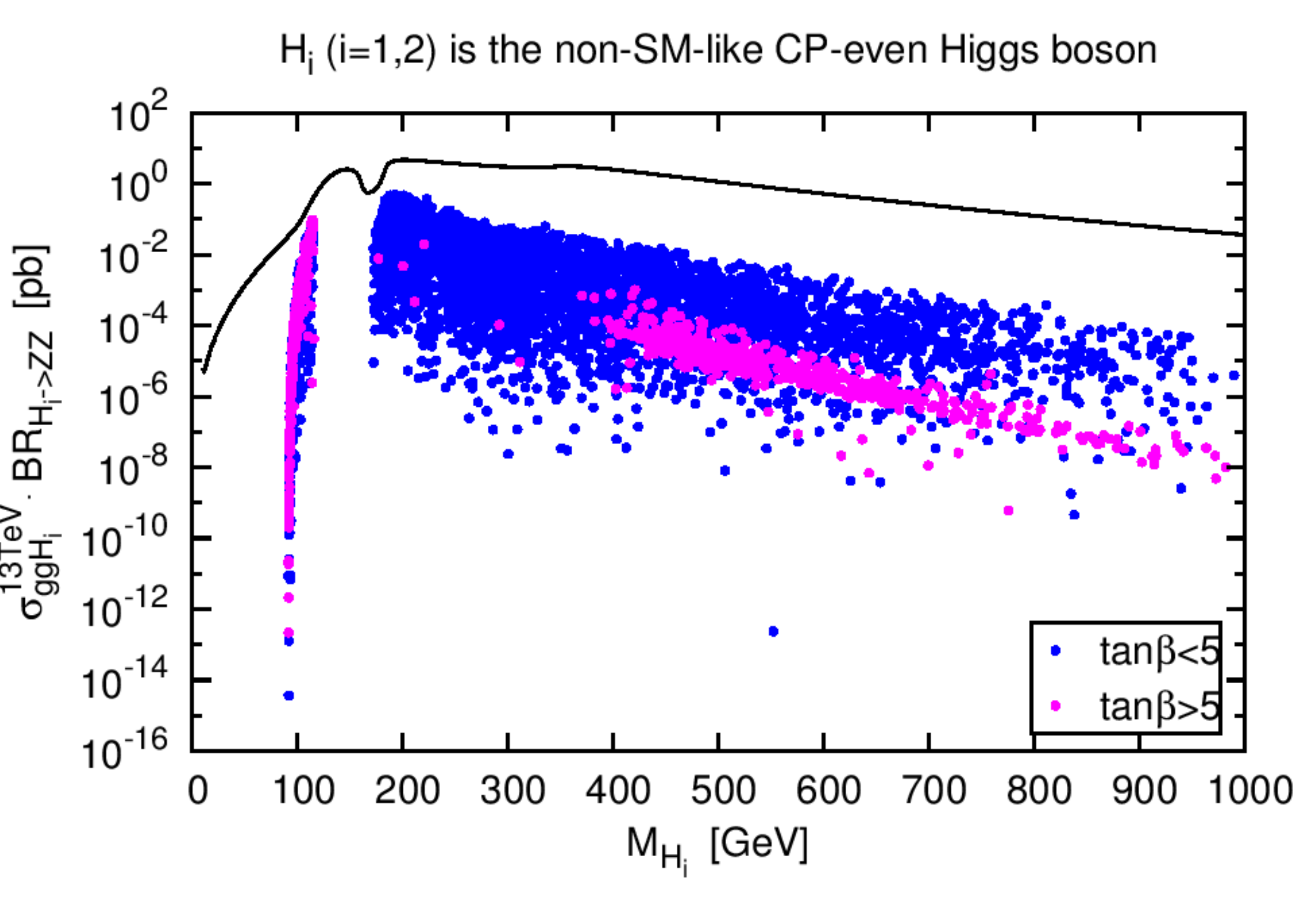}
\includegraphics[width=7.9cm]{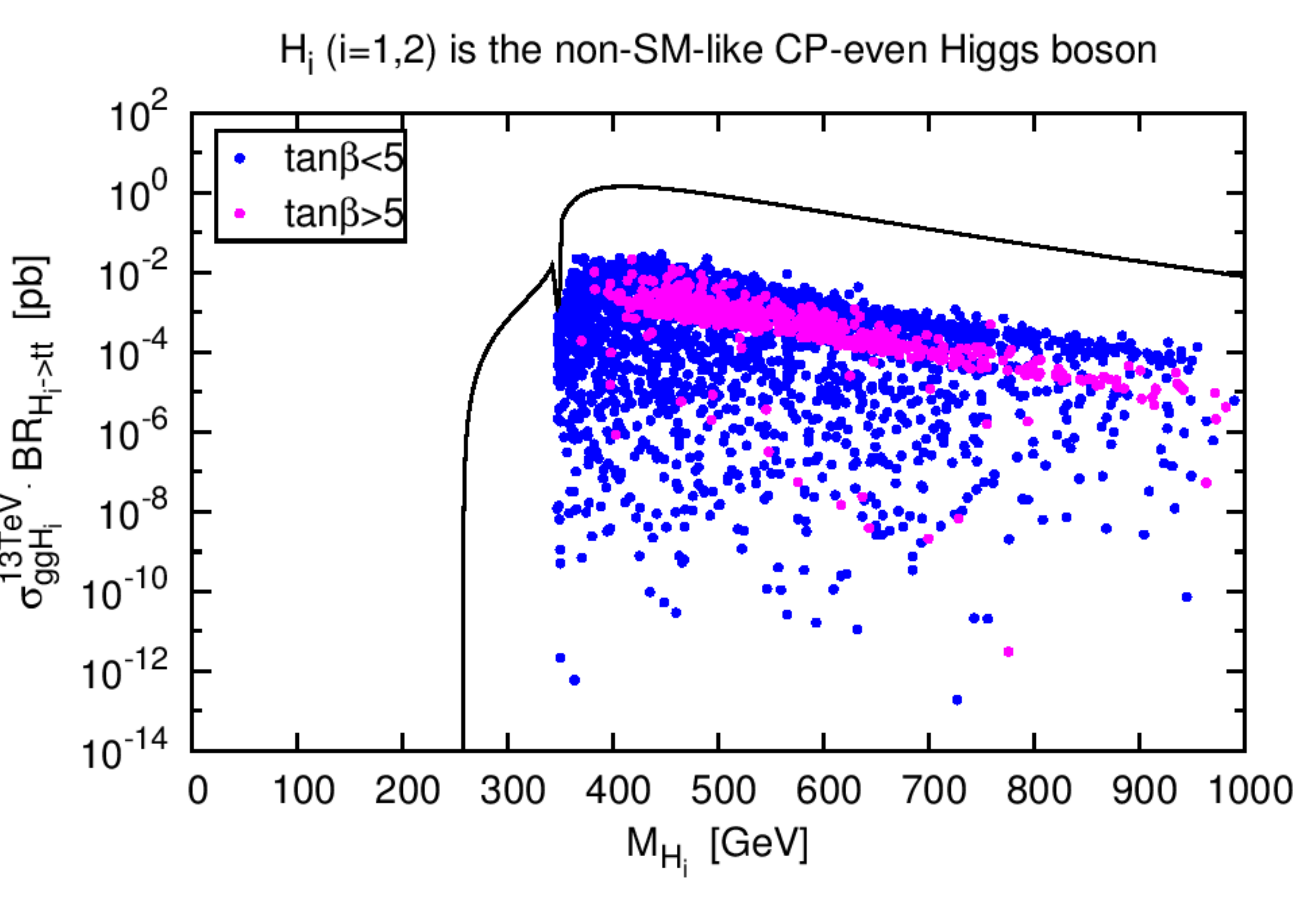} \\
\caption{Production rates in pb for the non-SM-like CP-even Higgs boson
  $H_i$ ($i=1,2$) into the $\gamma\gamma$ (upper left), the
  $b\bar{b}$ (upper right), the $ZZ$ (lower left) and $t\bar{t}$
  (lower right) final states for $\tan\beta <5$ (blue) and $\tan\beta
  >5$ (pink), as a function of $M_{H_i}$ at a c.m.~energy
  $\sqrt{s}=13$~TeV. The full black line shows the 
  production rate for a SM Higgs boson with same mass.}
\label{fig:hiredrates}
\end{center}
\end{figure}

\underline{Signal rates for the non-SM-like $H_i$ ($i=1,2$):}
Figure~\ref{fig:hiredrates} shows the production rates in pb at a
c.m.~energy of $\sqrt{s}=13$~TeV for the
non-SM-like CP-even Higgs boson $H_i$ ($i=1,2$) in the $\gamma\gamma$,
$b\bar{b}$, $ZZ$ and $t\bar{t}$ final states. The inclusive production
cross section has been approximated by the dominant gluon fusion
production mechanism. The production rates of $H_i$, $\sigma_{XX}
(H_i)$,  have been  calculated in the narrow width approximation by
multiplying the production cross section in gluon fusion at
13~TeV\footnote{Gluon fusion production increases by $\sim 12\%$ for a
  Higgs mass of 90~GeV up to $\sim 19\%$ for a Higgs mass of 300~GeV
  when increasing the c.m.~energy from 13 to 14~TeV.} 
with the branching ratios into the various SM particle final states,
\beq
\sigma_{XX} (H_i) \equiv \sigma_{ggH_i}^{\text{13 TeV}} BR_{H_i\to XX}
\; , \quad X=\gamma,b,\tau,W^\pm,Z,t \;.
\eeq
The NMSSM gluon fusion production cross section has been obtained
according to Eq.~(\ref{eq:ggfuscxn}), and the branching ratios have
been taken from {\tt NMSSMTools} in order to be consistent with the
NMSSM mass values, which we have obtained at 2-loop order from {\tt
  NMSSMTools}. The
black line corresponds to the SM Higgs boson production rate with the
same mass as $H_i$. Again the points with mass values below $\sim
115$~GeV correspond to $H_i=H_1$ and hence $H_2$ being SM-like,
while for mass values $\gsim 170$~GeV we have $H_1$ taking the SM role
and $H_i=H_2$.  For large values of $\tan\beta$, the dominant
production channel is associated production of the Higgs bosons
with a $b\bar{b}$ pair. We have computed these cross sections by
multiplying the SM cross section, computed at NNLO with the 
  code {\tt SusHi} \cite{sushi} at the same mass value as the NMSSM
  Higgs boson, with the NMSSM Higgs coupling squared to the
$b$-quark pair in terms of the SM coupling. With the exception of
small mass values, for large $\tan\beta$
values the cross sections are roughly a factor 10 larger than the
gluon fusion result. If not stated otherwise, we do not show separate
plots for this case, but 
assume implicitly that for $\tan\beta >5$ larger rates are possible
through associated production, thus ameliorating the discovery prospects. \s

\underline{{\it a) Signal rates for $H_{i=1}$:}}
For small Higgs masses $H_1$ is singlet-like, but can be more
doublet-like in the mass regions close to 125~GeV.  The photonic final
state rate can then even exceed the corresponding SM result, {\it
  cf.}~Fig.~\ref{fig:hiredrates} (upper left). For the photonic final 
state in inclusive SM Higgs boson production at leading order, ATLAS studies
\cite{atlastdr} have been performed for Higgs masses between 80 and
150~GeV at $\sqrt{s}=14$~TeV and 100~fb$^{-1}$ integrated 
luminosity. From these it may be 
concluded, that taking into account higher order corrections to the
production cross section \cite{lhchiggswg}, which increase the cross
section by about a  factor 2, a SM Higgs boson could be discovered
down to about 80 GeV if both ATLAS and CMS accumulate 300~fb$^{-1}$
each. A light scalar NMSSM Higgs boson may hence be 
discovered in the photonic final state for scenarios where its rates
are of the order of the SM ones. \s

Further possible discovery modes might be the
$b\bar{b}$ and/or $\tau\tau$ final states in the regions, where the rates are
compatible with the ones of a SM Higgs boson, {\it i.e.} for $M_{H_i}$
close to 115 GeV. The $\tau\tau$ final states are not shown in
Fig.~\ref{fig:hiredrates}, but exhibit the same pattern as the $b\bar{b}$
final state with an additional suppression factor of 10. The $Z$ and
$W^\pm$ boson final states reach maximum signal rates of 
about 0.1~pb for the former and 1~pb for the latter. Again the
$W^\pm$ final states, not depicted in Fig.~\ref{fig:hiredrates}, show the
same behaviour as the $Z$ boson final states. They are enhanced by a
factor 10 compared to these, however with missing energy in the final
state, so that only transverse Higgs masses can be reconstructed. \s

\underline{{\it b) Signal rates for $H_{i=2}$:}}
We now turn to possible discovery channels for $H_2$ being the
non-SM-like Higgs boson, hence investigate the mass regions above 
$170$~GeV in Fig.~\ref{fig:hiredrates}. 
The next-to-lightest Higgs boson $H_2$ with $H_1\equiv h$ is mostly
singlet-like for small $\tan\beta$ values and mostly doublet-like for
$\tan\beta \ge 5$. The SM-like Higgs $H_1$ needs to be doublet-like
with a large $H_u$ component, in order to 
have a substantial coupling to top quarks and hence a gluon fusion
cross section large enough to lead to signal rates compatible with the
LHC data. Due to the unitarity of the mixing matrix ${\cal R}^S$ rotating the
current to the mass eigenstates, $H_2$ is hence doublet-like but with a
large $H_d$ component, so that its couplings to down-type
fermions are enhanced compared to the SM. The coupling $G_{H_i VV}$ to massive
vector bosons $V=Z,W^\pm$ on the other hand is suppressed. This can be
understood by looking at the coupling, which normalised to the SM
coupling is given by
\beq
\frac{G_{H_i VV}}{g_{H^{\text SM} VV}}  \equiv g_{H_i VV} = ({\cal R}^S_{i1} \cos\beta + {\cal R}^S_{i2} \sin\beta) \;, \label{eq:gaugecoupl}
\eeq
where ${\cal R}^S_{i1}$ (${\cal R}^S_{i2}$) quantifies the $H_d$ ($H_u$)
component of $H_i$. The $H_d$ component can be
substantial for large $\tan\beta$ but its contribution in the coupling
is suppressed by the factor $\cos\beta$. The up-type
part of the coupling comes with $\sin\beta$ and also for small
$\tan\beta$ values would be small, as the $H_u$ component is taken by
$H_1\equiv h$. As can be inferred from the plot, for Higgs mass values
above $\sim 400$~GeV the rates into $b\bar{b}$ are enhanced for large
$\tan\beta$. This is due to an enhanced branching ratio into bottom
pairs. The total decay width in this mass range is dominated by the
decay into $V=W,Z$, and the $H_2$ branching ratio
into $b\bar{b}$ can be approximated by
\beq
\mbox{BR}^{\text{NMSSM}}_{H_2 \to b\bar{b}} &=& \frac{g_{H_2 d\bar{d}}^2
  \, \Gamma^{\text{SM}}_{H^{\text{SM}}\to  b\bar{b}}}{
g_{H_2 VV}^2 \Gamma^{\text{SM}}_{H^{\text{SM}}\to  WW} +
g_{H_2 VV}^2 \Gamma^{\text{SM}}_{H^{\text{SM}}\to  ZZ} +   ...
} \nonumber \\
&\approx&
\frac{g_{H_2 d\bar{d}}^2}{g_{H_2 VV}^2} \, \mbox{BR}_{H^{\text{SM}} \to
  b\bar{b}}^{\text{SM}} \;,
\eeq
with $M_{H_2} = M_{H^\text{SM}}$ and where $g_{H_2 d\bar{d}}$ denotes
the $H_2$ coupling to down-type fermions in terms of the SM
coupling. The enhanced coupling to down-type quarks and the suppressed
coupling to massive gauge bosons explain the observed behaviour. \s

For large values of $\tan\beta$ and mass values above 400~GeV, 
the heavier $H_2$ may therefore be discovered in its $b\bar{b}$ and
$\tau\tau$ final state, which exhibits the same behaviour as the $b\bar{b}$ final
state, the latter suffering from large backgrounds though, so that the  
$\tau$ production rate, suppressed by roughly a factor 10
compared to the former, may be more promising for discovery. In
particular, the production in association with a $b$-quark pair (not shown here)
increases the signal rates by another factor $\sim 10$ compared to 
production in gluon fusion. The gauge boson and top
quark final states are more challenging on the other hand, even in
associated production, due to the (above discussed) suppression of the
couplings to gauge bosons and top quarks. For $\tan\beta \lsim 5$ the
$H_2$ searches have to rely on a combined search in the $\tau\tau$
final states (most interesting for $H_2$ masses up to $\sim 250$~GeV)
and the vector-boson and top-pair channels. The associated production
with $b$-quarks is not effective for these $\tan\beta$ values. 
\s

\begin{figure}[t]
\begin{center}
\includegraphics[width=7.9cm]{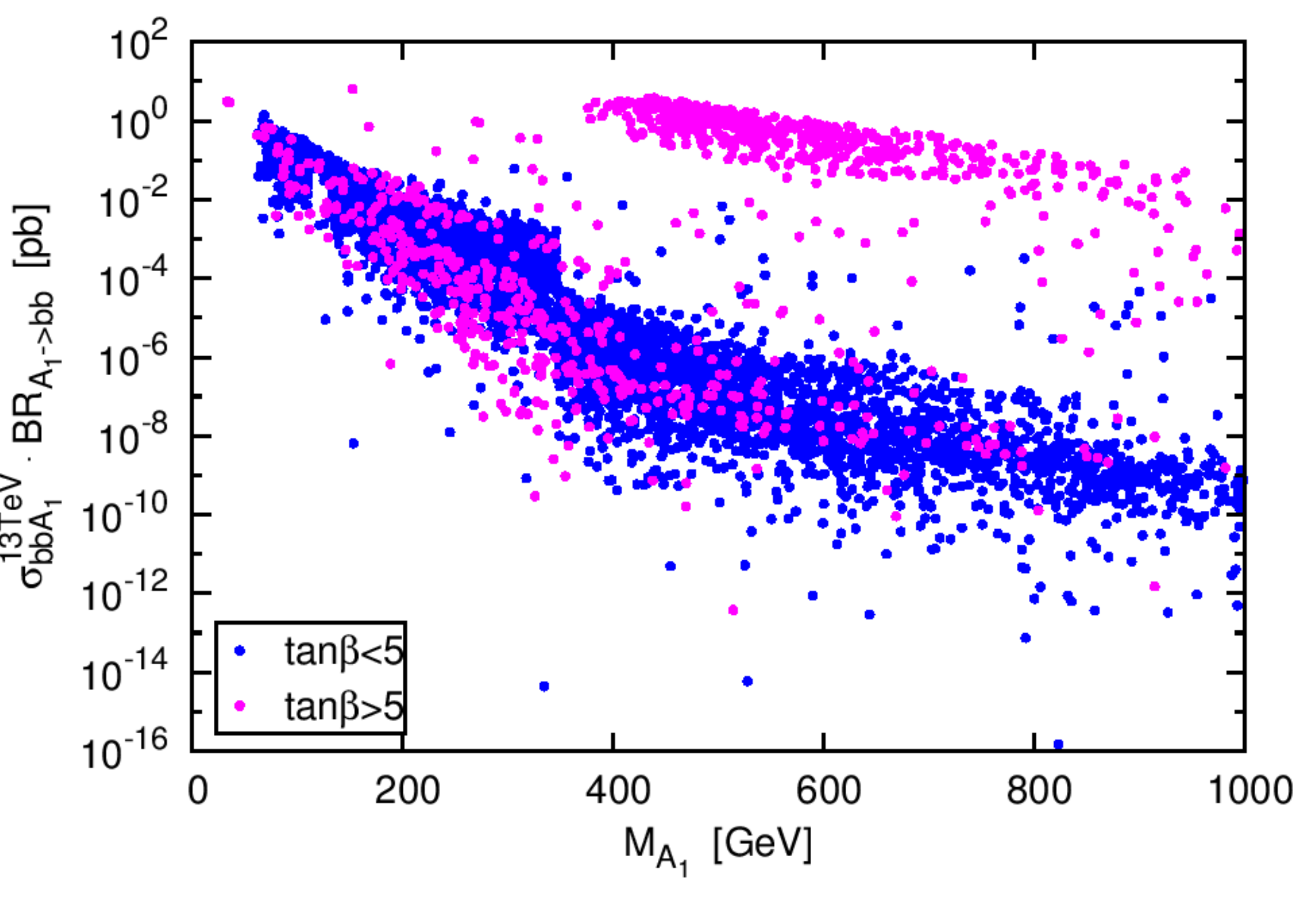} 
\includegraphics[width=7.9cm]{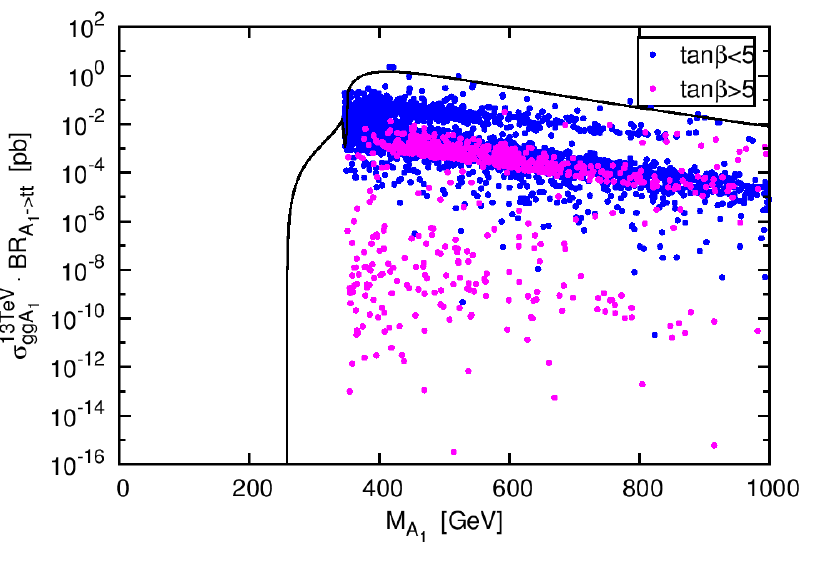} 
\caption{Production rates in pb for the lightest CP-odd Higgs boson
  $A_1$ in associated production with $b\bar{b}$ in the
  $b\bar{b}$ final state (left) and in inclusive production with subsequent decay into
  $t\bar{t}$ final states (right) $\tan\beta <5$ (blue) and $\tan\beta
  >5$ (pink), as a function of $M_{A_1}$, at $\sqrt{s}=13$~TeV. The
  full black line is the production rate for a SM Higgs boson with same mass.}
\label{fig:ha1redrates}
\end{center}
\end{figure}
\underline{Production rates of $A_1$:}
Figure~\ref{fig:ha1redrates} displays the signal rates for the
lightest CP-odd Higgs boson $A_1$ produced in association with a
$b$-quark pair in the $b\bar{b}$ final state (left) and in gluon
fusion in the top quark decay (right). The rates into
$\tau\tau$ and $\mu\mu$ final states show the same pattern as the ones
into $b\bar{b}$ but are suppressed by a factor of 10 and $\sim
10^4$ each. Massive gauge boson final states are forbidden
due to the CP-odd nature of $A_1$. The decay rates into photons 
both in associated and in inclusive production are below $10^{-4}$
already in the low mass range. In this mass range $A_1$ may be
discovered in its decays into $\tau$'s, as the $b$-quark final states
are notoriously difficult due to large backgrounds. This discovery
channel is also interesting for large $M_{A_1}$ at $\tan\beta >5$. 
For $M_{A_1} \gsim 400$~GeV the pseudoscalar, which in most scenarios
is singlet-like, can also be MSSM-like with a large $a_d$ component,
in particular for $\tan\beta > 5$, {\it
  cf.}~Fig.~\ref{fig:mhima1singl}. Associated production 
with a $b$-quark pair here leads to cross sections of up to
4~pb. Otherwise the top quark final state has to be exploited for
heavy $A_1$ production, though challenging with cross sections of at
most a few pb and a complicated final state. 
We note, that for small $\tan\beta$ values, there is a step in
Fig.~\ref{fig:ha1redrates} in the areas  
covered by the scatter points around $M_{A_1}= 350$~GeV. This is due
to the opening of the decays into the top pair final state, which
steeply increase at the top pair threshold and therefore cause a
rapid fall in the branching ratios into the other final states. For large
$\tan\beta$ the decay into top quarks does not play a role, so that
there is no such behaviour for the scatter points at large $\tan\beta$
values. \s

\begin{figure}[t]
\begin{center}
\includegraphics[width=7.9cm]{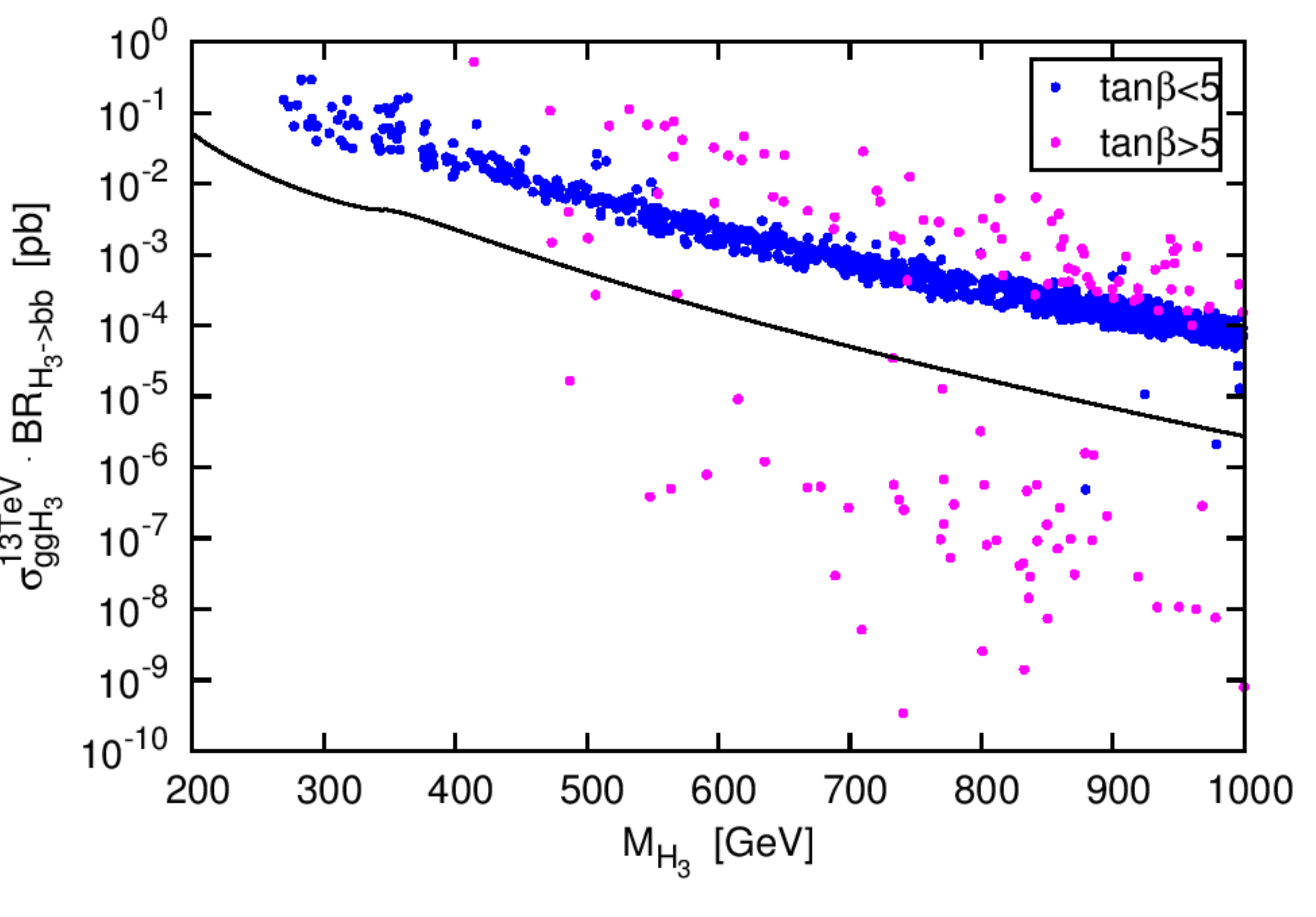}
\includegraphics[width=7.9cm]{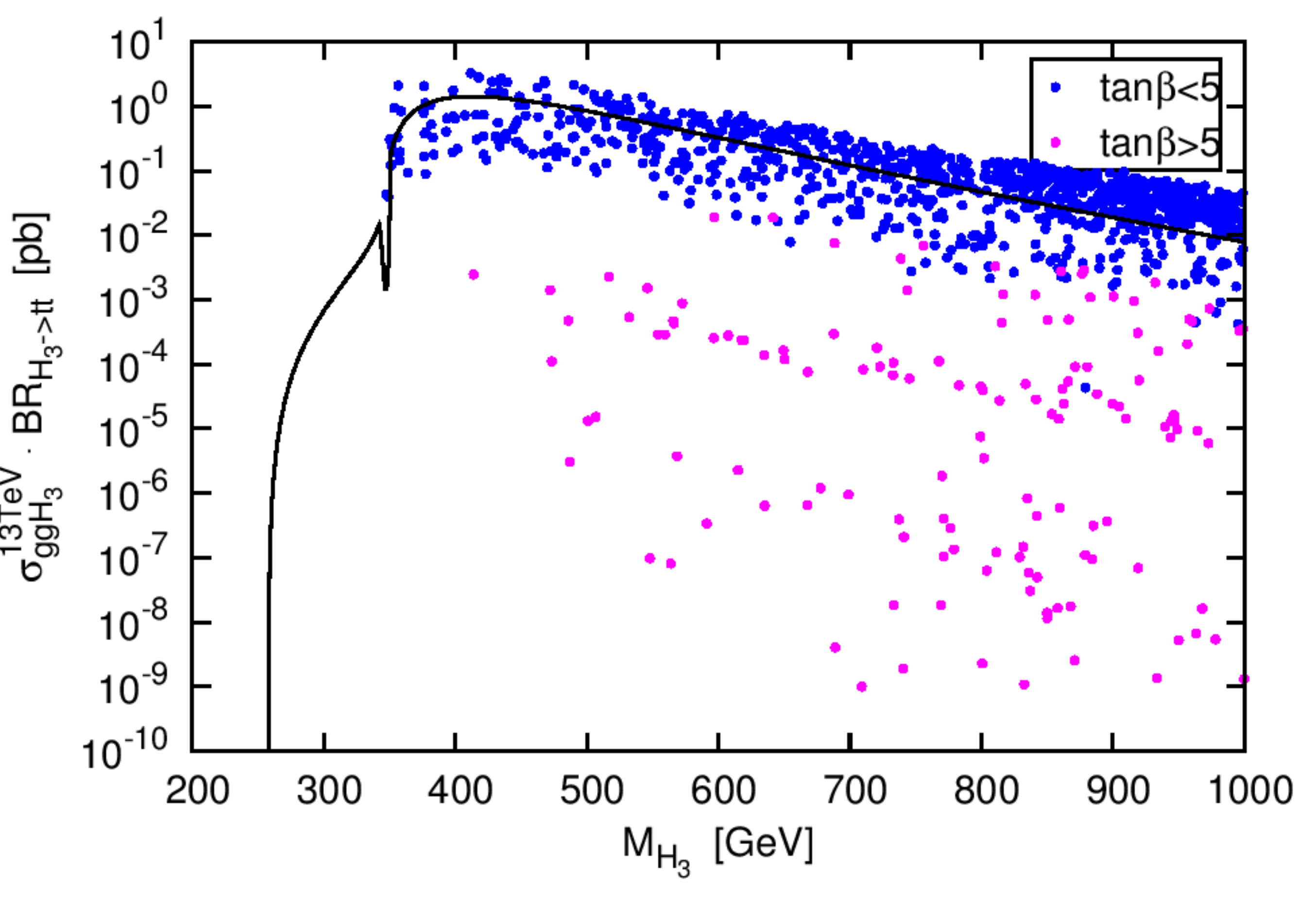} \\
\includegraphics[width=7.9cm]{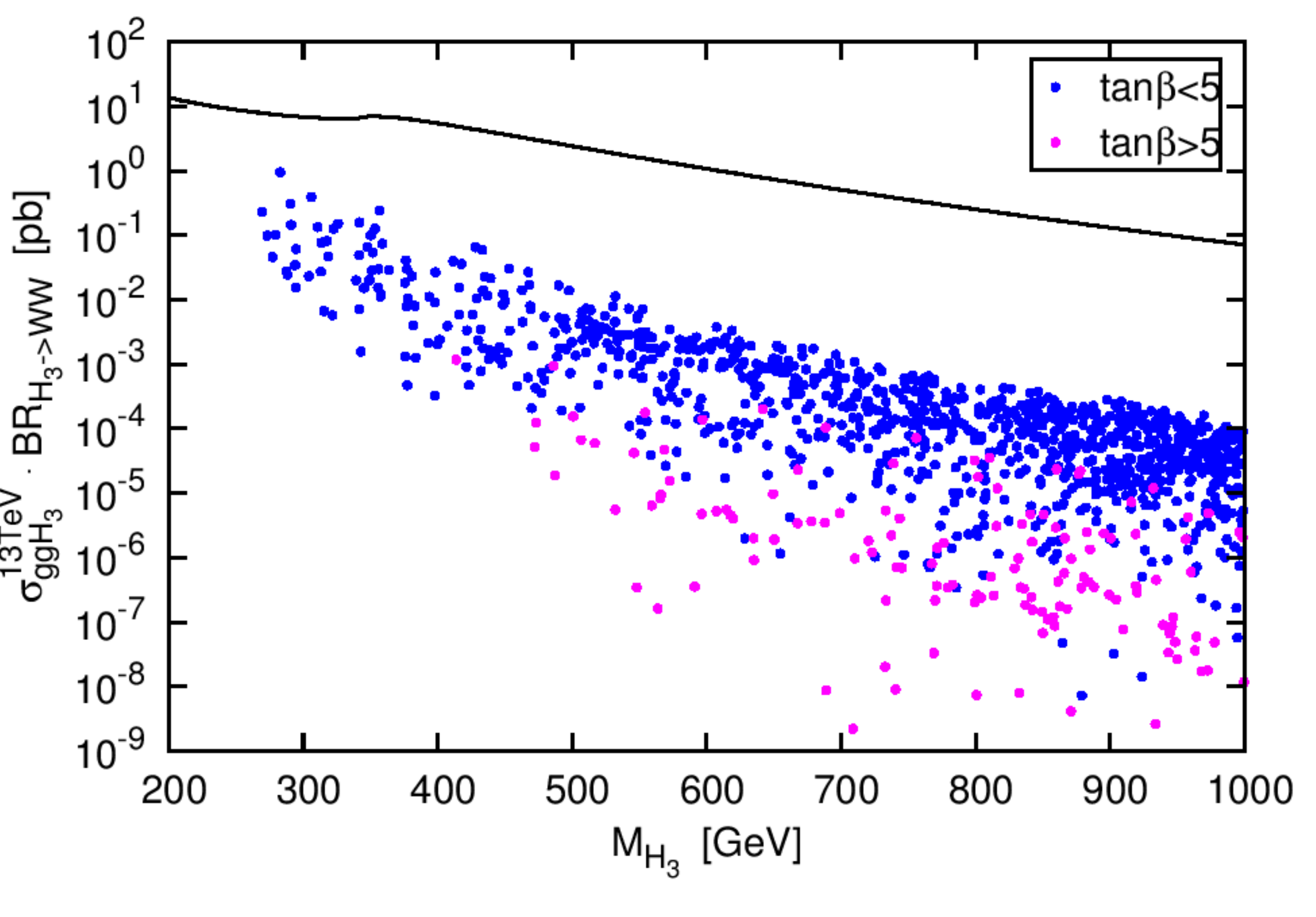}
\includegraphics[width=7.9cm]{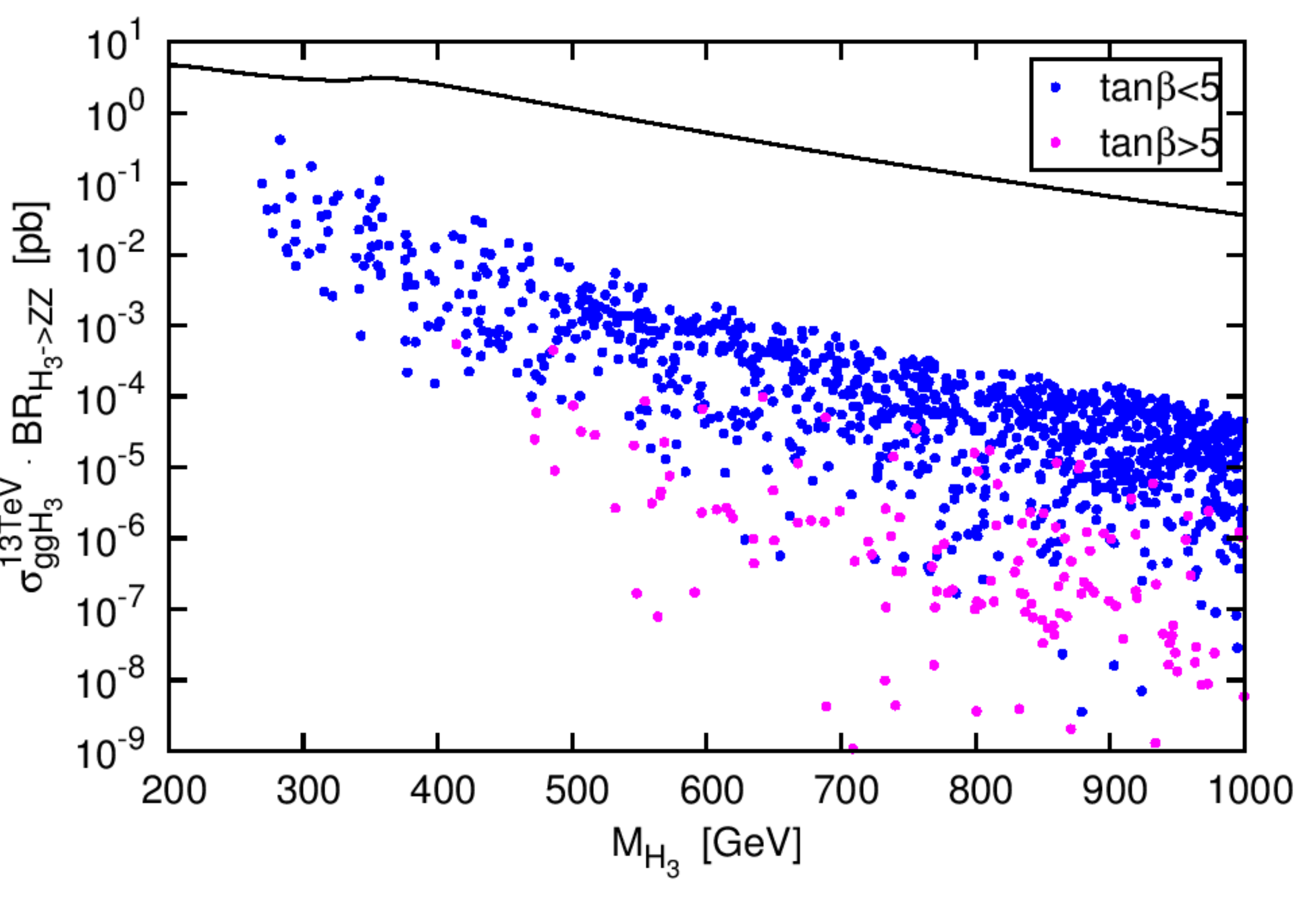} 
\caption{Production rates in pb for the heaviest CP-even Higgs boson
  $H_3$ into the $b\bar{b}$ (upper left), the
  $t\bar{t}$ (upper right), the $WW$ (lower left) and $ZZ$ (lower
  right) final states for $\tan\beta <5$ (blue) and $\tan\beta 
  >5$ (pink), as a function of $M_{H_3}$, at 
  $\sqrt{s}=13$~TeV. The full black line is the 
  production rate for a SM Higgs boson with same mass.}
\label{fig:h3prodrates}
\end{center}
\end{figure}

\underline{Heavy $H_3$ production:}
Apart from large $\tan\beta$ values with a SM-like $H_1$ boson, the
$H_3$ is mostly doublet-like, however with a large $H_d$ component so
that the couplings to gauge bosons are suppressed, as discussed
previously, likewise the couplings to top quarks. The branching
ratios into bottom and top quark pairs, however, can be enhanced due to the small
coupling to gauge bosons, as the total width in this heavy mass region
is dominated by the decays into the latter.
Figure~\ref{fig:h3prodrates} (upper) shows the
production rates for the heaviest CP-even, $H_3$, into the $b\bar{b}$
and the $t\bar{t}$ final states. In the $b\bar{b}$ final
state the rates are at most 0.5~pb in the lower mass region, where the
SM-like Higgs boson is mostly $H_1$ and below 0.01~pb above about 400~GeV,
where the SM-like resonance can also be $H_2$. In $\tau$ pair final
states they are a factor 10 lower. For large $\tan\beta$
values the rates can be enhanced by about roughly a factor 10 in $b\bar{b}H_3$
production. In the top quark pair final states the cross sections 
are somewhat larger ranging from $\sim 5$~pb at the threshold to
${\cal O}(0.01)$~pb for $\tan\beta<5$. For large $\tan\beta$ the 
$t\bar{t}$ production rates are below 0.05~pb. The rates into $WW$
range for $\tan\beta<5$ maximally between 1~pb and $10^{-4}$~pb
in $250 \mbox{ GeV} \lsim M_{H_3} \lsim 1 \mbox{ TeV}$, {\it
  cf.}~Fig.\ref{fig:h3prodrates} (lower left).  For the $ZZ$ final state
they are somewhat smaller. For large $\tan\beta$ values these final
states are not interesting. The photonic final state rates, finally,
are below $10^{-4}$~pb. \s

\underline{Heavy $A_2$ production:} The heavy pseudoscalar $A_2$
production rates in the $b\bar{b}$ and $t\bar{t}$ final states, as
well as the photonic mode show almost the same pattern as the ones for
$H_3$ with the same absolute values, and the same conclusions as for
the $H_3$ searches apply. The pseudoscalar cannot decay
into gauge bosons, so that these final states cannot be exploited
here. \s

The above results show that the discovery of {\it all} NMSSM Higgs bosons
is not straightforward and additional alternative discovery modes need to be
exploited, which shall be discussed in the following sections. 
Before doing so, let us briefly review
what the current experimental status is. So far, 
dedicated searches in the NMSSM have been performed by CMS for a very
light pseudoscalar with mass between 5.5 and 14~GeV, decaying into
a $\mu$ pair, for which exclusion limits have been derived
\cite{cmsamusearch}. The ATLAS experiment has investigated the decay
chain for a heavy CP-even Higgs boson into a light pseudoscalar Higgs
pair, that subsequently decays into photons, $H \to aa \to
\gamma\gamma + \gamma\gamma$ \cite{atlasagamsearch}. The signal has
been studied using simulated samples over a range of Higgs boson
masses between 110 and 150 GeV and for three $a$ boson masses,
$M_a = 100, 200$ and 400 MeV. For a c.m.~energy of 7~TeV and
4.9~fb$^{-1}$ integrated luminosity, the derived 95\% confidence level (CL)
exclusion limits on the cross section
times branching ratio are 0.1 pb in the Higgs boson mass range between
115 and 140 GeV and $\sim 0.2$ pb in the region outside. 
Recently CMS has published results on the search
for the resonant production of Higgs pairs in the decay channel $X \to
HH \to \gamma\gamma + b\bar{b}$ using 19.7 fb$^{-1}$ integrated
luminosity collected at $\sqrt{s}=8$~TeV \cite{cmsXtohhsearch}. For the
investigated mass range between $m_X=260$ and 1100 GeV upper limits
at 95\% CL on the cross section have been extracted between about 4
and 0.4~fb. The ATLAS experiment has performed searches for resonant
and non-resonant Higgs boson pair production in the $hh \to
\gamma\gamma b\bar{b}$ final state at a c.m.~energy of 8~TeV and an
integrated luminosity of 20.3~fb$^{-1}$ \cite{atlaspairprod}. Assuming
SM branching ratios a 95\% CL upper limit of 2.2~pb is extracted on the
cross section times branching ratio of the non-resonant
production. The corresponding limit observed for a narrow resonance
lies between 0.8 and 3.5~pb depending on its mass. 
Extrapolating these results to the high-energy LHC with up
to 300 fb$^{-1}$ integrated luminosity per experiment, decay
cross sections for Higgs-to-Higgs decays with subsequent Higgs decays
into the $(\gamma\gamma)(b\bar{b})$ final states down to ${\cal
  O}$(fb) should be large enough for detection. 

\section{Discovering Natural NMSSM at the
  LHC \label{sec:naturalnmssm}}
In the previous section we have discussed the signal rates that can be
expected in the NMSSM parameter space that is left over after applying
all constraints set by the Higgs search results from experiment and by
the relic density. In this section now we investigate that subspace of
the NMSSM, that we found to give the best discovery prospects at the
high-energy LHC for all of the neutral NMSSM Higgs bosons and that can
hence strongly be constrained at the next run of the LHC. Such part of
the NMSSM parameter space is given by what we call the 
Natural NMSSM. It is characterized by an approximate Peccei-Quinn
symmetry\footnote{This part of the NMSSM parameter space is favoured by 
low fine-tuning considerations for $\lambda>0.55$. For a recent analysis
see \cite{ournmssmpapers}.}, hence small
$\kappa$ values, by rather small 
$|\mu_{\text{eff}}|$ values and by small $\tan\beta$. In particular we choose the
following part of the NMSSM parameter space
\beq
0.6 \le \lambda \le 0.7 \,, \; -0.3 \le \kappa \le 0.3 \,, \; 1.5 \le \tan\beta
 \le 2.5\, ,\; 100\mbox{ GeV} \le |\mu_{\text{eff}}| \le 185 \mbox{ GeV} \, .
\eeq
In this parameter region the second lightest Higgs boson $H_2$ is
SM-like.\footnote{In order to have $H_1 \equiv h$, higher
  $\mu_{\text{eff}}$ values  than the ones chosen here would be
  required.}  The heavier CP-even 
  and CP-odd Higgs bosons, $H_3$ and $A_2$, are predominantly a
  superposition of the components of the Higgs doublets (MSSM-like
  states). The lightest scalar and pseudoscalar Higgs states, $H_1$ and
  $A_1$, are singlet dominated. In the following we will use the
  convenient notation:
\begin{itemize}
\item As before, $h$ denotes the SM-like Higgs boson, and here $H_2
  \equiv h$.
\item The doublet-like heavy Higgs bosons will be denoted as $H_3
  \equiv H$ and $A_2 \equiv A$.
\item The singlet dominated lightest CP-even and CP-odd Higgs bosons
  are called $H_1 \equiv H_s$ and $A_1 \equiv A_s$, respectively.
\end{itemize}
At tree-level the
NMSSM with approximate Peccei-Quinn symmetry leads to a hierarchical
structure of the Higgs spectrum which consists of the heaviest CP-even,
the heaviest CP-odd and the charged Higgs bosons 
being almost degenerate. Their mass scale is set by $\mu_{\text{eff}}\tan\beta$
\cite{Nevzorov:2004ge1}, hence
\beq
M_{H} \approx M_A \approx M_{H^\pm} \approx \mu_{\text{eff}} \tan\beta \;.
\eeq
Since $|\kappa| < \lambda$, the masses of the singlet dominated
CP-even and CP-odd Higgs states are smaller than 
$\mu_{\text{eff}}\tan\beta$, with the upper bound given by
\beq
M_{A_s}^2 + 3 M_{H_s}^2 \approx 12 \left( \frac{\kappa}{\lambda} \mu_{\text{eff}}
\right)^2 + \Delta  \;.
\label{eq:appmassrel}
\eeq
The additional contribution $\Delta$ will be quantified later. 
We now turn to the discussion of the discovery prospects for the Higgs spectrum in
this set-up. In the part of the NMSSM parameter space that we are
considering the almost degenerate heavy CP-even, CP-odd and charged Higgs
bosons have masses below about 530~GeV, so that they should still be
light enough to be observed at the 13 TeV LHC. Due to the substantial
mixing, because of  the large value of $\lambda$, between the SM-like
Higgs state and the singlet dominated  CP-even state the production
cross sections of the lightest and second lightest CP-even Higgs states
are in general large enough to produce these particles. In case the
non-SM-like CP-even Higgs boson is almost a singlet it may
still be produced via the decays of the heavier Higgs states,
because of the large $\lambda$ value. The same holds for the
singlet-dominated lightest CP-odd Higgs state. We list in the Appendix the
Higgs couplings involved in the Higgs-to-Higgs decays for the
approximations made here. Their inspection gives information on the
possible decays to be expected. In summary, all Higgs bosons of the
Natural NMSSM should in general be accessible, so that this scenario
may be constrained at the next round of the LHC run. \s

We confront these approximate considerations with the results from our
parameter scan in the subspace given by the natural NMSSM. The approximate 
$\mu_{\text{eff}} \tan\beta$ mass value of the heavy Higgs states is
slightly modified by loop corrections and they range between about
230~GeV and 530~GeV. Taking into account higher order mass corrections the
mass relation for the singlet states, Eq.~(\ref{eq:appmassrel}), 
approximately holds if we choose $\Delta = 18690$~(GeV)$^2$. 
Furthermore, $H$ and $A$ are indeed dominantly
doublet-like, while the lightest CP-even and CP-odd states $H_s$ and $A_s$ are
singlet-like. The CP-even singlet mass ranges between 27~GeV and
117~GeV. The upper bound is given  by the LHC exclusion limits on the
one hand and the fact, that the 125~GeV Higgs boson must be SM-like,
on the other hand. The CP-odd singlet mass lies between 29~GeV and
300~GeV. Note that for $H_s$ there are only a few points below 62~GeV
and for $A_s$ even less. The reason is that the SM-like $h$ could
decay into these final states and this would drive its reduced signal
rates away from the measured values. Thus the Natural NMSSM scenario
implies the existence of a CP-even Higgs state $H_s$ that tends to
have a mass of $62 \mbox{ GeV} \lsim M_{H_s} \lsim 117 \mbox{ GeV}$,
and of a CP-odd state $A_s$ with $62 \mbox{ GeV} \lsim M_{A_s} \lsim
300 \mbox{ GeV}$.\s

\begin{figure}[!b]
\begin{center}
\includegraphics[width=7.9cm]{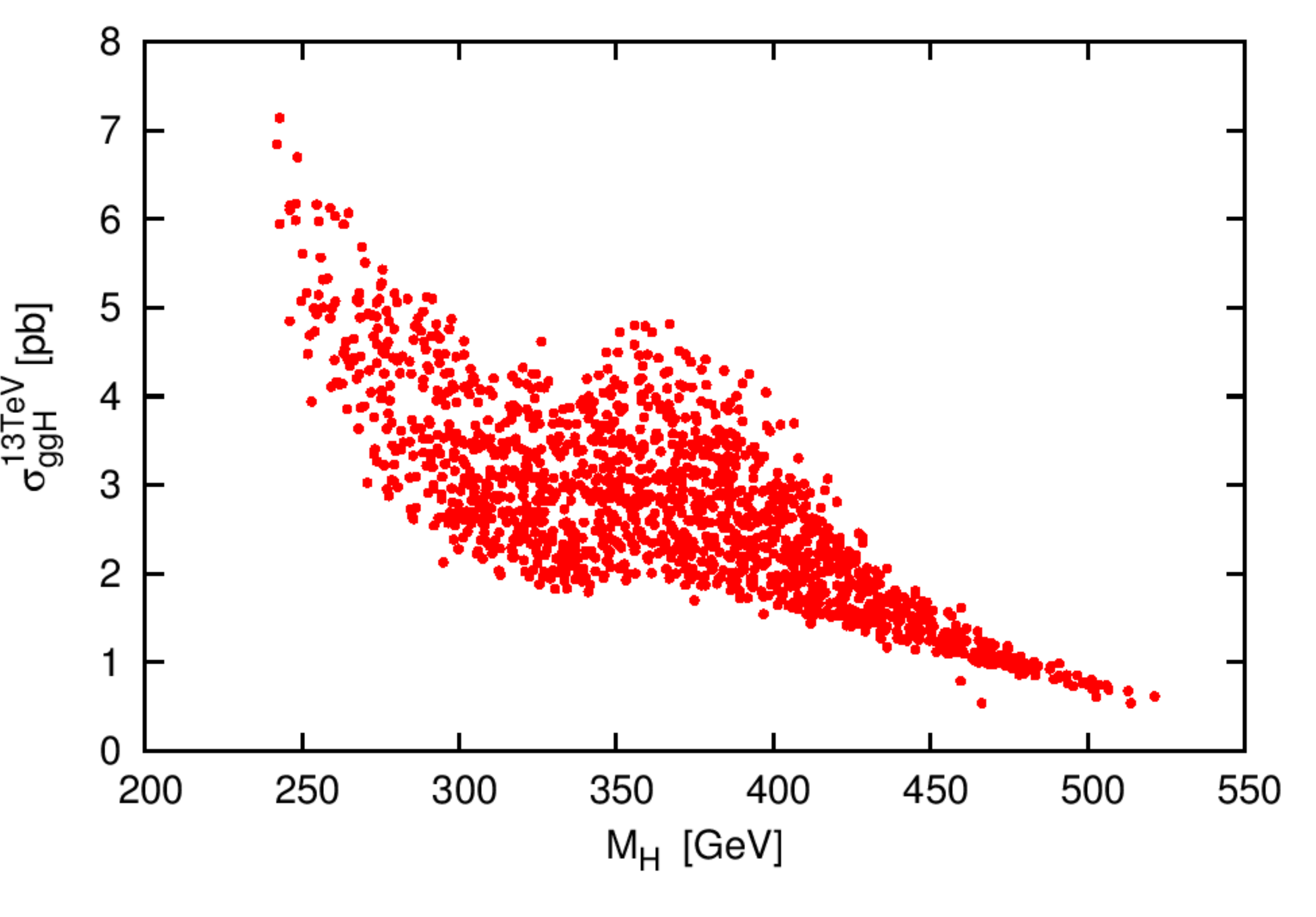}
\includegraphics[width=7.9cm]{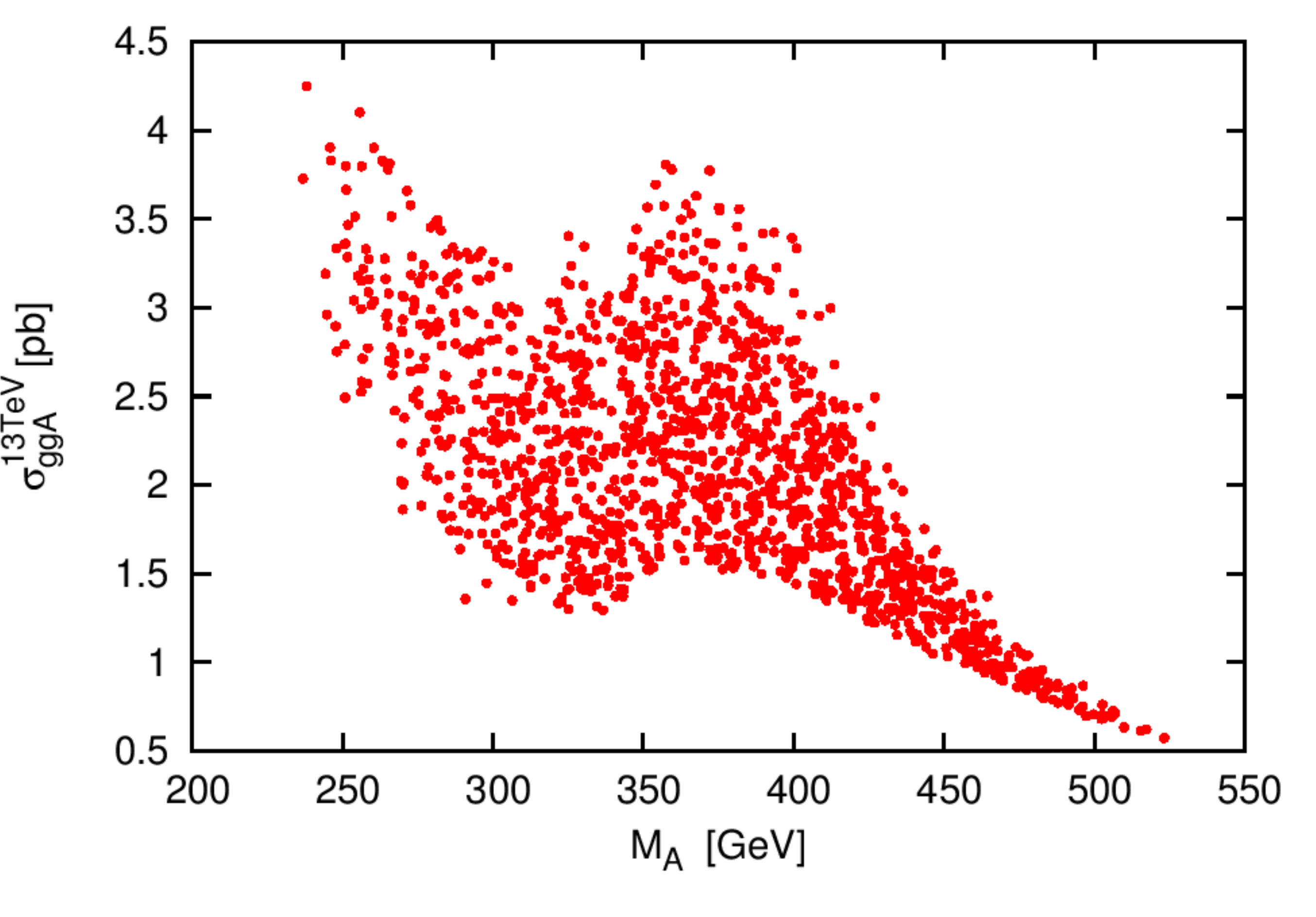}
\caption{The gluon fusion production cross section at
  $\sqrt{s}=13$~TeV for $H$ (left) and $A$ (right) as a function of
  their mass. \label{fig:handaprod}} 
\end{center}
\end{figure}
We find that the $H$ and $A$ gluon fusion
production cross sections range between $\sim 7.5$~pb ($H$),
respectively, $\sim 4.5$~pb
($A$) at the low mass end and 0.6-0.8~pb at the high mass end, as shown in
Fig.~\ref{fig:handaprod}. As $H$ has a small up-type component
admixture, the cross section cannot be as large as in
the SM for a Higgs boson of same mass. Still, in the Natural NMSSM, 
the size of the cross sections in most cases is large enough to discover these
particles in the standard final states like $b\bar{b}$ (modulo the
challenge due to large backgrounds), $\tau\tau$ or even
$t\bar{t}$. For $H$ the massive gauge boson final states add to the
search channels. As can be inferred from Figs.~\ref{fig:h3prodrates}
of the enlarged scan, in the mass and $\tan\beta$ range we consider
here, for most of the cases the production rates in the various final
states are above 1~fb, depending on the final state even well above 1~fb. \s 

\begin{figure}[!h]
\begin{center}
\includegraphics[width=7.9cm]{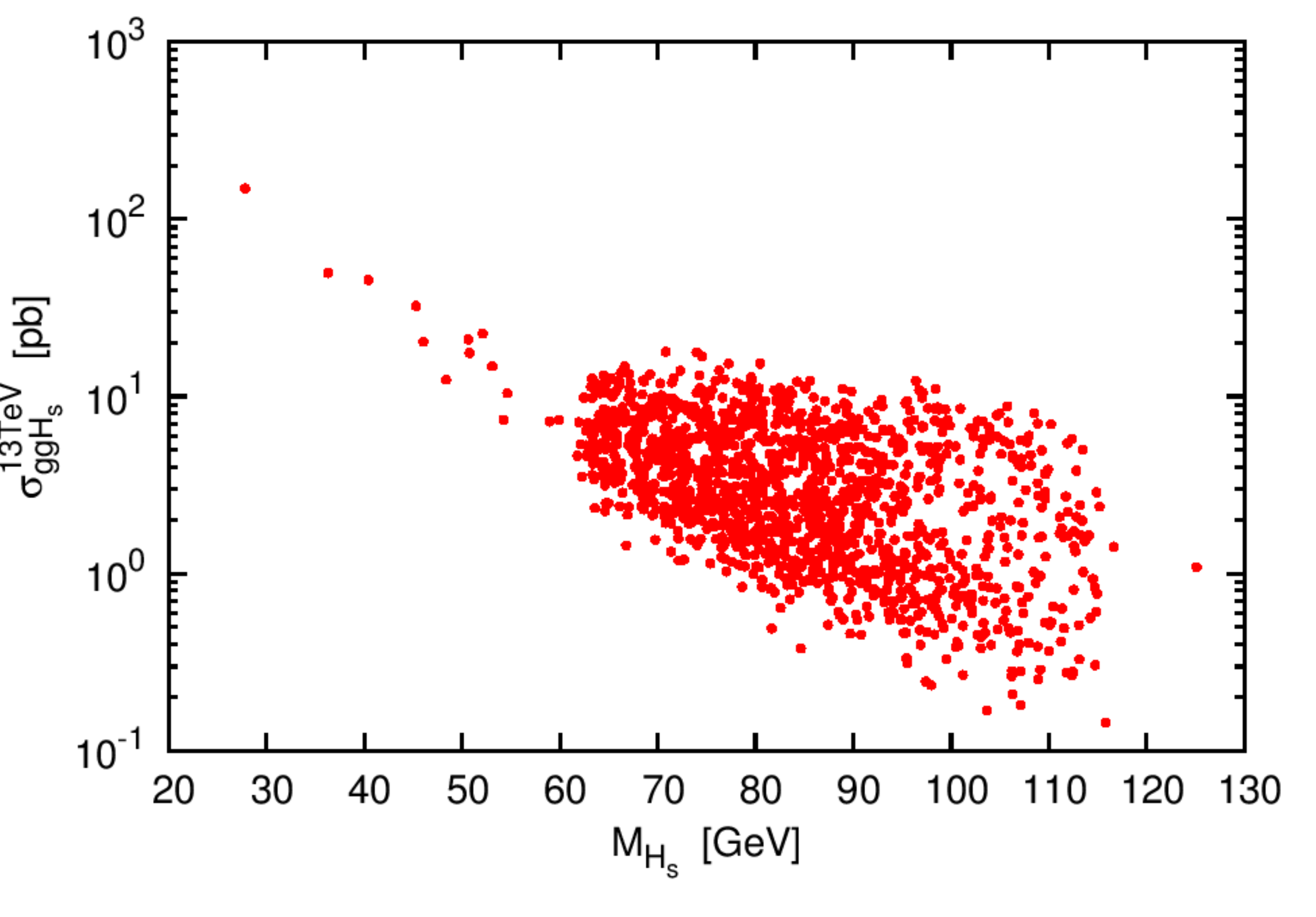}
\includegraphics[width=7.9cm]{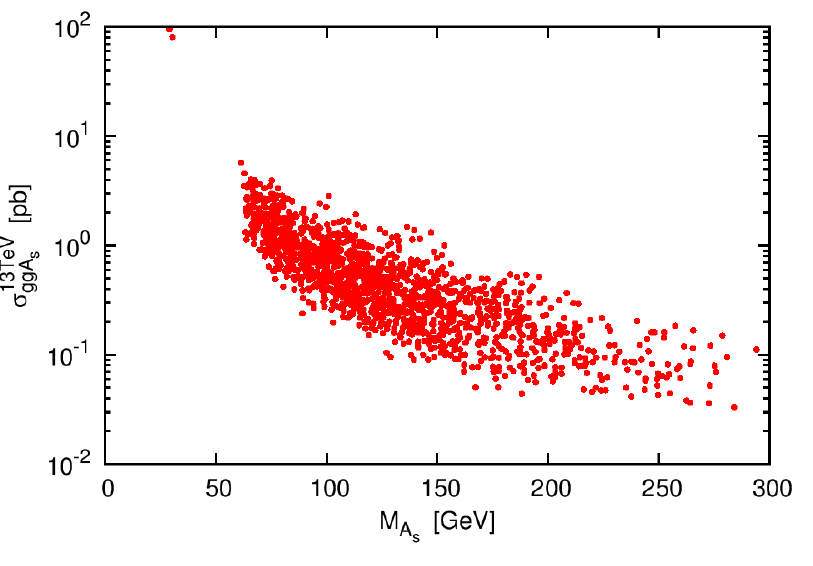}
\caption{The gluon fusion production cross section at
  $\sqrt{s}=13$~TeV for $H_s$ (left) and $A_s$ (right) as a function of
  their mass. \label{fig:ggfussinglet}} 
\end{center}
\end{figure}
Figure~\ref{fig:ggfussinglet} shows the gluon fusion production cross
sections for $H_s$ (left) and $A_s$ (right) at a c.m.~energy of
$\sqrt{s}=13$~TeV. For small $H_s$ masses the cross sections are
rather large with several 10's of pb. Above $\sim 90$~GeV the maximum values are
below 10~pb and can go down to ${\cal O}(0.1\mbox{ pb})$. 
With the exception of cross sections around 100~pb for very small
$A_s$ masses, the pseudoscalar production cross sections reach at most
6~pb in the lower mass range. However, already for masses around
150~GeV, the cross sections can be as small as 0.1~pb. \s

\begin{figure}[h!]
\begin{center}
\includegraphics[width=7.9cm]{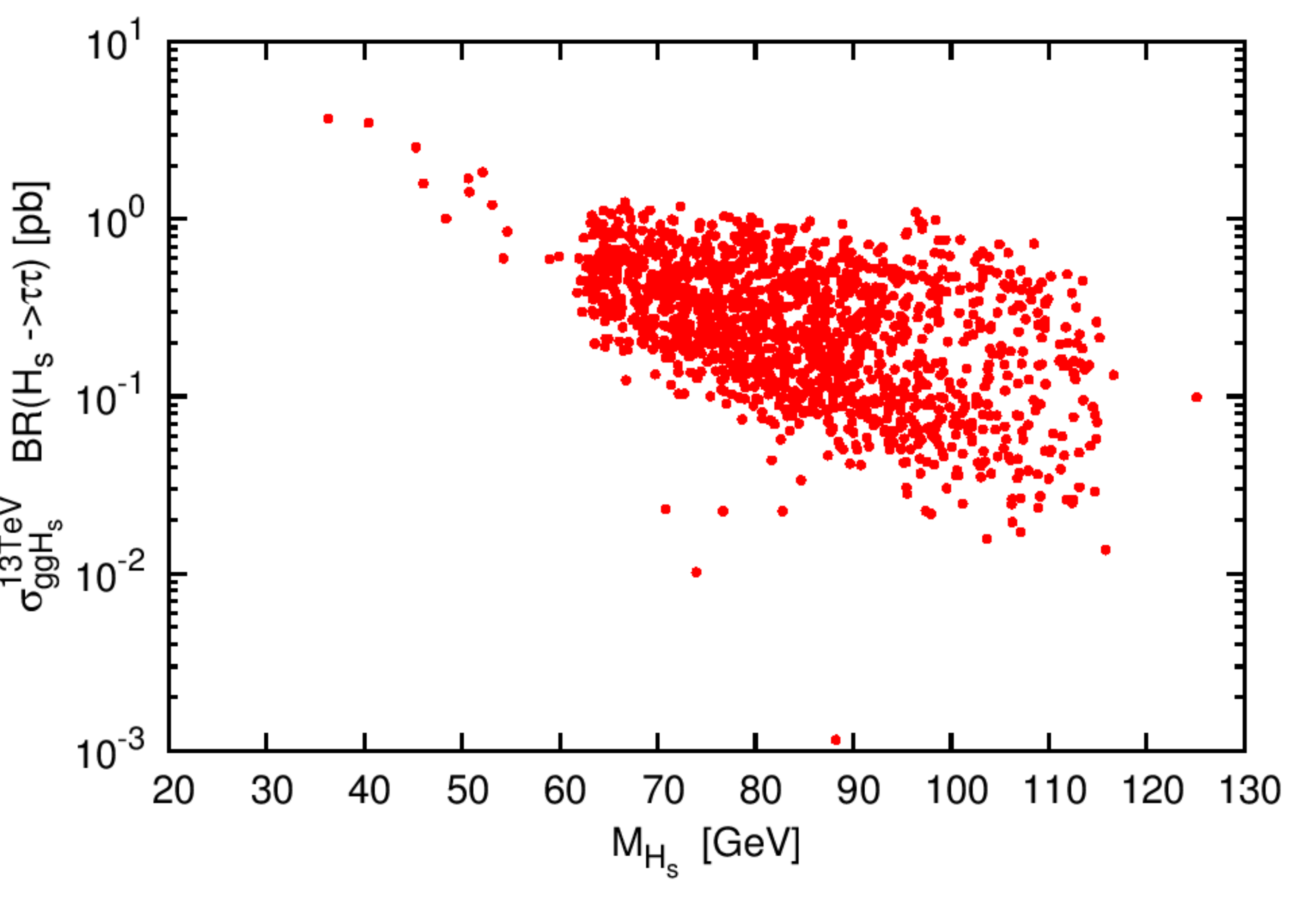} 
\includegraphics[width=7.9cm]{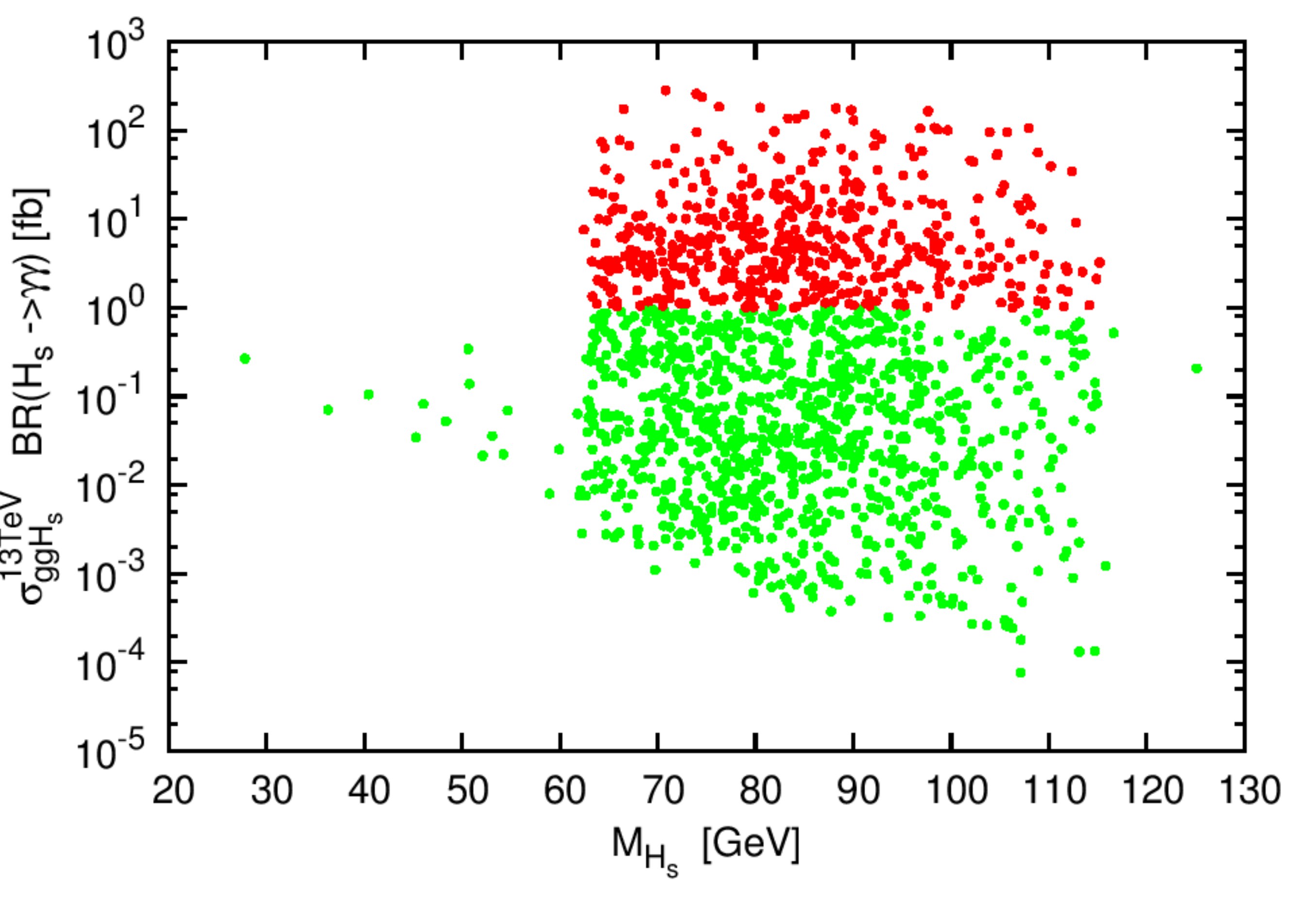} \\
\includegraphics[width=7.9cm]{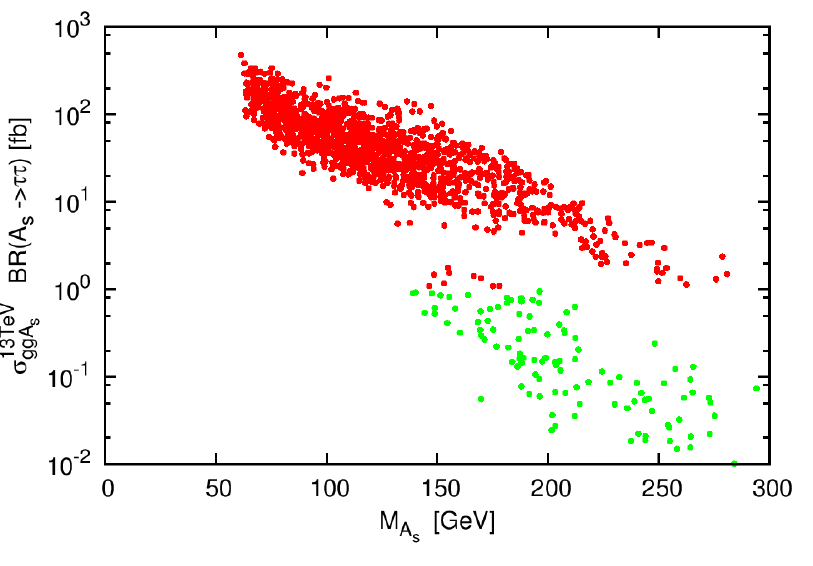} 
\includegraphics[width=7.9cm]{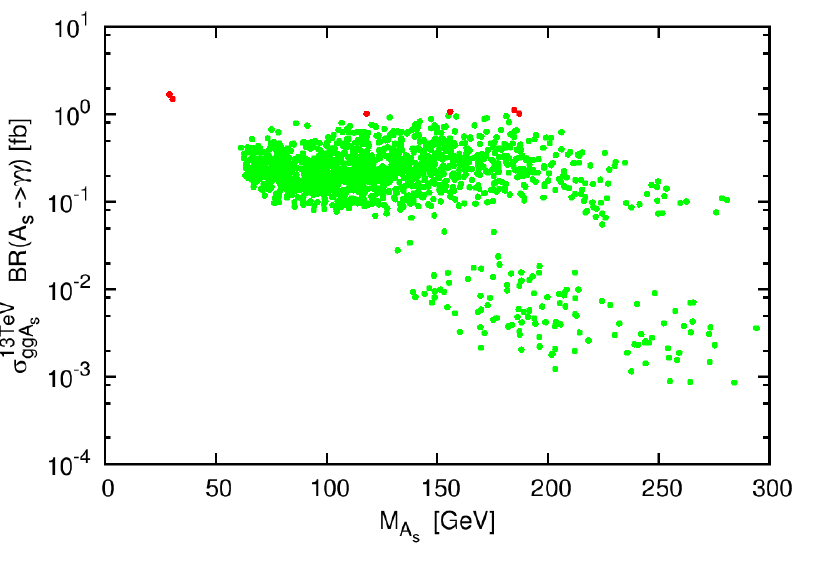}
\caption{Production rates for $H_s$ (upper) and $A_s$ (lower) production
  in gluon fusion with subsequent decay into the $\tau\tau$ (left) and
  $\gamma\gamma$ (right) final states as a
  function of the involved singlet mass, at $\sqrt{s}=13$~TeV. Red (green)
  points mark cross sections above 
  (below) 1~fb.\label{fig:directHsAs}}
\end{center}
\end{figure}

If the light Higgs bosons are very singlet-like their production rates
may become very small, in particular in the pseudoscalar case, as can
be inferred from 
Figs.~\ref{fig:directHsAs}, which show the production in the photon
and $\tau$-pair final states.\footnote{For the $b$-quark pair final state, the
cross sections, not shown here, are always above 1~fb and can reach
values of up to ${\cal O}(10 \mbox{ pb})$, for very light $H_s$ even
several tens of pb.}
In this case and/or if the
production rates also for the heavy Higgs bosons are small, further
alternative production channels should be exploited to increase the
discovery potential of the entire Higgs spectrum. This is subject
to the following two subsections, that discuss
Higgs-to-Higgs and Higgs-to-gauge-Higgs decays in the Natural NMSSM.

\subsection{Singlet-like Higgs Production from Higgs-to-Higgs
  Decays} 
In case not all neutral Higgs bosons can be discovered in direct production
with subsequent decay, alternative search channels might be given by
production from SUSY states decaying into Higgs bosons, or they
may be searched for in Higgs decays into a Higgs and gauge boson pair
as well as in Higgs-to-Higgs decays,  
{\it i.e.}
\beq
\sigma (gg\to \phi_i) \times BR(\phi_i \to \phi_j \phi_k) \times
BR(\phi_j \to XX) \times BR(\phi_k \to YY) \;, \label{eq:htohh}
\eeq
where $\phi_{i,j,k}$ generically denotes one of the five neutral Higgs
bosons\footnote{Of course in Eq.~(\ref{eq:htohh}) only these
  Higgs-to-Higgs decays are considered that are allowed by the quantum
numbers.} and with $M_{\phi_i} > M_{\phi_j} + M_{\phi_k}$. \s

\begin{figure}[!ht]
\begin{center}
\includegraphics[width=7.9cm]{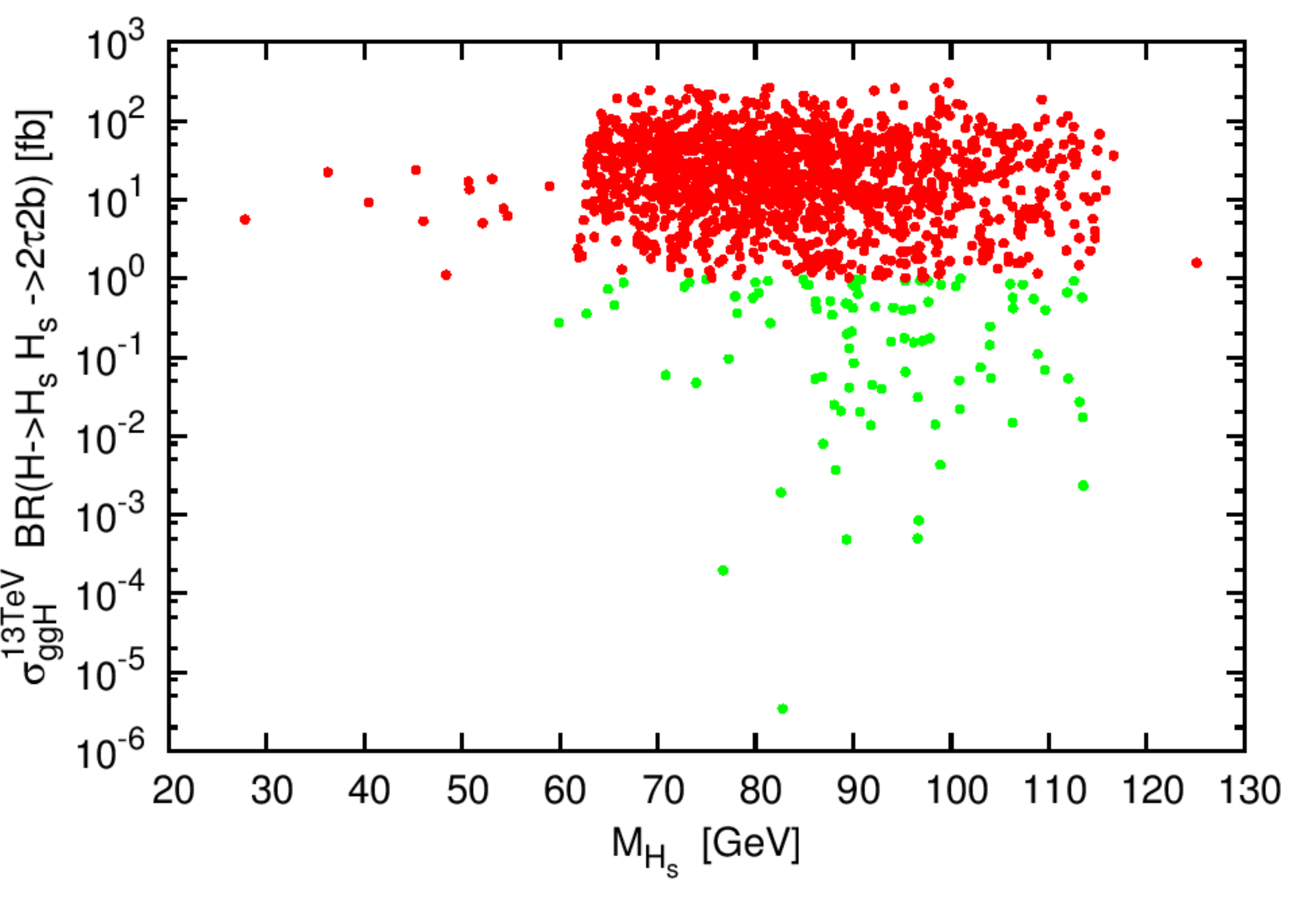}
\includegraphics[width=7.9cm]{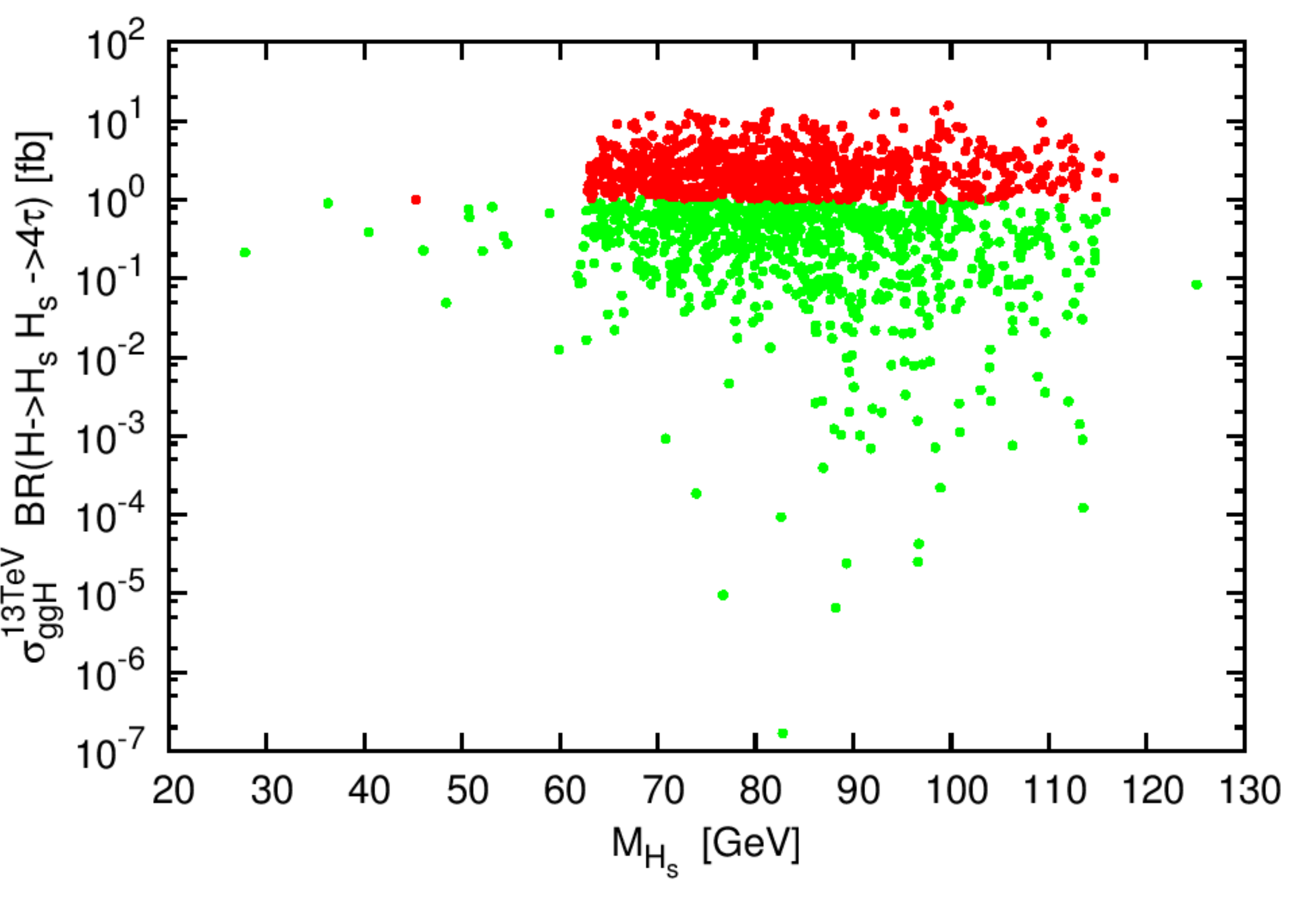} \\
\includegraphics[width=7.9cm]{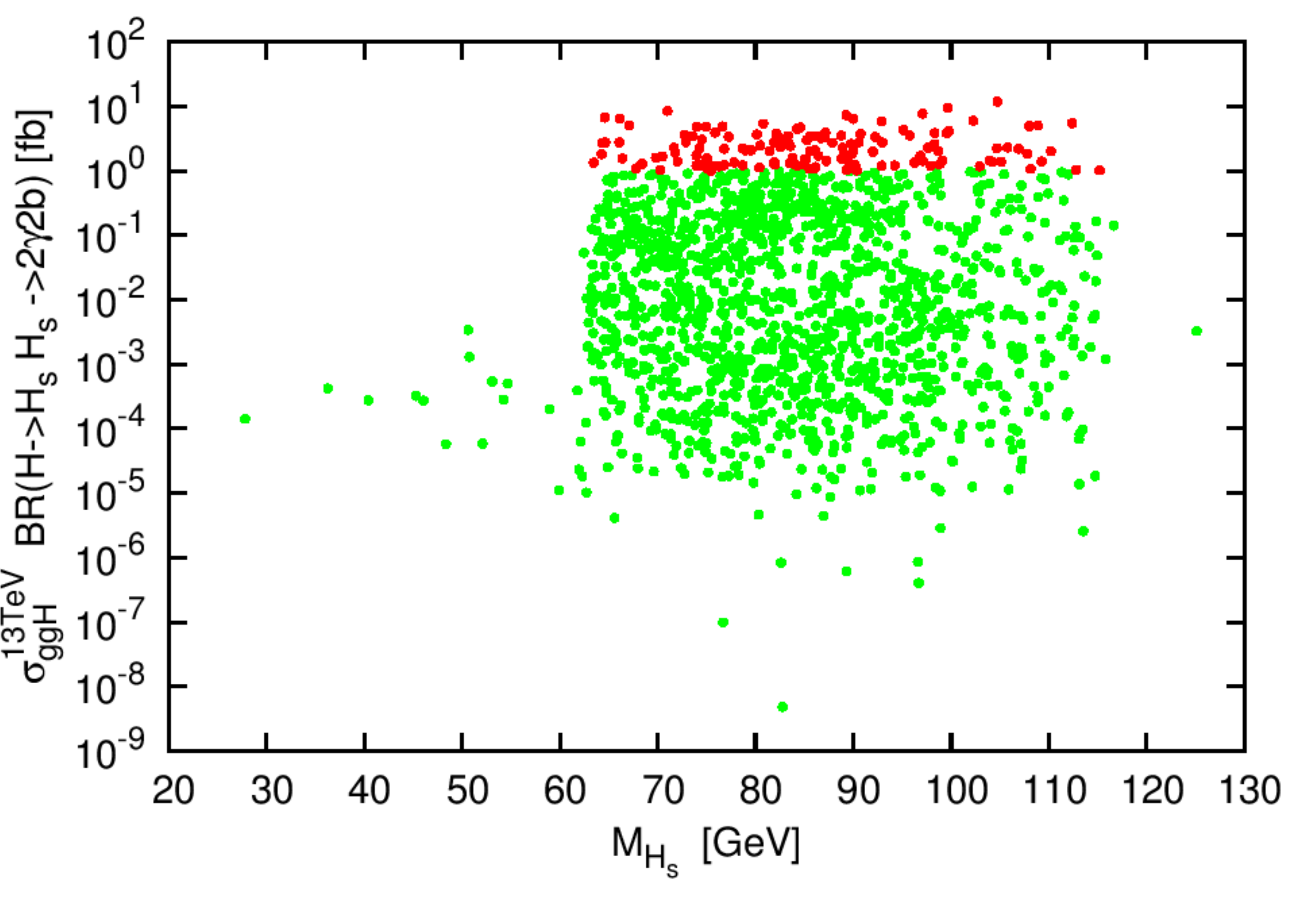} 
\includegraphics[width=7.9cm]{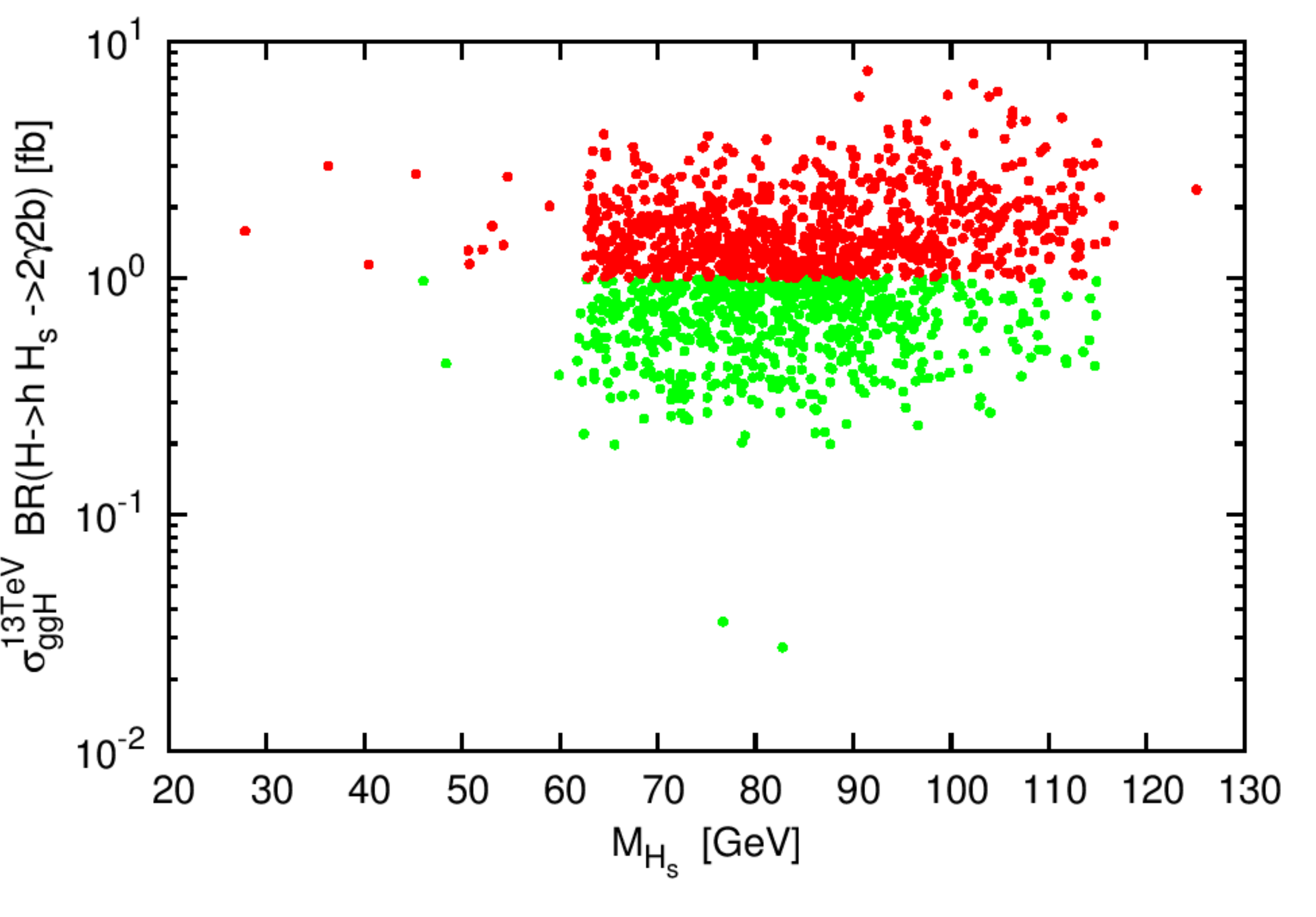}\\
\includegraphics[width=7.9cm]{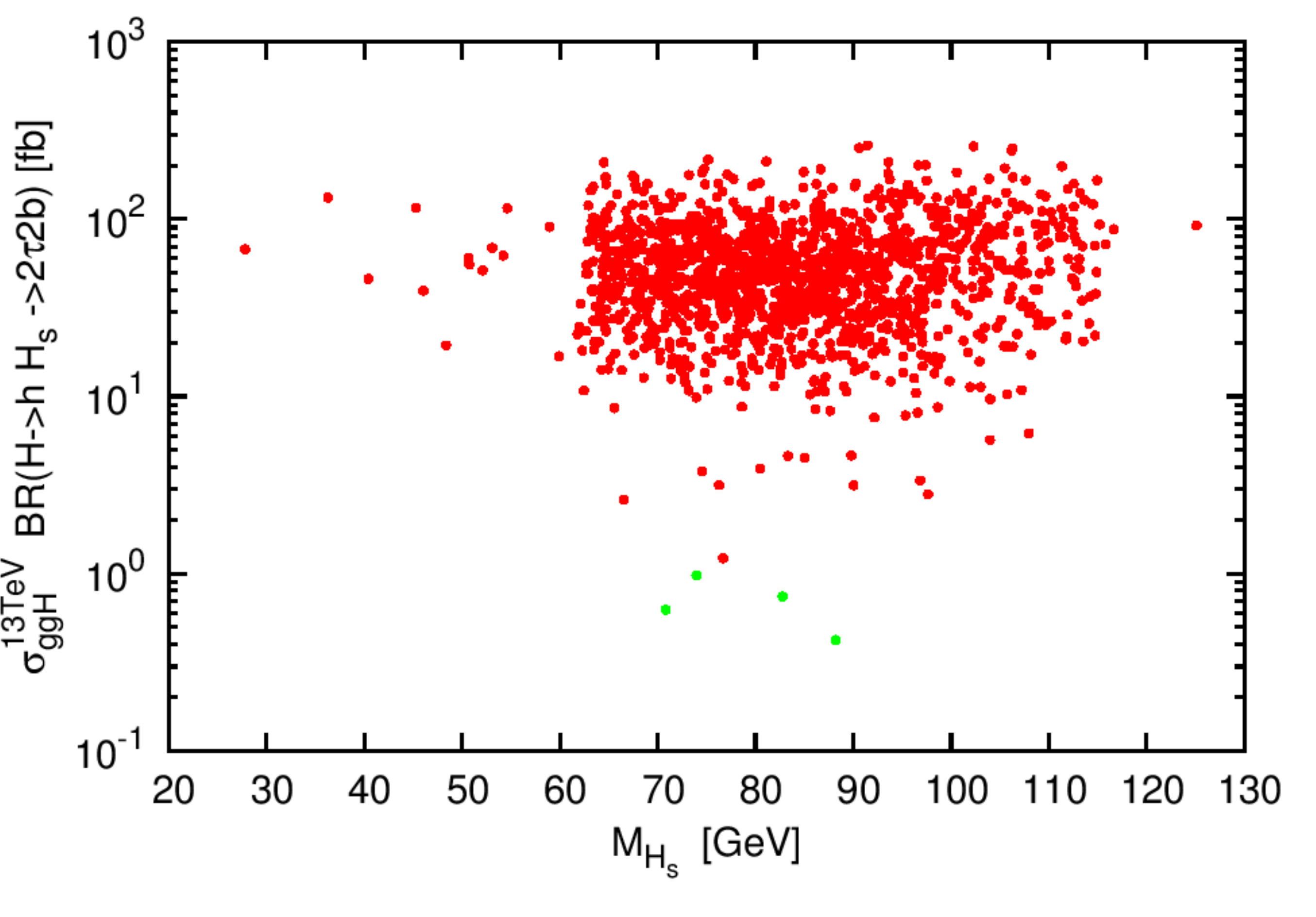}
\includegraphics[width=7.9cm]{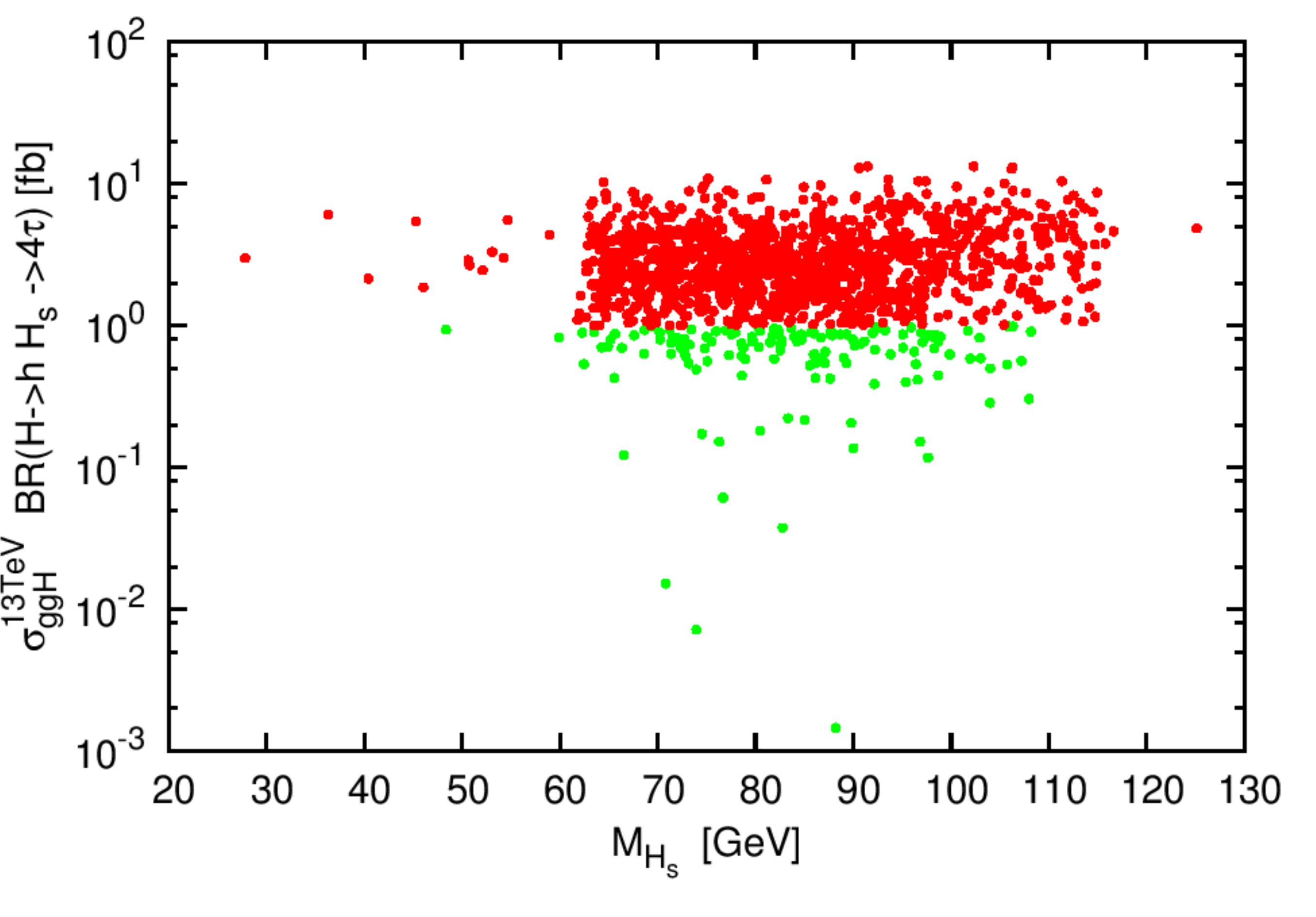} \\
\caption{Cross sections for $H_s H_s$ production (upper two rows) and
  $hH_s$ production (lower two rows) from $H$
  decay, in the $(2\tau)(2b)$ (upper/lower left), $4\tau$ (upper/lower right) and
  $(2\gamma)(2b)$ (middle) final states at $\sqrt{s}= 13$~TeV as a function
  of the singlet mass $M_{H_s}$. Red (green) points mark cross sections above
  (below) 1~fb.\label{fig:h3tohshsvarious}}
\end{center}
\end{figure}

The production rates for singlet Higgs pairs $H_sH_s$ and a singlet
plus SM-like Higgs, $H_s h$, from the heavy CP-even Higgs boson
$H$ are shown in Figs.~\ref{fig:h3tohshsvarious}, including their subsequent decay
into SM particles. Red (green) points refer to cross sections above
(below) 1~fb. The figures show, that in the $(2\tau)(2b)$ final state the
cross sections are above 1~fb, with the exception of a few points in
the $H_s H_s$ production. They can even reach values of several hundred
fb. The $4\tau$ final state is suppressed by a factor 10 compared to
the former, but still a good fraction of scenarios, in particular for
$h H_s$ production, reaches cross sections larger than 1~fb. As
expected, the $(2\gamma) (2b)$ final state rates are smaller, but
also here we have scenarios with rates exceeding the fb level. In principle $H$ could
also decay into a pair of pseudoscalar singlets. However, in this case
all the decay rates turned out to be tiny so that we do not display
them here. \s

\begin{figure}[!h]
\begin{center}
\includegraphics[width=7.9cm]{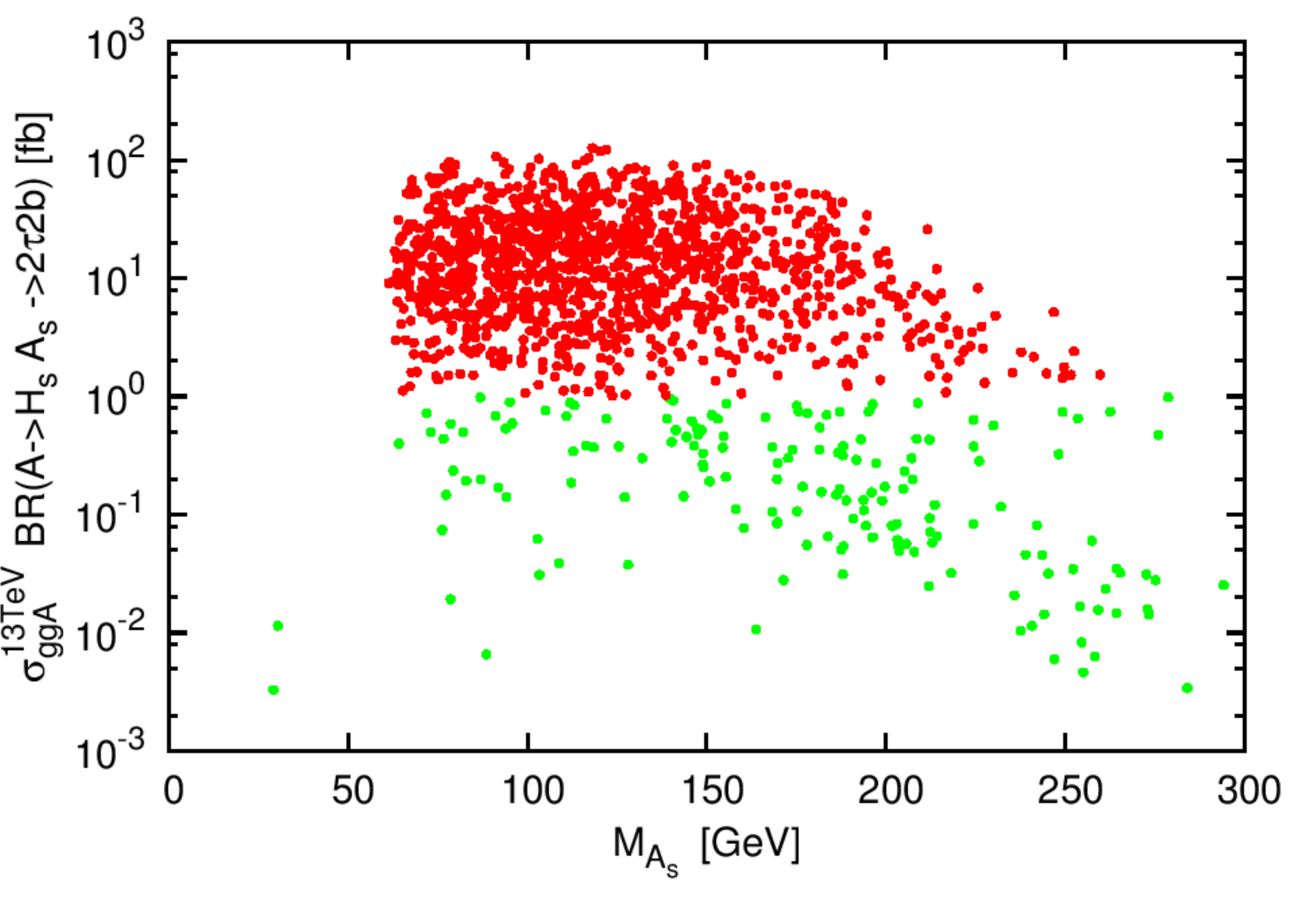} 
\includegraphics[width=7.9cm]{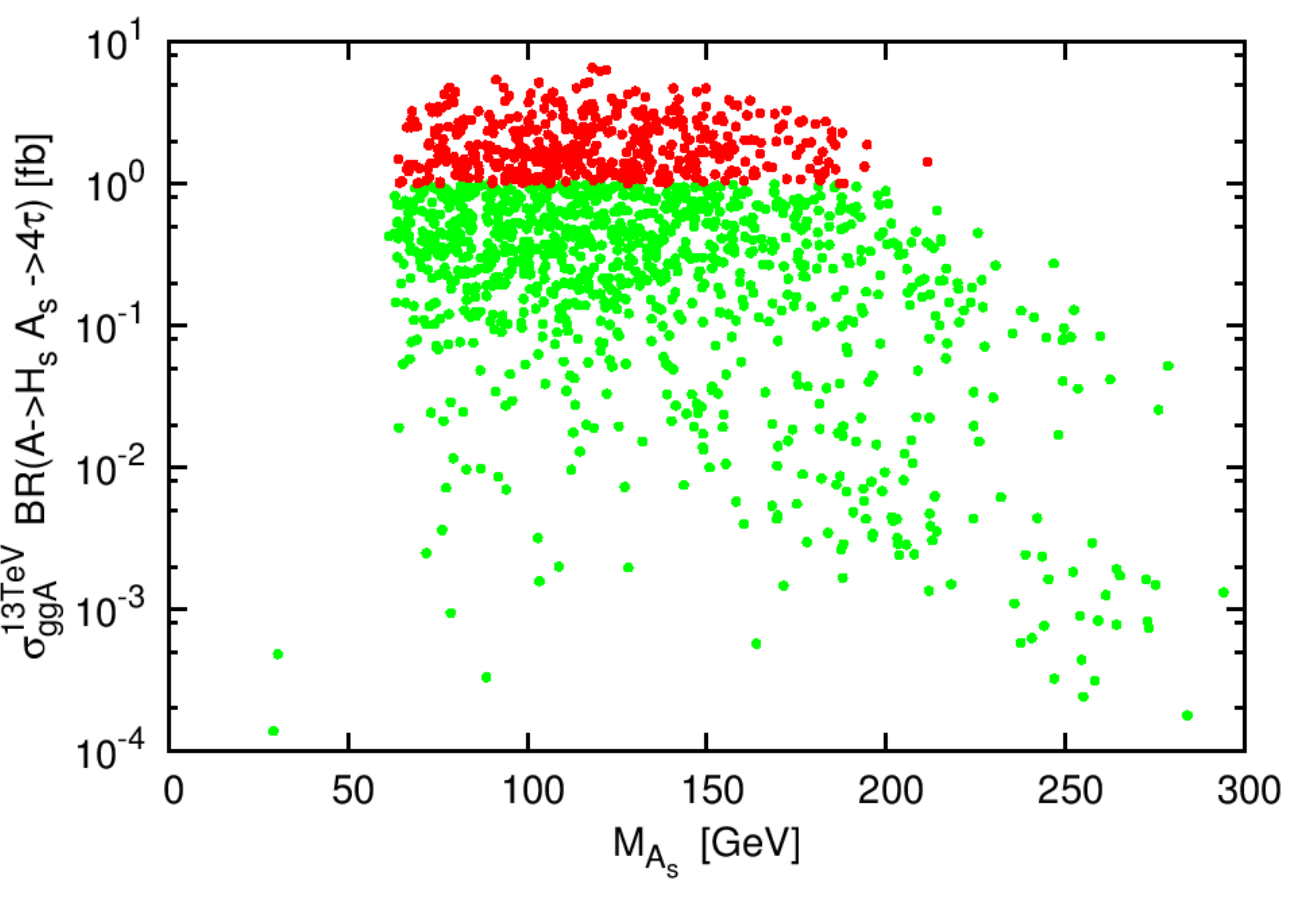} \\
\includegraphics[width=7.9cm]{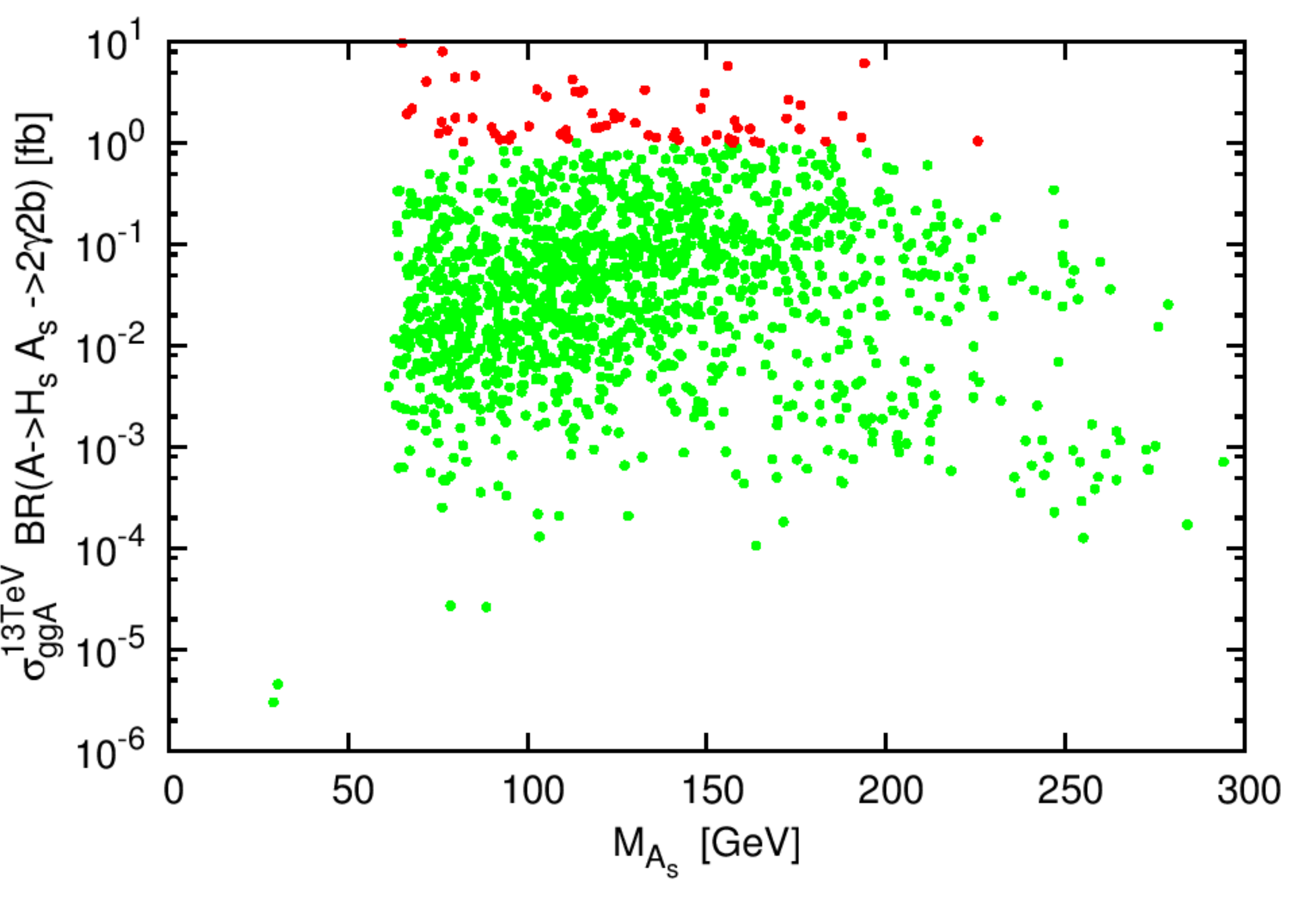} 
\includegraphics[width=7.9cm]{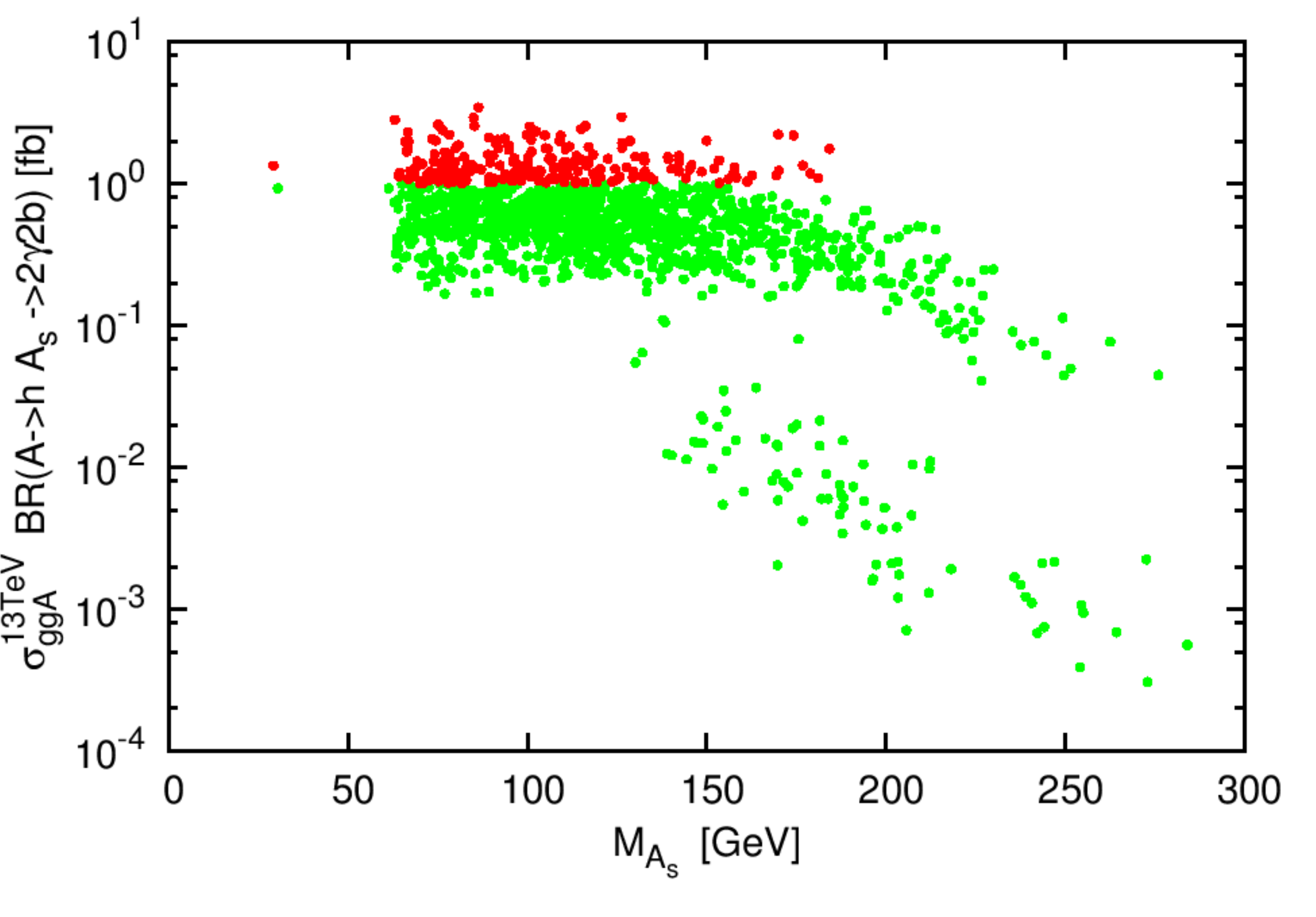} \\
\includegraphics[width=7.9cm]{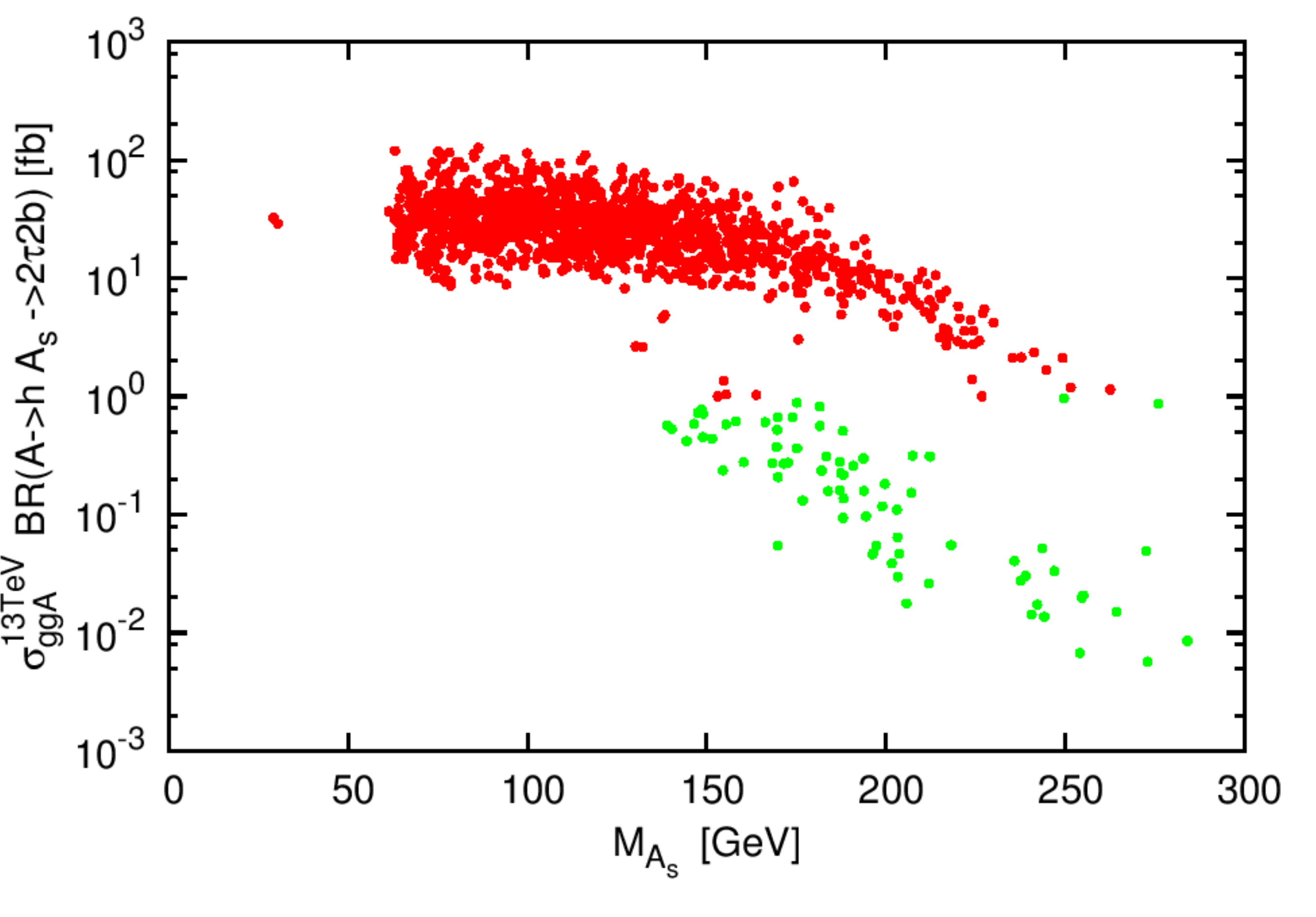} 
\includegraphics[width=7.9cm]{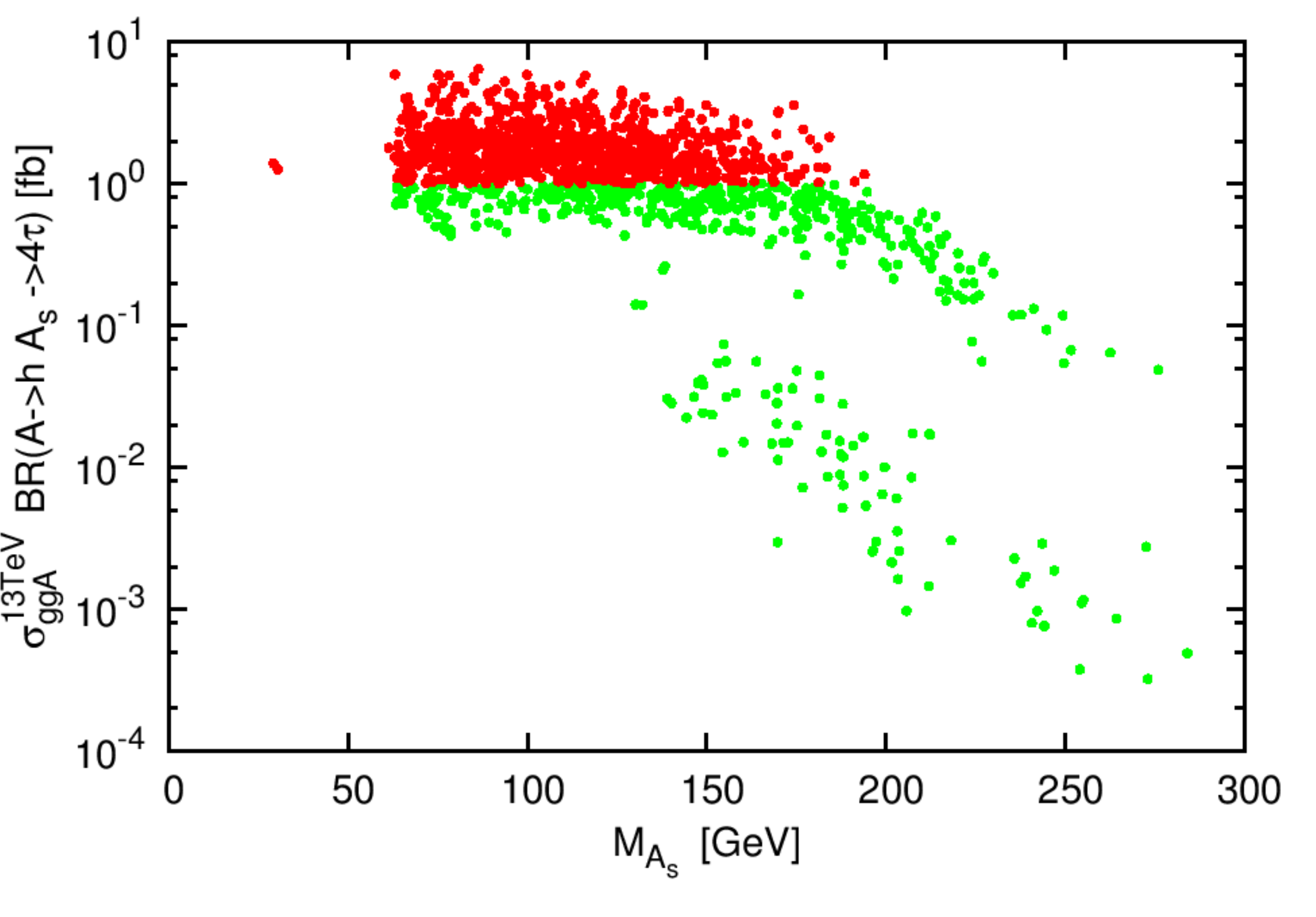} 
\caption{Cross sections for $A_s H_s$ production (upper two rows) and
  $h A_s$ production (lower two rows) from $A$
  decay, in the $(2\tau)(2b)$ (upper/lower left), $4\tau$ (upper/lower right) and
  $(2\gamma)(2b)$ (middle) final states at $\sqrt{s}= 13$~TeV as a function
  of the singlet mass $M_{A_s}$. Red (green) points mark cross sections above
  (below) 1~fb.\label{fig:atohHsandAsvarious}}
\end{center}
\end{figure}

Singlet Higgs bosons can also be produced from heavy pseudoscalar
decays into $H_s A_s$ or $h A_s$. The production rates into the
$(2\tau)(2b)$, $4\tau$ and $(2\gamma)(2b)$ final states are
shown in Figs.~\ref{fig:atohHsandAsvarious}. Again we have a
non-negligible fraction of scenarios that lead to rates exceeding 1~fb in the
pure fermionic final states. In the $(2\gamma)(2b)$ final states the
majority of points is below 1~fb. \s

Finally note, that we have shown here only the simplest final state
combinations. As will be evident from the benchmark discussion, however, there
can also be more complicated and more exotic final states arising from
two Higgs-to-Higgs decays in one decay chain. 

\subsection{Singlet-like Higgs Production from Higgs-to-Gauge-Higgs
  Decays} 
\begin{figure}[h!]
\begin{center}
\includegraphics[width=7.9cm]{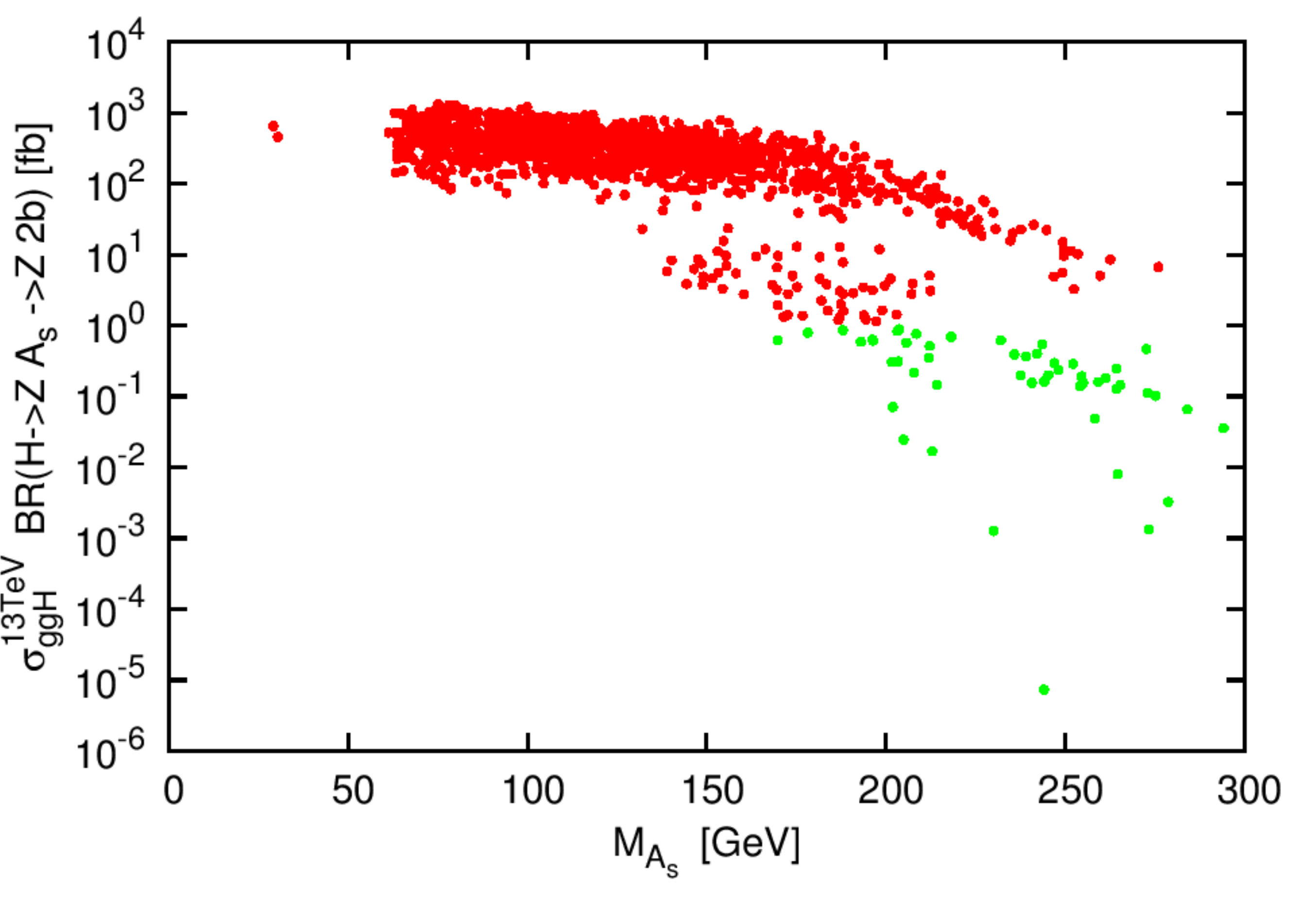} 
\includegraphics[width=7.9cm]{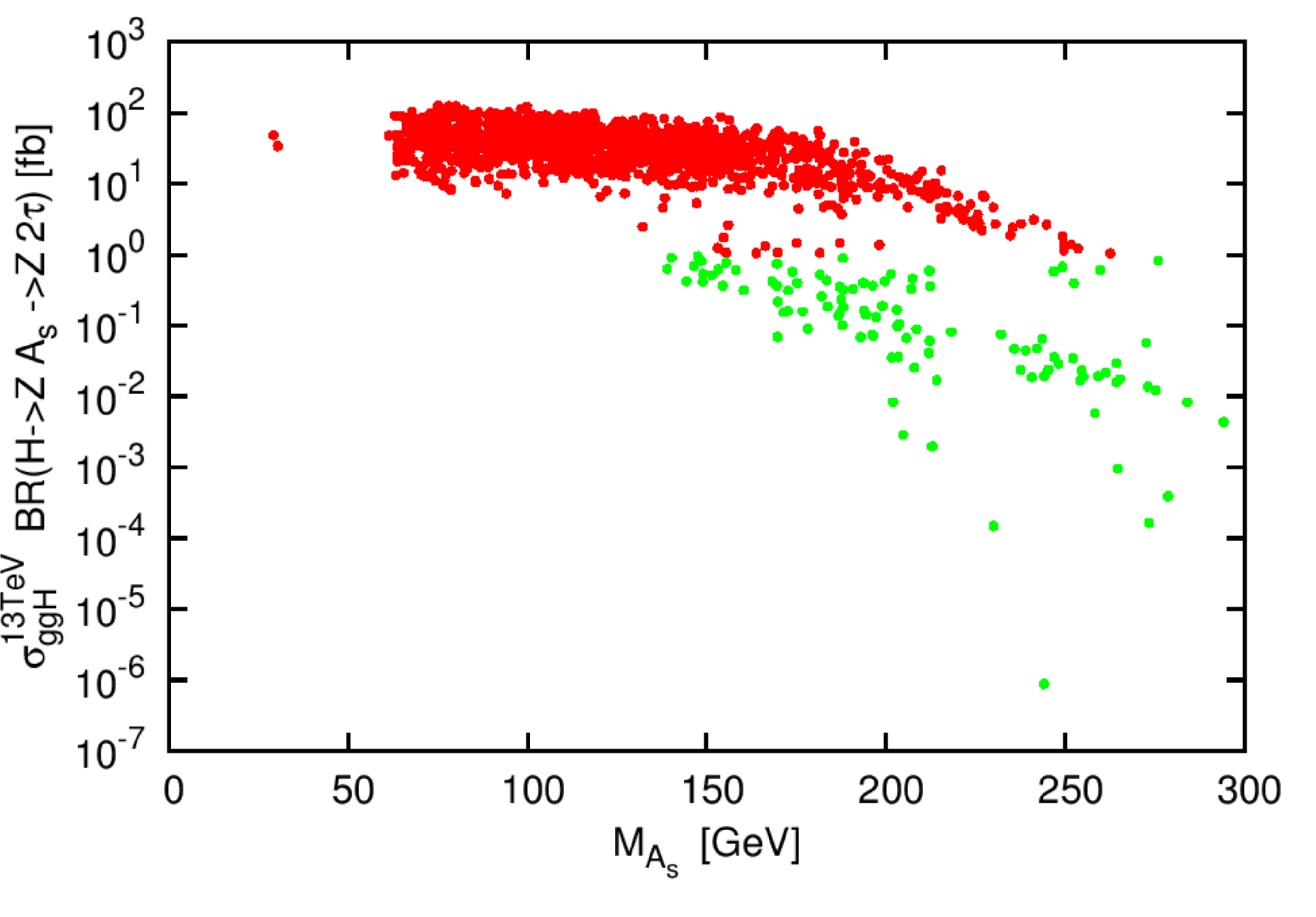} \\
\includegraphics[width=7.9cm]{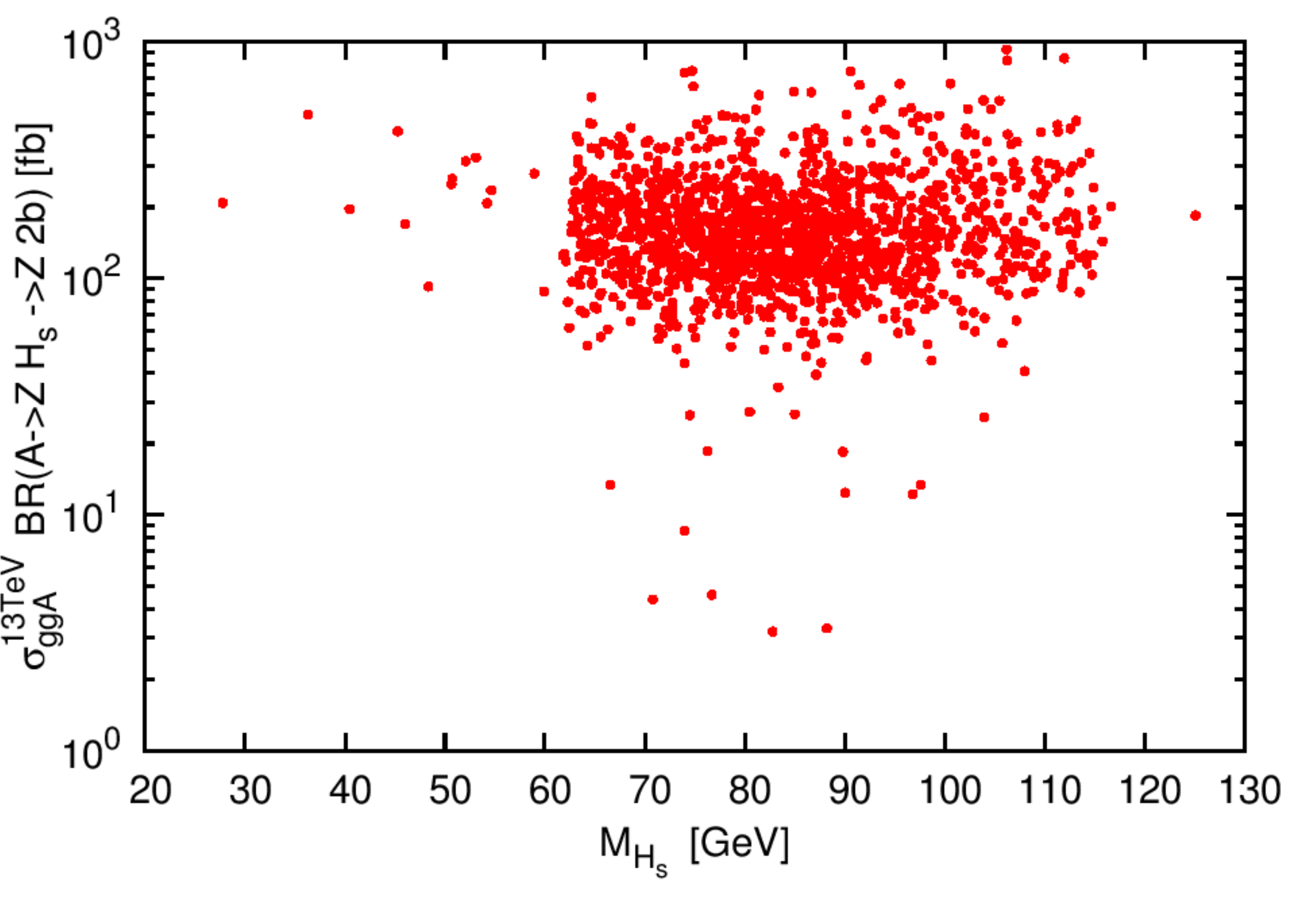}
\includegraphics[width=7.9cm]{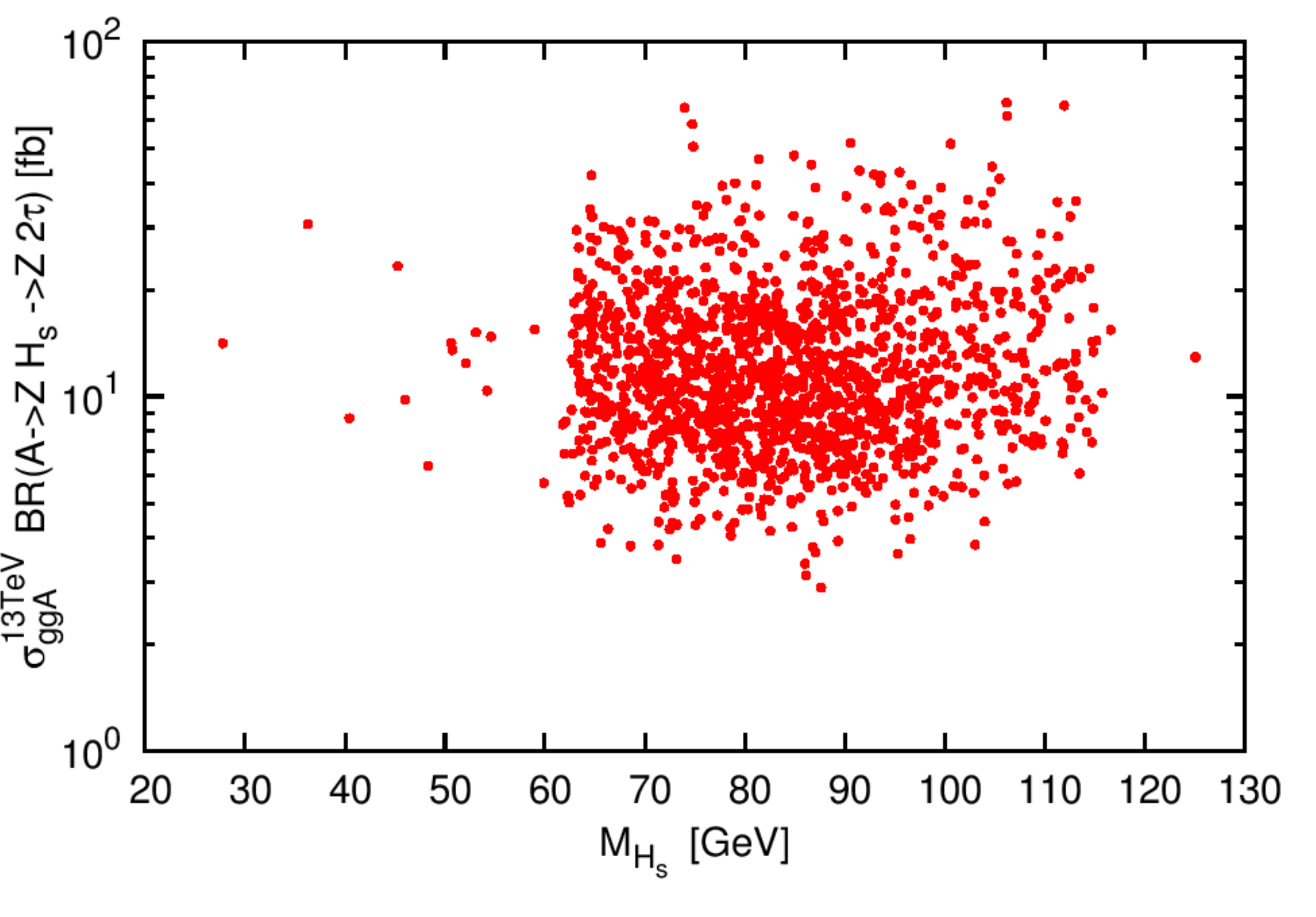}
\caption{Cross sections for $Z A_s$ production from $H$ decay (upper)
  and for $Z H_s$ production from $A$ decay (lower) in the $Z+2b$
  (left) and $Z+2\tau$ (right) final state, respectively, as a
  function of the involved singlet mass, at $\sqrt{s}=13$~TeV. Red (green)
  points mark cross sections above 
  (below) 1~fb.\label{fig:gaugehiggs}}
\end{center}
\end{figure}

Singlet Higgs boson production is also possible via heavy Higgs decays
into a singlet Higgs and a massive gauge boson. While these cross
sections do not involve trilinear Higgs couplings, they lead to
interesting decay rates. The cross sections for the
$H \to Z A_s$ and $A \to Z H_s$ decays in the $Z+2b$ and $Z+2\tau$
final states are shown in Fig.~\ref{fig:gaugehiggs}. The rates will go
down by another factor 30 due to the $Z$ boson decay into
fermions. Still, almost all parameter points have cross sections above
1~fb up to pb in the $Z +2b$ final state, reduced by a factor 10 in
the $Z + 2\tau$ final state, and with the exception of some points in
the chain via the $Z A_s$ final state for pseudoscalar singlet masses
above $\sim 150$~GeV. These decay chains can hence be used as
additional and complementary production channels for singlet Higgs
bosons. \s 

In summary, in general the complete Higgs spectrum of the Natural
NMSSM should be accessible at the high-energy LHC by exploiting the
entity of Higgs production channels through direct production in gluon
fusion\footnote{We remind the reader that for simplification we only
  took into account the dominant gluon fusion production. In addition
  to this also the other LHC production mechanisms will contribute and
  even ameliorate the discovery prospects.} with subsequent decay into
SM particles and the production from 
heavy Higgs boson decays into lighter Higgs pairs and/or Higgs plus
gauge boson final states. Decays of heavy SUSY particles into lighter
Higgs bosons, not discussed here, may possibly add to the discovery
contours. This part of the NMSSM parameter space will therefore either 
strongly be constrained at the next LHC run to come, or lead to the
discovery of beyond the SM physics revealed by the Higgs sector.

\section{Benchmarks for Higgs-to-Higgs Decays \label{sec:scenarios}}
Higgs decays, as given in Eq.~(\ref{eq:htohh}), into Higgs pairs that
subsequently decay into SM 
particles are interesting for the discovery of the heavier Higgs boson
$\phi_i$ provided its production cross section is large enough and the
branching ratio into the lighter Higgs pair dominates over its
branching ratios into the SM final states. For the lighter Higgs bosons
$\phi_j, \phi_k$ this production mechanism becomes interesting in case
their direct production is strongly suppressed due to them being very
singlet-like. Furthermore, Higgs-to-Higgs decays are interesting in themselves as
they give access to the trilinear Higgs self-couplings, which can then be used to
reconstruct the Higgs potential as the last step in the experimental
verification of the Higgs mechanism \cite{selfcouplings}. \s

In this section, we concentrate on Higgs-to-Higgs decays and present
exemplary scenarios that are compatible with all constraints discussed
above and that arise from our scan of the parameter space defined in
Eqs.~(\ref{eq:cond1})-(\ref{eq:cond6}), {\it i.e.}~not restricted to 
the Natural NMSSM. 
The presented scenarios not only lead to sufficiently large cross sections for
double Higgs production from heavier Higgs decays but some of them
also entail exotic multi-photon and/or multi-fermion final
states. The possibility of such final states should be taken 
into consideration by the experiments in order not to miss possible
Higgs signatures, that can appear in multi-Higgs models like
the NMSSM without particular fine-tuning of the parameters. We also
include scenarios featuring a light CP-even Higgs boson with a mass
below 10~GeV. The presented scenarios highlight specific features of
NMSSM Higgs decay chains and can be viewed as benchmark scenarios for
future LHC searches. \s 

In the following we will again use the convenient notation introduced
in section~\ref{sec:naturalnmssm}. Note, however, that the SM-like
Higgs boson $h$ can now be $H_1$ or $H_2$. The more doublet-like
scalar $H$ can be given by $H_2$ or $H_3$, depending on the scenario. 
The pseudoscalar doublet state $A$, as before, is given by $A_2$, and
the singlet like CP-odd $A_s$ by $A_1$. The singlet-like scalar $H_s$,
however, can be $H_1$, $H_2$ or even $H_3$. \s

The benchmark scenarios that we present can be classified as follows:
\begin{itemize}
\item[A)] $H_2=h$, $H_1=H_s$, $\tan\beta$ small, light spectrum $\lsim
  350$~GeV: \\
The scenarios A are characterized by the second lightest Higgs being
SM-like, the lightest Higgs boson singlet-like and small
$\tan\beta$. The overall spectrum is rather light with maximal Higgs
mass values of $M_H=341$~GeV and $M_A=347$~GeV. 
\begin{itemize}
\item[A.1:] The CP-odd singlet is close in mass to the SM-like Higgs
  with $M_{A_s} = 119$~GeV. The heavy Higgs bosons have masses around
  341 and 347~GeV. 
\item[A.2:] The CP-odd singlet is light with $M_{A_s}=79$~GeV, the
  heavy Higgs bosons with masses around 326~GeV are somewhat lighter
  than in scenario A.1. 
\end{itemize}
\item[B)] $H_1=h$, $H_2=H_s$, $\tan\beta$ small: \\
The scenarios B are characterized by the lightest Higgs being
SM-like, the second lightest Higgs boson singlet-like and small
$\tan\beta$. 
\begin{itemize}
\item[B.1:] The CP-odd singlet is light with $M_{A_s} = 73$~GeV and
  the overall spectrum as well with the heavy $H$ mass of 323~GeV and
  $M_A=312$~GeV. 
\item[B.2:] The pseudoscalar singlet with $M_{A_s}=107$~GeV is still
  lighter than the SM-like Higgs, but the overall spectrum is heavier
  with $M_H=556$~GeV and $M_A=578$~GeV.
\item[B.3:] The singlet mass $M_{A_s}=133$~GeV is now heavier than
  $M_h$. The maximum mass is given by $M_H=463$~GeV, closely followed
  by $M_A=457$~GeV. 
\end{itemize}
\item[C)] $\tan\beta> 5$: \\
Here we explicitly chose a scenario where $\tan\beta=17$ is large. The
lightest Higgs boson is SM-like $H_1=h$. The spectrum is peculiar in
the sense that the heavy scalar with a mass of $\sim 3.5$~TeV is very
heavy. Additionally, $H_3$ is singlet-like. The heavy pseudoscalar and
second lightest scalar are doublet-like with masses of ${\cal
  O}(600 \mbox{ GeV})$. The pseudoscalar singlet is much heavier than
$h$, $M_{A_s}=312$~GeV. 
\item[D)] The SM-like $h$, given by $H_2$, can decay into Higgs pairs: \\
The scenarios D allow for decays of the SM-like Higgs $h$ into lighter
singlet Higgs pairs. 
\begin{itemize}
\item[D.1:] The CP-even singlet is very light, $M_{H_s}<10$~GeV, and
  $h$ decays into CP-even singlet pairs are non-negligible. The pseudoscalar
  singlet is heavier than 125~GeV, and heavy Higgs bosons have masses
  up to 793~GeV. The $\tan\beta$ value is small and below 5.
\item[D.2:] Here the CP-odd singlet is the lightest Higgs, and the SM-like 
  Higgs $h$ decays into $A_s$ pairs. The scalar singlet $H_s$ with 112~GeV is now
  closer in mass to $h$. The heavy Higgs masses are very large and of
  ${\cal O} (1.3 \mbox{ TeV})$. We have a medium $\tan\beta = 6.4$.
\end{itemize}
\end{itemize}
In all scenarios the lightest pseudoscalar is singlet-like. \s

In the subsequent subsections the details on these benchmark scenarios
will be given. The discovery prospects of the various Higgs bosons
shall be highlighted with particular focus on the Higgs-to-Higgs
decays and the related signatures. 

\subsection{Scenarios A - $H_2 \equiv h$ and small $\tan\beta$} 

\begin{table}[b!]
\begin{center}
{\small \begin{tabular}{|l|l|l|l|}
\hline \hline
\textbf{A.1 (Point ID 3877)} & \multicolumn{3}{l|}{\textbf{Scenario}}
\\ \hline \hline 
$M_{H_1},M_{H_2},M_{H_3}=M_{H_s},M_{h},M_{H}$ & 90.3 GeV & 126.8 GeV & 341.3 GeV \\ \hline
$M_{A_1},M_{A_2}=M_{A_s},M_{A}$ & 118.5 GeV & \multicolumn{2}{l|}{346.7 GeV} \\ \hline 
$|S_{H_1 h_s}|^2, |P_{A_1 a_s}|^2$ & 0.97 & \multicolumn{2}{l|}{0.94} \\ \hline \hline 
$\mu_{\tau\tau}$, $\mu_{bb}$ & 1.09 & \multicolumn{2}{l|}{1.08} \\ \hline
$\mu_{ZZ}$, $\mu_{WW}$, $\mu_{\gamma\gamma}$  & 0.85 & 0.85 & 0.88 \\
\hline \hline
$\tan\beta$, $\lambda$, $\kappa$ & 1.66 & 0.64 & 0.11 \\ \hline
$A_{\lambda}$, $A_ {\kappa}$, $\mu_{\text{eff}}$ & 338.0 GeV & -71.2 GeV & 162.8 GeV \\ \hline
$A_ t$, $A_b$, $A_{\tau}$ & 181.1 GeV & -1530.0 GeV & 87.2 GeV \\ \hline
$M_1$, $M_2$, $M_3$ & 440.0 GeV & 813.7 GeV & 1710.2 GeV \\ \hline
$M_{Q_3}=M_{t_R}$, $M_{b_R}$ & 1827.5 GeV  & \multicolumn{2}{l|}{3 TeV} \\ \hline
$M_{L_3}=M_{\tau_R}$, $M_{\text{SUSY}}$ & 1663.7 GeV &
\multicolumn{2}{l|}{3 TeV} \\ \hline \hline
\end{tabular}}
\caption{The parameters defining scenario A.1, together with the Higgs
boson masses, singlet components and reduced signal rates of $h$. \label{table:scenA1}}
\end{center}
\end{table}
\begin{table}[!ht]
\begin{center}
{\small \begin{tabular}{|l|l|l|l|}
\hline \hline
\textbf{A.2 (Point ID 2212)} & \multicolumn{3}{l|}{\textbf{Scenario}}
\\ \hline \hline 
$M_{H_s},M_{h},M_{H}$ & 98.6 GeV & 125.6 GeV & 325.9 GeV \\ \hline
$M_{A_s},M_{A}$ & 78.6 GeV & \multicolumn{2}{l|}{325.5 GeV} \\ \hline
$|S_{H_1h_s}|^2, |P_{A_1 a_s}|^2$ & 0.89 & \multicolumn{2}{l|}{0.96} \\ \hline \hline
$\mu_{\tau\tau}$, $\mu_{bb}$ & 1.05 & \multicolumn{2}{l|}{0.93} \\ \hline
$\mu_{ZZ}$, $\mu_{WW}$, $\mu_{\gamma\gamma}$  & 0.86 & 0.87 & 0.90 \\ 
\hline \hline
$\tan\beta$, $\lambda$, $\kappa$ & 1.69 & 0.56 & 0.12 \\ \hline
$A_{\lambda}$, $A_ {\kappa}$, $\mu_{\text{eff}}$ & -259.2 GeV & -22.8 
GeV & -147.4 GeV \\ \hline
$A_ t$, $A_b$, $A_{\tau}$ & 1927.4 GeV & -948.9 GeV & 1621.4 GeV \\ \hline
$M_1$, $M_2$, $M_3$ & 755.9 GeV & 646.7 GeV & 2424.9 GeV \\ \hline
$M_{Q_3}=M_{t_R}$, $M_{b_R}$ & 2468.3 GeV  & \multicolumn{2}{l|}{3 TeV} \\ \hline
$M_{L_3}=M_{\tau_R}$, $M_{\text{SUSY}}$ & 1623.0 GeV &  
\multicolumn{2}{l|}{3 TeV} \\ \hline \hline
\end{tabular}}
\caption{The parameters defining scenario A.2, together with the Higgs
boson masses, singlet components and reduced signal rates of $h$. \label{table:scenA2}}
\end{center}
\end{table}

The NMSSM-specific and soft SUSY breaking parameters defining scenarios
A.1 and A.2 are given in Table~\ref{table:scenA1} and
\ref{table:scenA2}, respectively. The relevant NMSSM Higgs
signal rates are summarised in
Table~\ref{tab:ratesscenA1} for A.1 and Table~\ref{tab:ratesscenA2} for
A.2. They feature a SM-like Higgs boson given by the next-to-lightest
CP-even Higgs. In both scenarios the couplings of the singlets $H_s$ and $A_s$
to gluons amount to ${\cal O}(10-20\%)$ of the corresponding SM Higgs
coupling of same mass leading to gluon fusion production cross
sections of $1-2.4$~pb. As the $b\bar{b}$ final state is difficult to
detect, they could be searched for in $\tau\tau$ or even
$\gamma\gamma$ final states. The heavy Higgs bosons $H$ and $A$ are
produced with ${\cal O}(4\mbox{ pb})$ cross sections and have sizeable
decay rates into Higgs pairs involving $A_s$ and $H_s$, and also into $Z$ and
a Higgs singlet, so that alternative search channels are given by
Higgs cascade decays. \s

\begin{table}[!ht]
\begin{center}
\vspace*{-0.2cm}
{\small
\begin{tabular}{|l|l|l|l|}
\hline \hline
\textbf{A.1 (Point ID 3877)} & \multicolumn{3}{l|}{\textbf{Signal Rates}} \\
\hline \hline 
$\sigma(ggH_s)$  & \multicolumn{3}{l|}{2.37 pb} \\ \hline
$\sigma(ggH_s)\text{BR}(H_s \rightarrow b\bar{b})$  & \multicolumn{3}{l|}{2.04 pb} \\ \hline
$\sigma(ggH_s)\text{BR}(H_s \rightarrow \tau\tau)$  & \multicolumn{3}{l|}{204.82 fb} \\ \hline
$\sigma(ggH_s)\text{BR}(H_s\rightarrow \gamma \gamma)$  &
\multicolumn{3}{l|}{2.74 fb} \\ \hline \hline
$\sigma(ggH)$  & \multicolumn{3}{l|}{4.29 pb} \\ \hline
$\sigma(ggH)\text{BR}(H\rightarrow b\bar{b})$  & \multicolumn{3}{l|}{40.88 fb} \\ \hline
$\sigma(ggH)\text{BR}(H\rightarrow \tau\tau)$  &
\multicolumn{3}{l|}{5.10 fb} \\ \hline 
$\sigma(ggH)\text{BR}(H\rightarrow WW)$  & \multicolumn{3}{l|}{49.13
  fb} \\ \hline
$\sigma(ggH)\text{BR}(H\rightarrow ZZ)$  & \multicolumn{3}{l|}{22.41 fb} \\ \hline
$\sigma(ggH)\text{BR}(H\rightarrow \tilde{\chi}_1^0 \tilde{\chi}_1^0)$  &
\multicolumn{3}{l|}{1.27 pb} \\ \hline 
$\sigma(ggH)\text{BR}(H\rightarrow H_sH_s)$  & \multicolumn{3}{l|}{458.74 fb} \\ \hline
$\sigma(ggH)\text{BR}(H\rightarrow H_sH_s\rightarrow bb + bb)$  & \multicolumn{3}{l|}{341.12 fb} \\ 
$\sigma(ggH)\text{BR}(H\rightarrow H_sH_s\rightarrow bb + \tau\tau)$
& \multicolumn{3}{l|}{68.34 fb} \\ 
$\sigma(ggH)\text{BR}(H\rightarrow H_sH_s\rightarrow \tau\tau +
\tau\tau)$  & \multicolumn{3}{l|}{3.42 fb} \\  
$\sigma(ggH)\text{BR}(H\rightarrow H_sH_s\rightarrow bb +
\gamma\gamma)$  & \multicolumn{3}{l|}{0.92 fb} \\ \hline 
$\sigma(ggH)\text{BR}(H\rightarrow hH_s)$  & \multicolumn{3}{l|}{505.60 fb} \\ \hline
$\sigma(ggH)\text{BR}(H\rightarrow hH_s\rightarrow bb + bb)$  & \multicolumn{3}{l|}{274.92 fb} \\ 
$\sigma(ggH)\text{BR}(H\rightarrow hH_s\rightarrow bb + \tau\tau)$  & \multicolumn{3}{l|}{56.46 fb} \\ 
$\sigma(ggH)\text{BR}(H\rightarrow hH_s\rightarrow \tau\tau + \tau\tau)$  & \multicolumn{3}{l|}{2.90 fb} \\ 
$\sigma(ggH)\text{BR}(H\rightarrow hH_s \rightarrow bb +
\gamma\gamma)$  & \multicolumn{3}{l|}{1.34 fb} \\ \hline 
$\sigma(ggH)\text{BR}(H\rightarrow Z A_s)$  & \multicolumn{3}{l|}{1.07
  pb} \\ \hline
$\sigma(ggH)\text{BR}(H\rightarrow Z A_s \rightarrow ll + bb)$  &
\multicolumn{3}{l|}{31.67 fb} \\ 
$\sigma(ggH)\text{BR}(H\rightarrow Z A_s \rightarrow \tau\tau + bb)$  &
\multicolumn{3}{l|}{46.59 fb} \\
$\sigma(ggH)\text{BR}(H\rightarrow Z A_s \rightarrow \tau\tau + \tau\tau)$  &
\multicolumn{3}{l|}{3.32 fb} \\ \hline\hline
$\sigma(ggA_s)$  & \multicolumn{3}{l|}{914.07 fb} \\ \hline
$\sigma(ggA_s)\text{BR}(A_s \rightarrow b\bar{b})$  & \multicolumn{3}{l|}{804.77 fb} \\ \hline
$\sigma(ggA_s)\text{BR}(A_s \rightarrow \tau\tau)$  & \multicolumn{3}{l|}{84.15 fb} \\ \hline
$\sigma(ggA_s)\text{BR}(A_s\rightarrow \gamma \gamma)$  &
\multicolumn{3}{l|}{0.36 fb} \\ \hline \hline
$\sigma(ggA)$  & \multicolumn{3}{l|}{3.36 pb} \\ \hline
$\sigma(ggA)\text{BR}(A \rightarrow t\bar{t})$  & \multicolumn{3}{l|}{1.43 pb} \\ \hline
$\sigma(ggA) \text{BR}(A \rightarrow \tilde{\chi}_1^0 \tilde{\chi}_1^0)$  & 
\multicolumn{3}{l|}{686.00 fb} \\ \hline
$\sigma(ggA) \text{BR}(A\rightarrow h A_s)$  &
\multicolumn{3}{l|}{472.37 fb} \\ \hline 
$\sigma(ggA) \text{BR}(A\rightarrow h A_s \rightarrow bb + bb)$
& \multicolumn{3}{l|}{262.24 fb} \\ 
$\sigma(ggA) \text{BR}(A\rightarrow h A_s \rightarrow \tau\tau + bb)$
& \multicolumn{3}{l|}{55.00 fb} \\ 
$\sigma(ggA) \text{BR}(A\rightarrow h  A_s \rightarrow \tau\tau +
\tau\tau)$  & \multicolumn{3}{l|}{2.88 fb} \\ 
$\sigma(ggA) \text{BR}(A\rightarrow h A_s\rightarrow WW +
bb)$  & \multicolumn{3}{l|}{85.39 fb} \\ 
$\sigma(ggA) \text{BR}(A\rightarrow h A_s\rightarrow \gamma\gamma +
bb)$  & \multicolumn{3}{l|}{1.04 fb} \\ \hline 
$\sigma(ggA) \text{BR}(A\rightarrow H_s A_s)$  &
\multicolumn{3}{l|}{285.76 fb} \\ \hline 
$\sigma(ggA) \text{BR}(A \rightarrow H_s A_s \rightarrow bb +
bb)$  & \multicolumn{3}{l|}{216.95 fb} \\ 
$\sigma(ggA) \text{BR}(A \rightarrow H_s A_s \rightarrow \tau\tau +
bb)$  & \multicolumn{3}{l|}{44.42 fb} \\ 
$\sigma(ggA) \text{BR}(A \rightarrow H_s A_s \rightarrow \tau\tau +
\tau\tau)$  & \multicolumn{3}{l|}{2.27 fb} \\ 
$\sigma(ggA) \text{BR}(A \rightarrow H_s A_s \rightarrow \gamma\gamma +
bb)$  & \multicolumn{3}{l|}{0.39 fb} \\ \hline
$\sigma(ggA) \text{BR}(A\rightarrow Z H_s)$
& \multicolumn{3}{l|}{158.13 fb} \\ \hline
$\sigma(ggA) \text{BR}(A\rightarrow Z H_s \rightarrow ll + b\bar{b})$  &
\multicolumn{3}{l|}{4.59 fb} \\ 
$\sigma(ggA) \text{BR}(A\rightarrow Z H_s \rightarrow \tau\tau + b\bar{b})$
& \multicolumn{3}{l|}{6.66 fb} \\ 
$\sigma(ggA) \text{BR}(A\rightarrow Z H_s \rightarrow \tau\tau + \tau\tau)$  &
\multicolumn{3}{l|}{0.46 fb} \\ \hline
\end{tabular}}
\caption{The signal rates for A.1. \label{tab:ratesscenA1}} 
\end{center}
\end{table}

In scenario
A.1 the $H$ branching ratios both into $H_sH_s$ and $h H_s$ are larger
than 10\% and into $ZA_s$ even 25\%. This leads to interesting final
states with $b$-quark and $\tau$-pairs, 4$\tau$'s or even $(b\bar{b})
(\gamma\gamma)$. The 4$b$ final state may be difficult to exploit due
to the large background. The pseudoscalar $A$ decays with a branching
ratio of 0.14 into $hA_s$ and less frequently into $H_s A_s$ and $Z H_s$
($\mbox{BR}(A\to H_s A_s)=0.085$, $\mbox{BR}(A\to Z
H_s)=0.047$). Still the rates are large enough to provide discovery
channels for the singlets. All these cascade decays of course also add
to the discovery channels of the heavy $H$ and $A$ themselves. \s

\begin{table}
\begin{center}
{\small \begin{tabular}{|l|l|l|l|}
\hline \hline
\textbf{A.2 (Point ID 2212)} & \multicolumn{3}{l|}{\textbf{Signal Rates}} \\
\hline \hline 
$\sigma(ggH_s)$  & \multicolumn{3}{l|}{2.36 pb} \\ \hline
$\sigma(ggH_s) \text{BR}(H_s \rightarrow b\bar{b})$  &
  \multicolumn{3}{l|}{2.12 pb} \\ \hline
$\sigma(ggH_s) \text{BR}(H_s \rightarrow \tau\tau)$  &
 \multicolumn{3}{l|}{214 fb} \\ \hline 
$\sigma(ggH_s) \text{BR}(H_s \rightarrow \gamma\gamma)$  &
 \multicolumn{3}{l|}{0.4 fb} \\ \hline \hline
$\sigma(ggH)$  & \multicolumn{3}{l|}{4.34 pb} \\ \hline
$\sigma(ggH) \text{BR}(H\rightarrow bb)$  & \multicolumn{3}{l|}{66.98 fb} \\ \hline 
$\sigma(ggH) \text{BR}(H\rightarrow \tau\tau)$  & \multicolumn{3}{l|}{8.29
fb} \\ \hline 
$\sigma(ggH) \text{BR}(H\rightarrow WW)$  & \multicolumn{3}{l|}{151.38
 fb} \\ \hline 
$\sigma(ggH) \text{BR}(H\rightarrow ZZ)$  & \multicolumn{3}{l|}{68.47
  fb} \\ \hline 
$\sigma(ggH) \text{BR}(H\rightarrow \tilde{\chi}_1^0
\tilde{\chi}_1^0)$  & \multicolumn{3}{l|}{854.09 fb} \\ \hline 
$\sigma(ggH) \text{BR}(H\rightarrow h H_s)$  &
\multicolumn{3}{l|}{899.34 fb} \\ \hline 
$\sigma(ggH) \text{BR}(H\rightarrow h H_s \to bb + bb)$  &
\multicolumn{3}{l|}{506.58 fb} \\ 
$\sigma(ggH) \text{BR}(H\rightarrow h H_s \to bb + \tau\tau)$  &
\multicolumn{3}{l|}{104.25 fb} \\ 
$\sigma(ggH) \text{BR}(H\rightarrow h H_s \to \tau\tau + \tau\tau)$  &
\multicolumn{3}{l|}{5.36 fb} \\ 
$\sigma(ggH) \text{BR}(H\rightarrow h H_s \to bb + \gamma \gamma)$  &
\multicolumn{3}{l|}{2.02 fb} \\ 
$\sigma(ggH) \text{BR}(H\rightarrow h H_s \to WW + b\bar{b})$  & \multicolumn{3}{l|}{161.89 fb} \\ \hline
$\sigma(ggH) \text{BR}(H\rightarrow Z A_s)$  &
\multicolumn{3}{l|}{1.43 pb} \\ \hline  
$\sigma(ggH) \text{BR}(H\rightarrow Z A_s \to ll + b\bar{b})$  &
\multicolumn{3}{l|}{43.41 fb} \\ 
$\sigma(ggH) \text{BR}(H\rightarrow Z A_s \to \tau\tau + b\bar{b})$  &
\multicolumn{3}{l|}{62.10 fb} \\ 
$\sigma(ggH) \text{BR}(H\rightarrow Z A_s \to \tau\tau + \tau\tau)$  &
\multicolumn{3}{l|}{4.15 fb} \\ \hline  \hline
$\sigma(ggA_s)$  & \multicolumn{3}{l|}{1.36 pb} \\ \hline
$\sigma(ggA_s) \text{BR}(A_s \rightarrow b\bar{b})$  &
  \multicolumn{3}{l|}{1.22 pb} \\ \hline
$\sigma(ggA_s) \text{BR}(A_s \rightarrow \tau\tau)$  &
  \multicolumn{3}{l|}{116.8 fb} \\ \hline 
$\sigma(ggA_s) \text{BR}(A_s \rightarrow \gamma\gamma)$  &
  \multicolumn{3}{l|}{0.3 fb} \\ \hline\hline
$\sigma(ggA)$ & \multicolumn{3}{l|}{3.04 pb} \\ \hline 
$\sigma(ggA)\text{BR}(A\rightarrow \tilde{\chi}_1^0 \tilde{\chi}_1^0)$
& \multicolumn{3}{l|}{1.16 pb}  \\ \hline 
$\sigma(ggA)\text{BR}(A\rightarrow hA_s)$  & \multicolumn{3}{l|}{1.13 
pb} \\ \hline
$\sigma(ggA)\text{BR}(A\rightarrow hA_s\rightarrow bb + bb)$  & 
\multicolumn{3}{l|}{640.74 fb} \\ 
$\sigma(ggA)\text{BR}(A\rightarrow hA_s\rightarrow bb + \tau\tau)$  & 
\multicolumn{3}{l|}{128.37 fb} \\ 
$\sigma(ggA)\text{BR}(A\rightarrow hA_s\rightarrow \tau\tau + 
\tau\tau)$  & \multicolumn{3}{l|}{6.42 fb} \\ 
$\sigma(ggA)\text{BR}(A\rightarrow hA_s\rightarrow bb + WW)$  & 
\multicolumn{3}{l|}{203.85 fb} \\ 
$\sigma(ggA)\text{BR}(A\rightarrow hA_s\rightarrow bb + \gamma\gamma)$  
& \multicolumn{3}{l|}{2.60 fb} \\ \hline 
$\sigma(ggA)\text{BR}(A\rightarrow H_sA_s)$  & 
\multicolumn{3}{l|}{131.49 fb} \\ \hline
$\sigma(ggA)\text{BR}(A\rightarrow H_sA_s\rightarrow bb + bb)$  & 
\multicolumn{3}{l|}{106.17 fb} \\ 
$\sigma(ggA)\text{BR}(A\rightarrow H_sA_s\rightarrow bb + \tau\tau)$  & 
\multicolumn{3}{l|}{20.87 fb} \\ 
$\sigma(ggA)\text{BR}(A\rightarrow H_sA_s\rightarrow \tau\tau + 
\tau\tau)$  & \multicolumn{3}{l|}{1.02 fb} \\  \hline
$\sigma(ggA)\text{BR}(A\rightarrow ZH_s)$
& \multicolumn{3}{l|}{378.42 fb}  \\ \hline 
$\sigma(ggA)\text{BR}(A\rightarrow ZH_s \rightarrow ll + b\bar{b})$
& \multicolumn{3}{l|}{11.43 fb}  \\ 
$\sigma(ggA)\text{BR}(A\rightarrow ZH_s \rightarrow \tau\tau + b\bar{b})$
& \multicolumn{3}{l|}{16.63 fb}  \\ 
$\sigma(ggA)\text{BR}(A\rightarrow ZH_s \rightarrow \tau\tau + \tau\tau)$
& \multicolumn{3}{l|}{1.16 fb}  \\ \hline 
\hline
\end{tabular}}
\caption{The signal rates for A.2. \label{tab:ratesscenA2}} 
\end{center}
\end{table}

In contrast to scenario A.1, in A.2 the decay of the heavy
$H$ into $H_sH_s$ is not interesting, instead the branching
ratio into $h H_s$ is almost doubled compared to A.1, and also the one
into $ZA_s$ is somewhat enhanced. As for the $A$ decays into $hA_s$
and $ZH_s$, they are a factor 2.6 larger and the one into $H_s A_s$
only half as large compared to A.1. Correspondingly the final state
rates are changed. \s

These two scenarios are examples of an NMSSM Higgs spectrum which is
rather light and where all Higgs bosons can be discovered, both
directly or in Higgs-to-Higgs or Higgs-to-gauge-Higgs decays. They
also show the importance of looking into photon final states at mass
values below 125~GeV. Furthermore, the Higgs cascade decays give
access to the trilinear Higgs self-couplings $\lambda_{H H_s H_s}$,
$\lambda_{H H_s h}$, $\lambda_{A A_s h}$ and $\lambda_{A A_s
  H_s}$. Finally, the heavy scalar and pseudoscalar also have sizeable
rates into a pair of two lightest neutralinos. Though this final
state leads to missing energy signatures and does not allow for the
mass reconstruction it adds to possible search channels. 

\subsection{Scenarios B - $H_1 \equiv h$ and small $\tan\beta$}
In the scenarios B it is the lightest Higgs boson that is SM-like. \s

\noindent
\underline{\it B.1) Singlet-like $A_1$ production from heavy Higgs decays:} 
The definition of scenario B.1 is given in Table~\ref{table:scenB1}
and the signal rates in Table~\ref{tab:ratesscenB1}. The scenario is
very special as it not only involves a very singlet-like lightest
pseudoscalar $A_1 \equiv A_s$, but this also has a large branching ratio
into photons, BR$(A_s\to \gamma\gamma)=0.84$. Together with large
Higgs-to-Higgs branching ratios for $H_s$ and $H$, BR$(H_s\to A_s A_s)=0.97$
and BR$(H\to h H_s)=0.51$, this leads to spectacular signatures with
multi-photon and/or multi-$b$ as well multi-$\tau$ final states. \s 

\begin{table}[!t]
 \centering
 {\small \begin{tabular}{|l||l|l|l|}
   \hline \hline
\textbf{B.1 (Point ID Poi2a)} & \multicolumn{3}{l|}{\textbf{Scenario}} \\ \hline \hline
$M_{h},M_{H_s},M_{H}$ & 124.6 GeV & 181.7 GeV & 322.6 GeV \\ \hline
$M_{A_s},M_{A}$ & 72.5 GeV & \multicolumn{2}{l|}{311.7 GeV} \\ \hline 
$|S_{H_2 h_s}|^2, |P_{A_1 a_s}|^2$ & 0.90 & \multicolumn{2}{l|}{1} \\ \hline \hline
$\mu_{\tau\tau}$, $\mu_{bb}$ & 1.54 & \multicolumn{2}{l|}{1.01} \\ \hline
$\mu_{ZZ}$, $\mu_{WW}$, $\mu_{\gamma\gamma}$  & 0.93 & 0.93 & 1.01 \\
\hline \hline
$\tan\beta, \lambda, \kappa$ & 1.9 & 0.628 & 0.354 \\ \hline
$A_\lambda, A_\kappa, \mu_{\text{eff}}$ & 251.2~GeV & 53.8~GeV & 158.9
GeV \\ \hline
$M_1,M_2,M_3$ & 890~GeV & 576~GeV & 1219~GeV \\ \hline
$A_t,A_b,A_\tau$ & 1555~GeV & -1005~GeV & -840~GeV \\ \hline
$M_{Q_3}=M_{t_R}$, $M_{b_R}$ & 1075~GeV & \multicolumn{2}{l|}{1~TeV}
\\ \hline
$M_{L_3}=M_{\tau_R}$, $M_{\text{SUSY}}$ & 530~GeV & \multicolumn{2}{l|}{2.5~TeV} \\ \hline \hline
\end{tabular}}
\caption{The parameters defining scenario B.1, together with the Higgs
boson masses, singlet components and reduced signal rates of
$h$. \label{table:scenB1}} 
\end{table}

\begin{table}
\begin{center}
{\small \begin{tabular}{|l||l|l|l|}
  \hline \hline
\textbf{B.1 (Point ID Poi2a)} & \multicolumn{3}{l|}{\textbf{Decay Rates}}
\\ \hline \hline 
$\sigma(gg H_s)$  &
\multicolumn{3}{l|}{282.37 fb} \\ \hline
$\sigma(gg H_s)\text{BR}(H_s \rightarrow WW)$  &
\multicolumn{3}{l|}{5.09 fb} \\ \hline
$\sigma(gg H_s)\text{BR}(H_s \rightarrow A_s A_s)$  &
\multicolumn{3}{l|}{274.75 fb} \\ \hline
$\sigma(ggH_s )\text{BR}(H_s \rightarrow A_s A_s \rightarrow
b\bar{b} + b\bar{b})$  & \multicolumn{3}{l|}{5.87 fb} \\
$\sigma(ggH_s )\text{BR}(H_s \rightarrow A_s A_s \rightarrow
\gamma\gamma + b\bar{b})$  & \multicolumn{3}{l|}{67.33 fb} \\
$\sigma(ggH_s )\text{BR}(H_s \rightarrow A_s A_s \rightarrow
\gamma\gamma + \gamma\gamma)$  & \multicolumn{3}{l|}{193.22 fb} \\
\hline \hline
$\sigma(gg H)$  & \multicolumn{3}{l|}{3.17 pb} \\ \hline
$\sigma(gg H) \text{BR} (H\to WW) $ & \multicolumn{3}{l|}{264.73  fb} \\ \hline
$\sigma(gg H) \text{BR} (H\to ZZ) $ & \multicolumn{3}{l|}{119.52  fb} \\ \hline
$\sigma(gg H) \text{BR} (H\to b\bar{b}) $ & \multicolumn{3}{l|}{297.37
 fb} \\ \hline
$\sigma(gg H) \text{BR} (H\to \tau\tau) $ & \multicolumn{3}{l|}{37.65
 fb} \\ \hline
$\sigma(gg H) \text{BR} (H\to \tilde{\chi}_1^0 \tilde{\chi}_1^0) $ &
\multicolumn{3}{l|}{383.33 fb} \\ \hline
$\sigma(gg H) \text{BR} (H\to \tilde{\chi}_1^+ \tilde{\chi}_1^-) $ &
\multicolumn{3}{l|}{403.14 fb} \\ \hline
$\sigma(gg H) \text{BR} (H\to hH_s) $ & \multicolumn{3}{l|}{1.609 pb} \\ \hline
$\sigma(gg H) \text{BR} (H\to hH_s \to bb + \tau\tau) $ & \multicolumn{3}{l|}{1.44
  fb} \\ \hline
$\sigma(gg H) \text{BR} (H\to hH_s \to h + A_s A_s \to bb + 4\gamma)
$ & \multicolumn{3}{l|}{712.47 fb} \\ 
$\sigma(gg H) \text{BR} (H\to hH_s \to h + A_s A_s \to \gamma\gamma +
4 b)$ & \multicolumn{3}{l|}{248.02 fb} \\ 
$\sigma(gg H) \text{BR} (H\to hH_s \to h + A_s A_s \to \tau\tau + 4\gamma)
$ & \multicolumn{3}{l|}{74.60 fb} \\ 
$\sigma(gg H) \text{BR} (H\to hH_s \to h + A_s A_s \to \gamma\gamma + 4\tau)
$ & \multicolumn{3}{l|}{2.47 fb} \\ 
$\sigma(gg H) \text{BR} (H\to hH_s \to h + A_s A_s \to 6\gamma)
$ & \multicolumn{3}{l|}{2.69 fb} \\ 
$\sigma(gg H) \text{BR} (H\to hH_s \to h + A_s A_s \to \tau\tau +
\gamma\gamma + b\bar{b})$ & \multicolumn{3}{l|}{49.55 fb} \\ 
\hline 
$\sigma(gg H) \text{BR}(H \to A_s A_s)$ & \multicolumn{3}{l|}{5.59 fb} \\ 
\hline
$\sigma(gg H) \text{BR}(H \to A_s A_s \to 4\gamma)$ &
\multicolumn{3}{l|}{3.93 fb} \\  \hline \hline
$\sigma(gg A_s)$  & \multicolumn{3}{l|}{0.08 fb} \\ \hline \hline
$\sigma(gg A)$  & \multicolumn{3}{l|}{2.51 pb} \\ \hline
$\sigma(gg A) \text{BR} (A\to\tau\tau) $ & \multicolumn{3}{l|}{14.42
  fb} \\ \hline
$\sigma(gg A) \text{BR} (A\to \tilde{\chi}_1^0 \tilde{\chi}_1^0) $ &
\multicolumn{3}{l|}{963.87 fb} \\ \hline
$\sigma(gg A) \text{BR} (A\to \tilde{\chi}_1^+ \tilde{\chi}_1^-) $ &
\multicolumn{3}{l|}{273.57 fb} \\ \hline 
$\sigma(gg A) \text{BR} (A\to H_s A_s) $ & \multicolumn{3}{l|}{525.56 fb} \\ \hline
$\sigma(gg A) \text{BR} (A\to H_s A_s \to A_sA_s + A_s \to 6
\gamma) $ & \multicolumn{3}{l|}{301.58 fb} \\ 
$\sigma(gg A) \text{BR} (A\to H_s A_s \to A_s A_s + A_s \to bb + 4\gamma)
$ & \multicolumn{3}{l|}{157.64 fb} \\ 
$\sigma(gg A) \text{BR} (A\to H_s A_s \to A_s A_s+ A_s \to
4b + \gamma\gamma) $ & \multicolumn{3}{l|}{27.47 fb} \\
$\sigma(gg A) \text{BR} (A\to H_s A_s \to A_s A_s +A_s \to \tau\tau + 4\gamma)
$ & \multicolumn{3}{l|}{14.99 fb} \\ 
$\sigma(gg A) \text{BR} (A\to H_s A_s \to A_s A_s +A_s \to
\tau\tau + bb + \gamma\gamma) $ & \multicolumn{3}{l|}{5.22 fb} \\ 
$\sigma(gg A) \text{BR} (A\to H_s A_s \to A_s A_s +A_s \to
4\tau + \gamma\gamma) $ & \multicolumn{3}{l|}{0.25 fb} \\
\hline
$\sigma(gg A) \text{BR} (A\to h A_s) $ & \multicolumn{3}{l|}{29.96 fb} \\ \hline
$\sigma(gg A) \text{BR} (A\to h A_s \to \gamma\gamma + b\bar{b}) $ &
\multicolumn{3}{l|}{16.25 fb} \\ 
$\sigma(gg A) \text{BR} (A\to h A_s \to \gamma \gamma + \tau \tau)
$ & \multicolumn{3}{l|}{1.70 fb} \\ 
$\sigma(gg A) \text{BR} (A\to h A_s \to b\bar{b} + b\bar{b}) $ &
\multicolumn{3}{l|}{2.83 fb} \\ \hline
$\sigma(gg A) \text{BR} (A\to Z H_s)$ & \multicolumn{3}{l|}{554.38 fb}
\\ \hline
$\sigma(gg A) \text{BR} (A\to Z H_s \to bb + A_s A_s \to
bb + 4\gamma) $ & \multicolumn{3}{l|}{57.36 fb} \\
$\sigma(gg A) \text{BR} (A\to Z H_s \to bb + A_s A_s \to
4b + \gamma\gamma) $ & \multicolumn{3}{l|}{19.99 fb} \\
$\sigma(gg A) \text{BR} (A\to Z H_s \to Z + A_s A_s \to
bb + \tau\tau + \gamma\gamma) $ & \multicolumn{3}{l|}{6.35 fb} \\
$\sigma(gg A) \text{BR} (A\to Z H_s \to ll/\tau\tau + A_s A_s \to
ll/\tau\tau + 4\gamma) $ & \multicolumn{3}{l|}{12.78 fb} \\
$\sigma(gg A) \text{BR} (A\to Z H_s \to ll/\tau\tau + A_s A_s \to
ll\tau\tau/4\tau + \gamma\gamma) $ & \multicolumn{3}{l|}{0.42 fb} \\
\hline\hline
\end{tabular}}
\caption{The signal rates for B.1. \label{tab:ratesscenB1}} 
\end{center}
\end{table}

In detail, due to the singlet nature of $A_s$ its gluon fusion
production cross section is extremely small with only 0.08~fb, so that
alternative production mechanisms must be exploited. The $H_s$
coupling to gluons is large enough to lead to a sizeable cross section of
282~fb. (It is smaller by almost a factor 10 compared to the scenarios
A because $H_s$ here is the second lightest Higgs and hence
the available phase space is smaller.) As $H_s$ has a large branching
ratio into $A_s$ pairs and 
$A_s$ itself into photon pairs, the thus produced four photon final
state rate amounts to almost 200~fb. The heavy doublet Higgs $H$ is
produced in gluon fusion with 3.2~pb and its branching ratio into
$h H_s$ is 0.51. With the large $H_s$ decay rate into $A_s A_s$ this
leads to very peculiar signatures with up to 6 photons in the final
state. The largest rate is given by the $(4\gamma) (b\bar{b})$ signature
with 712~fb. Additionally, we have $A$ production in gluon fusion at
2.5~pb, which leads via the decay into $H_s A_s$ (BR$(A\to H_s A_s)$=0.21)
again to multiphoton final states with or without additional $b$-quark
or $\tau$-pairs. Here the 6$\gamma$ final state even amounts to
302~fb. With a branching ratio BR$(A \to ZH_s)=0.22$ also this decay
leads to interesting final states with {\it e.g.}~$(4\gamma) (b\bar{b})$
production at 58~fb. Finally, also the decay into $h A_s$ may be
exploited in its $(\gamma\gamma) (b\bar{b})$ final state with a cross
section of 16~fb. These and more possible combinations and final
states are summarised in Table~\ref{tab:ratesscenB1}. It shows that in
this scenario with a rather light overall spectrum, all Higgs bosons
can be discovered, and that there are spectacular signatures possible
that can be helpful for the discovery. Additionally, in the cascade
decays the trilinear couplings $\lambda_{H_s A_s A_s}$, $\lambda_{HH_s
h}$, $\lambda_{HA_s A_s}$, $\lambda_{AA_s H_s}$ and $\lambda_{A A_s
h}$ are accessible. Note 
finally, that the multiphoton (plus fermion) final states discussed
here cannot occur in the MSSM and are unique to an extension beyond, 
as the NMSSM. \s

\noindent
\underline{\it B.2) Heavy Higgs spectrum:}
Scenario B.2, defined in Table~\ref{table:scenB2} is an example for a
spectrum where it is challenging to 
find all NMSSM Higgs states. This is because of the heavier $H$ and
$A$ with masses around 560~GeV, a not so light $A_s$ as {\it e.g.}~in
scenario A.2 and a very singlet-like $H_s$. The latter has a gluon
fusion production cross section of 19~fb, leading to a $\tau\tau$
final state at only 0.3~fb, {\it cf.}~Table~\ref{tab:ratesscenB2}. The
only alternative production via 
Higgs-to-Higgs decays is given by the $H \to h H_s$ decay leading to
small 0.5~fb in the $(\tau\tau)(b\bar{b})$ final state. The
pseudoscalar singlet $A_s$ on the other hand has a larger gluon fusion
production cross section now than in B.1 and can be searched for in
standard final states. Additionally the decay of $A$ into $hA_s$ can be
used to produce $A_s$ in the $(\tau\tau)(b\bar{b})$ final state {\it
  e.g.}~at 21~fb. Also the $H\to ZA_s \to (\tau\tau) (b\bar{b})$
production leading to 14~fb, may be used.
Both $H$ and $A$ are heavy enough to decay into top
pairs and may be discovered in these decay channels with rates into
$t\bar{t}$ between 550~fb for $A$ and 623~fb for $H$. In this scenario
only the trilinear couplings $\lambda_{HH_s h}$ and $\lambda_{AA_s h}$
are accessible. \s
\begin{table}[!h]
 \centering
{\small \begin{tabular}{|l|l|l|l|}
\hline \hline
\textbf{B.2 (Point ID 1142)} & \multicolumn{3}{l|}{\textbf{Scenario}}
\\ \hline \hline 
$M_{h},M_{H_s},M_{H}$ & 126.8 GeV & 176.2 GeV & 556.3 GeV \\ \hline
$M_{A_s},M_{A}$ & 106.6 GeV & \multicolumn{2}{l|}{577.8 GeV} \\ \hline
$|S_{H_2h_s}|^2, |P_{A_1 a_s}|^2$ & 0.99 & \multicolumn{2}{l|}{0.92} \\ \hline \hline
$\mu_{\tau\tau}$, $\mu_{bb}$ & 1.03 & \multicolumn{2}{l|}{1.21} \\ \hline
$\mu_{ZZ}$, $\mu_{WW}$, $\mu_{\gamma\gamma}$  & 0.93 & 0.93 & 0.89 \\ 
\hline \hline
$\tan\beta$, $\lambda$, $\kappa$ & 1.61 & 0.62 & -0.22 \\ \hline
$A_{\lambda}$, $A_ {\kappa}$, $\mu_{\text{eff}}$ & -709.2 GeV & -169.3 
GeV & -236.5 GeV \\ \hline
$A_ t$, $A_b$, $A_{\tau}$ & -899.8 GeV & -1436.3 GeV & 857.5 GeV \\ \hline
$M_1$, $M_2$, $M_3$ & 651.4 GeV & 307.0 GeV & 1340.6 GeV \\ \hline
$M_{Q_3}=M_{t_R}$, $M_{b_R}$ & 2578.9 GeV  & \multicolumn{2}{l|}{3 TeV} \\ \hline
$M_{L_3}=M_{\tau_R}$, $M_{\text{SUSY}}$ & 1863.7 GeV & 
\multicolumn{2}{l|}{3 TeV} \\ \hline \hline
\end{tabular}}
\caption{The parameters defining scenario B.2, together with the Higgs
boson masses, singlet components and reduced signal rates of
$h$. \label{table:scenB2}} 
\end{table}
 
\begin{table}[!h]
 \centering
{\small \begin{tabular}{|l|l|l|l|}
\hline \hline
\textbf{B.2 (Point ID 1142)} & \multicolumn{3}{l|}{\textbf{Signal Rates}} \\
\hline \hline
$\sigma(ggH_s)$ & \multicolumn{3}{l|}{19.03 fb} \\ \hline
$\sigma(ggH_s) \text{BR}(H_s \rightarrow WW)$ &
\multicolumn{3}{l|}{15.46 fb} \\ \hline
$\sigma(ggH_s) \text{BR}(H_s \rightarrow b\bar{b})$ &
\multicolumn{3}{l|}{2.68 fb} \\ \hline 
$\sigma(ggH_s) \text{BR}(H_s \rightarrow \tau\tau)$ &
\multicolumn{3}{l|}{0.30 fb} \\ \hline \hline
$\sigma(ggH)$ & \multicolumn{3}{l|}{1.15 pb} \\ \hline
$\sigma(ggH) \text{BR}(H \rightarrow t\bar{t})$ &
\multicolumn{3}{l|}{623.39 fb} \\ \hline
$\sigma(ggH) \text{BR}(H \rightarrow \tilde{\chi}_1^0 \tilde{\chi}_1^0)$ &
\multicolumn{3}{l|}{10.84 fb} \\ \hline
$\sigma(ggH) \text{BR}(H \rightarrow Z A_s)$ &
\multicolumn{3}{l|}{322.93 fb} \\ \hline
$\sigma(ggH) \text{BR}(H \rightarrow Z A_s \rightarrow ll + b\bar{b})$ &
\multicolumn{3}{l|}{9.61 fb} \\ 
$\sigma(ggH) \text{BR}(H \rightarrow Z A_s \rightarrow \tau\tau + b\bar{b})$ &
\multicolumn{3}{l|}{14.03 fb} \\ \hline 
$\sigma(ggH) \text{BR}(H \rightarrow h H_s)$ &
\multicolumn{3}{l|}{26.62 fb} \\ \hline
$\sigma(ggH) \text{BR}(H \rightarrow h H_s \rightarrow \tau\tau + b\bar{b})$ &
\multicolumn{3}{l|}{0.50 fb} \\ 
$\sigma(ggH) \text{BR}(H \rightarrow h H_s \rightarrow WW + b\bar{b}
\rightarrow (l\nu)(l\nu) b\bar{b})$ &
\multicolumn{3}{l|}{0.16 fb} \\ \hline
\hline
$\sigma(ggA_s)$  & \multicolumn{3}{l|}{1.56 pb} \\ \hline
$\sigma(ggA_s) \text{BR}(A_s \rightarrow b\bar{b})$  &
  \multicolumn{3}{l|}{1.38 pb} \\ \hline
$\sigma(ggA_s) \text{BR}(A_s \rightarrow \tau\tau)$  &
  \multicolumn{3}{l|}{140 fb} \\ \hline \hline
$\sigma(ggA)$  & \multicolumn{3}{l|}{824.23 fb} \\ \hline
$\sigma(ggA) \text{BR}(A \rightarrow t\bar{t})$  &
\multicolumn{3}{l|}{551.88 fb} \\ \hline 
$\sigma(ggA)\text{BR}(A\rightarrow hA_s)$  & \multicolumn{3}{l|}{184.87 
fb} \\ \hline
$\sigma(ggA)\text{BR}(A\rightarrow hA_s\rightarrow bb + bb)$  & 
\multicolumn{3}{l|}{100.05 fb} \\ 
$\sigma(ggA)\text{BR}(A\rightarrow hA_s\rightarrow bb + \tau\tau)$  & 
\multicolumn{3}{l|}{20.72 fb} \\ 
$\sigma(ggA)\text{BR}(A\rightarrow hA_s\rightarrow \tau\tau + 
\tau\tau)$  & \multicolumn{3}{l|}{1.07 fb} \\ 
$\sigma(ggA)\text{BR}(A\rightarrow hA_s\rightarrow bb + \gamma\gamma)$  
& \multicolumn{3}{l|}{0.36 fb} \\ \hline \hline
\end{tabular}}
\caption{The signal rates for B.2. \label{tab:ratesscenB2}} 
\vspace*{0.7cm}
\end{table}

\noindent
\underline{\it B.3) Heavy Higgs spectrum, B.2 reversed:} 
The scenario B.3, {\it cf.}~Table~\ref{table:scenB3}, is reversed
compared to B.2 in the sense that now $H_2 \equiv H_s$ is somewhat
less singlet-like and $A_1 \equiv A_s$ is very singlet-like, so that
here $A_s$ production will be challenging. Its gluon fusion production
cross section amounts only to 32~fb leading to a $\tau\tau$ rate of
3~fb as can be inferred from Table~\ref{tab:ratesscenB3}. It can also
be produced in heavy $A$ cascade decays via $hA_s$ 
and $H_s A_s$. The former leads to 3.9~fb in the
$(\tau\tau)(b\bar{b})$ final state, the latter to 1.7~fb. The singlet
$H_s$ now can be looked for in standard final states as {\it
  e.g.}~$\tau\tau$ with 18~fb or with larger rates in gauge boson final
states. It can also be produced from $A\to ZH_s$ decays. The heavy
$H$ and $A$ again can decay into top pairs, at rates around 680~fb. The
pseudoscalar doublet $A$ additionally decays into a lightest neutralino
pair at sizeable rate. The trilinear couplings accessible in this
scenario are $\lambda_{HH_s h}$, $\lambda_{AA_s h}$ and 
$\lambda_{A A_s H_s}$. 

\begin{table}[h!]
\begin{center}
{\small \begin{tabular}{|l|l|l|l|}
\hline \hline
\textbf{B.3 (Point ID 210)} & \multicolumn{3}{l|}{\textbf{Scenario}} \\ \hline \hline
$M_{h},M_{H_s},M_{H}$ & 124.1 GeV & 184.3 GeV & 463.1 GeV \\ \hline
$M_{A_s},M_{A}$ & 133.4 GeV & \multicolumn{2}{l|}{457.2 GeV} \\ \hline
$|S_{H_2h_s}|^2, |P_{A_1} a_s|^2$ & 0.96 & \multicolumn{2}{l|}{0.99} \\ \hline \hline
$\mu_{\tau\tau}$, $\mu_{bb}$ & 1.44 & \multicolumn{2}{l|}{1.99} \\ \hline
$\mu_{ZZ}$, $\mu_{WW}$, $\mu_{\gamma\gamma}$  & 0.90 & 0.90 & 0.97 \\ 
\hline \hline
$\tan\beta$, $\lambda$, $\kappa$ & 2.22 & 0.60 & 0.30 \\ \hline
$A_{\lambda}$, $A_ {\kappa}$, $\mu_{\text{eff}}$ & 348.7 GeV & 4.5 GeV & 
191.8 GeV \\ \hline
$A_ t$, $A_b$, $A_{\tau}$ & -1130.2 GeV & -6.5 GeV & 1951.6 GeV \\ \hline
$M_1$, $M_2$, $M_3$ & 136.4 GeV & 273.4 GeV & 1789.0 GeV \\ \hline
$M_{Q_3}=M_{t_R}$, $M_{b_R}$ & 2838.3 GeV  & \multicolumn{2}{l|}{3 TeV} \\ \hline
$M_{L_3}=M_{\tau_R}$, $M_{\text{SUSY}}$ & 1659.3 GeV & 
\multicolumn{2}{l|}{3 TeV} \\ \hline \hline
\end{tabular}}
\caption{The parameters defining scenario B.3, together with the Higgs
boson masses, singlet components and reduced signal rates of
$h$. \label{table:scenB3}} 
\end{center}
\end{table}

\begin{table}[h!]
\begin{center}
{\small \begin{tabular}{|l|l|l|l|}
\hline \hline
\textbf{B.3 (Point ID 210)} & \multicolumn{3}{l|}{\textbf{Signal
    Rates}} \\ \hline \hline  
$\sigma(ggH_s)$  &
\multicolumn{3}{l|}{390.38 fb} \\ \hline 
$\sigma(ggH_s) \text{BR}(H_s \rightarrow b\bar{b})$  &
\multicolumn{3}{l|}{160.37 fb} \\ \hline 
$\sigma(ggH_s) \text{BR}(H_s \rightarrow \tau\tau)$  &
\multicolumn{3}{l|}{18.46 fb} \\ \hline 
$\sigma(ggH_s) \text{BR}(H_s \rightarrow WW)$  &
\multicolumn{3}{l|}{176.63 fb} \\ \hline 
$\sigma(ggH_s) \text{BR}(H_s \rightarrow ZZ)$  &
\multicolumn{3}{l|}{29.00 fb} \\ \hline \hline
$\sigma(ggH)$  &
\multicolumn{3}{l|}{1.326 pb} \\ \hline 
$\sigma(ggH)\text{BR}(H\rightarrow t\bar{t})$  &
\multicolumn{3}{l|}{684.96 fb} \\ \hline \hline
$\sigma(ggH)\text{BR}(H\rightarrow hH_s)$  & \multicolumn{3}{l|}{184.85 
fb} \\ \hline
$\sigma(ggH)\text{BR}(H\rightarrow hH_s\rightarrow bb + bb)$  & 
\multicolumn{3}{l|}{50.46 fb} \\ 
$\sigma(ggH)\text{BR}(H\rightarrow hH_s\rightarrow bb + \tau\tau)$ & 
\multicolumn{3}{l|}{11.08 fb} \\ 
$\sigma(ggH)\text{BR}(H\rightarrow hH_s\rightarrow \tau\tau + 
\tau\tau)$  & \multicolumn{3}{l|}{0.61 fb} \\ 
$\sigma(ggH)\text{BR}(H\rightarrow hH_s \rightarrow bb + \gamma\gamma)$  
& \multicolumn{3}{l|}{0.24 fb} \\ \hline \hline
$\sigma(ggH)\text{BR}(H\rightarrow ZA_s)$  & \multicolumn{3}{l|}{36.41 
fb} \\ \hline
$\sigma(ggH)\text{BR}(H\rightarrow ZA_s \rightarrow ll + b\bar{b})$  &
\multicolumn{3}{l|}{1.09 fb} \\ 
$\sigma(ggH)\text{BR}(H\rightarrow ZA_s \rightarrow \tau\tau + b\bar{b})$  &
\multicolumn{3}{l|}{1.62 fb} \\ \hline \hline
$\sigma(ggA_s)$  & \multicolumn{3}{l|}{31.49 fb} \\ \hline
$\sigma(ggA_s)\text{BR}(A_s \rightarrow b\bar{b})$  &
\multicolumn{3}{l|}{28.03 fb} \\ \hline
$\sigma(ggA_s)\text{BR}(A_s \rightarrow \tau\tau)$  & \multicolumn{3}{l|}{3.01 fb} \\ \hline
$\sigma(ggA_s)\text{BR}(A_s\rightarrow \gamma \gamma)$  &
\multicolumn{3}{l|}{0.13 fb} \\ \hline \hline
$\sigma(ggA)$  & \multicolumn{3}{l|}{1.26 pb} \\ \hline
$\sigma(ggA)\text{BR}(A \rightarrow t\bar{t})$  & \multicolumn{3}{l|}{680.53 fb} \\ \hline
$\sigma(ggA) \text{BR}(A \rightarrow \tilde{\chi}_1^0 \tilde{\chi}_1^0)$  & 
\multicolumn{3}{l|}{109.32 fb} \\ \hline
$\sigma(ggA) \text{BR}(A\rightarrow h A_s)$  &
\multicolumn{3}{l|}{31.19 fb} \\ \hline
$\sigma(ggA) \text{BR}(A\rightarrow h A_s \rightarrow b\bar{b}+ b\bar{b})$  &
\multicolumn{3}{l|}{18.41 fb} \\ 
$\sigma(ggA) \text{BR}(A\rightarrow h A_s \rightarrow \tau\tau+ b\bar{b})$  &
\multicolumn{3}{l|}{3.91 fb} \\ 
$\sigma(ggA) \text{BR}(A\rightarrow h A_s \rightarrow \tau\tau+ \tau\tau)$  &
\multicolumn{3}{l|}{0.21 fb} \\ 
$\sigma(ggA) \text{BR}(A\rightarrow h A_s \rightarrow \gamma\gamma +
bb)$  & \multicolumn{3}{l|}{0.15 fb} \\ \hline
$\sigma(ggA) \text{BR}(A\rightarrow H_s A_s)$  &
\multicolumn{3}{l|}{20.94 fb} \\ \hline
$\sigma(ggA) \text{BR}(A \rightarrow H_s A_s\rightarrow b\bar{b} +
b\bar{b})$  & \multicolumn{3}{l|}{7.67 fb} \\ 
$\sigma(ggA) \text{BR}(A \rightarrow H_s A_s\rightarrow \tau\tau +
b\bar{b})$  & \multicolumn{3}{l|}{1.71 fb} \\ \hline
$\sigma(ggA) \text{BR}(A\rightarrow Z H_s)$
& \multicolumn{3}{l|}{90.21 fb} \\ \hline
$\sigma(ggA) \text{BR}(A\rightarrow Z H_s \rightarrow ll + bb)$
& \multicolumn{3}{l|}{1.25 fb} \\ 
$\sigma(ggA) \text{BR}(A\rightarrow Z H_s \rightarrow \tau\tau + bb)$
& \multicolumn{3}{l|}{1.90 fb} \\ \hline \hline
\end{tabular}}
\caption{The signal rates for B.3. \label{tab:ratesscenB3}} 
\end{center}
\end{table}

\subsection{Scenario C - $H_1 \equiv h$ and large $\tan\beta$}
Scenario C in Table~\ref{table:scenC} explicitly features a large value
of $\tan\beta$, {\it i.e.}~$\tan\beta= 17$. Along with this comes a
very heavy, additionally practically 100\% singlet-like, Higgs
boson of 3.5~TeV. This makes it very difficult if not impossible to
discover the complete NMSSM Higgs spectrum. Due to the large value of
$\tan\beta$ gluon fusion is not effective as production
mechanism. Instead the
signal rates for Higgs production in association with $b\bar{b}$ are
given in Table~\ref{tab:ratesscenC} for the doublet-like Higgs bosons
$H$ and $A$, both with masses around 600 GeV, and the singlet-like
pseudoscalar $A_s$, that is also rather heavy with a mass around
312~GeV. For comparison 
we also give the corresponding cross section for the SM-like Higgs
boson. The production cross sections are large enough to produce the
NMSSM Higgs bosons, apart from the heavy $H_s$, at sufficient rates for
discovery in the standard channels. In addition, the pseudoscalar
singlet $A_s$ can be searched for in the cascade decay via $A \to h
A_s$ or $H \to Z A_s$. All other Higgs-to-Higgs or Higgs-to-gauge-Higgs
decays for $H$ or $A$ are kinematically closed. Therefore in this
scenario the only trilinear Higgs coupling that may be measurable is
$\lambda_{AA_s h}$. 

\begin{table}[!h]
\begin{center}
{\small \begin{tabular}{|l|l|l|l|}
\hline \hline
\textbf{C (Point ID 2296)} & \multicolumn{3}{l|}{\textbf{Scenario}} \\ \hline \hline
$M_{h},M_{H},M_{H_s}$ & 124.1 GeV & 597.7
GeV & 3528.3 GeV \\ \hline 
$M_{A_s}, M_{A}$ & 311.8 GeV &
\multicolumn{2}{l|}{614.5 GeV} \\ \hline 
$|S_{H_3 h_s}|^2$, $|P_{A_1 a_s}|^2$ & 1 & \multicolumn{2}{l|}{0.93} \\ \hline \hline
$\mu_{\tau\tau}$, $\mu_{bb}$ & 0.97 & \multicolumn{2}{l|}{1.06} \\ \hline
$\mu_{ZZ}$, $\mu_{WW}$, $\mu_{\gamma\gamma}$  & 0.78 & 0.78 & 0.80 \\ 
\hline \hline
$\tan\beta$, $\lambda$, $\kappa$ & 17.06 & 0.08 & -0.63 \\ \hline
$A_{\lambda}$, $A_ {\kappa}$, $\mu_{\text{eff}}$ & -1766.2 GeV & -24.2 
GeV & -217.1 GeV \\ \hline
$A_ t$, $A_b$, $A_{\tau}$ & 1961.8 GeV & -1535.3 GeV & -1211.9 GeV \\ \hline
$M_1$, $M_2$, $M_3$ & 478.3 GeV & 369.2 GeV & 2847.8 GeV \\ \hline
$M_{Q_3}=M_{t_R}$, $M_{b_R}$ & 977.0 GeV  & \multicolumn{2}{l|}{3 TeV} \\ \hline
$M_{L_3}=M_{\tau_R}$, $M_{\text{SUSY}}$ & 2797.1 GeV & 
\multicolumn{2}{l|}{3 TeV} \\ \hline \hline
\end{tabular}}
\caption{The parameters defining scenario C, together with the Higgs
boson masses, singlet components and reduced signal rates of
$h$. \label{table:scenC}} 
\end{center}
\end{table}

\begin{table}[!h]
\begin{center}
{\small \begin{tabular}{|l|l|l|l|}
\hline \hline
\textbf{C (Point ID 2296)} & \multicolumn{3}{l|}{\textbf{Higgs Decays}} \\
\hline \hline 
$\sigma(bbH)$  & \multicolumn{3}{l|}{346.97 fb} \\ \hline
$\sigma(bbH)\text{BR}(H\rightarrow b\bar{b})$  & 
\multicolumn{3}{l|}{190.72 fb} \\ \hline
$\sigma(bbH)\text{BR}(H\rightarrow \tau\tau)$  & 
\multicolumn{3}{l|}{23.32 fb} \\ \hline
$\sigma(bbH)\text{BR}(H\rightarrow t\bar{t})$  & 
\multicolumn{3}{l|}{5.37 fb} \\ \hline
$\sigma(bbH)\text{BR}(H\rightarrow \tilde{\chi}_1^0 \tilde{\chi}_1^0)$  & 
\multicolumn{3}{l|}{7.00 fb} \\ \hline
$\sigma(bbH)\text{BR}(H\rightarrow \tilde{\chi}_1^+ \tilde{\chi}_1^-)$  & 
\multicolumn{3}{l|}{16.21 fb} \\ \hline
$\sigma(bbH)\text{BR}(H\rightarrow Z A_s)$  & 
\multicolumn{3}{l|}{101.84 fb} \\ \hline
$\sigma(bbH)\text{BR}(H\rightarrow Z A_s \rightarrow ll + b\bar{b})$  & 
\multicolumn{3}{l|}{3.08 fb} \\ 
$\sigma(bbH)\text{BR}(H\rightarrow Z A_s \rightarrow \tau\tau + b\bar{b})$  & 
\multicolumn{3}{l|}{4.61 fb} \\ \hline \hline
$\sigma(bbA_s)$  & 
\multicolumn{3}{l|}{404.91 fb} \\ \hline
$\sigma(bbA_s)\text{BR}(A_s\rightarrow b\bar{b})$  & 
\multicolumn{3}{l|}{364.21 fb} \\ \hline
$\sigma(bbA_s)\text{BR}(A_s\rightarrow \tau\tau)$  & 
\multicolumn{3}{l|}{40.17 fb} \\ \hline\hline
$\sigma(bbh)$  & 
\multicolumn{3}{l|}{643.60 fb} \\ \hline \hline
$\sigma(bbA)$  & 
\multicolumn{3}{l|}{282.80 fb} \\ \hline
$\sigma(bbA)\text{BR}(A\rightarrow b\bar{b})$  & 
\multicolumn{3}{l|}{151.41 fb} \\ \hline
$\sigma(bbA)\text{BR}(A\rightarrow \tau\tau)$  & 
\multicolumn{3}{l|}{18.60 fb} \\ \hline
$\sigma(bbA)\text{BR}(A\rightarrow t\bar{t})$  & 
\multicolumn{3}{l|}{5.08 fb} \\ \hline
$\sigma(bbA)\text{BR}(A\rightarrow \tilde{\chi}_1^0 \tilde{\chi}_1^0)$  & 
\multicolumn{3}{l|}{6.85 fb} \\ \hline 
$\sigma(bbA)\text{BR}(A\rightarrow h A_s)$  & 
\multicolumn{3}{l|}{76.27 fb} \\ \hline
$\sigma(bbA)\text{BR}(A\rightarrow h A_s \rightarrow bb + bb)$  & 
\multicolumn{3}{l|}{46.65 fb} \\ 
$\sigma(bbA)\text{BR}(A\rightarrow h A_s \rightarrow bb + 
\tau\tau)$  & \multicolumn{3}{l|}{9.98 fb} \\
$\sigma(bbA)\text{BR}(A\rightarrow h A_s \rightarrow \tau\tau + 
\tau\tau)$  & \multicolumn{3}{l|}{0.53 fb} \\ \hline \hline
\end{tabular}}
\caption{The signal rates for C. \label{tab:ratesscenC}} 
\end{center}
\end{table}
\vspace{1cm}

\subsection{Scenarios D}
The scenarios D are characterized by a SM-like Higgs given by $H_2$,
$H_2 \equiv h$, that can decay into lighter singlet pairs.  \s

\begin{table}[!t]
\begin{center}
{\small \begin{tabular}{|l|l|l|l|}
\hline \hline
\textbf{D.1 (Point ID 5416)} & \multicolumn{3}{l|}{\textbf{Scenario}}
\\ \hline \hline 
$M_{H_s},M_{h},M_{H}$ & 9.6 GeV &
124.2 GeV & 793.4 GeV \\ \hline 
$M_{A_s}, M_A$ & 273.2 GeV &
\multicolumn{2}{l|}{792.2 GeV} \\ \hline 
$|S_{H_1h_s}|^2$, $|P_{A_1 A_s}|^2$ & 0.98 & \multicolumn{2}{l|}{0.99} \\ \hline \hline
$\mu_{\tau\tau}$, $\mu_{bb}$ & 0.90 & \multicolumn{2}{l|}{0.89} \\ \hline
$\mu_{ZZ}$, $\mu_{WW}$, $\mu_{\gamma\gamma}$  & 0.92 & 0.92 & 0.92 \\ 
\hline \hline
$\tan\beta$, $\lambda$, $\kappa$ & 3.37 & 0.64 & 0.20 \\ \hline
$A_{\lambda}$, $A_ {\kappa}$, $\mu_{\text{eff}}$ & -709.0 GeV & 297.3 
GeV & -222.4 GeV \\ \hline
$A_ t$, $A_b$, $A_{\tau}$ & -1075.3 GeV & -1973.1 GeV & -143.7 GeV \\ \hline
$M_1$, $M_2$, $M_3$ & 307.7 GeV & 789.8 GeV & 2933.1 GeV \\ \hline
$M_{Q_3}=M_{t_R}$, $M_{b_R}$ & 2931.3 GeV  & \multicolumn{2}{l|}{3 TeV} \\ \hline
$M_{L_3}=M_{\tau_R}$, $M_{\text{SUSY}}$ & 2930.8 GeV & 
\multicolumn{2}{l|}{3 TeV} \\ \hline \hline
\end{tabular}}
\caption{The parameters defining scenario D.1, together with the Higgs
boson masses, singlet components and reduced signal rates of
$h$. \label{table:scenD.1}} 
\end{center}
\end{table}

\begin{table}[!h]
\begin{center}
{\small \begin{tabular}{|l|l|l|l|}
\hline \hline
\textbf{D.1 (Point ID 5416)} & \multicolumn{3}{l|}{\textbf{Signal Rates}} \\
\hline \hline 
$\sigma(ggh)$  & 
\multicolumn{3}{l|}{44.28 pb} \\ \hline
$\sigma(ggh)\text{BR}(h\rightarrow H_sH_s)$  & 
\multicolumn{3}{l|}{4.22 pb} \\ \hline
$\sigma(ggh)\text{BR}(h\rightarrow H_sH_s\rightarrow \tau\tau + 
\tau\tau)$  & \multicolumn{3}{l|}{3.58 pb} \\ 
$\sigma(ggh)\text{BR}(h\rightarrow H_sH_s\rightarrow \tau\tau + 
\mu\mu)$  & \multicolumn{3}{l|}{31.64 fb} \\ \hline \hline
$\sigma(ggH_s)$  & \multicolumn{3}{l|}{439.80 pb} \\ \hline
$\sigma(ggH_s)\text{BR}(H_s\rightarrow \mu\mu)$  &
\multicolumn{3}{l|}{1.79 pb} \\ \hline 
$\sigma(ggH_s)\text{BR}(H_s\rightarrow \tau\tau)$  &
\multicolumn{3}{l|}{405.09 pb} \\ \hline 
$\sigma(ggH_s)\text{BR}(H_s\rightarrow c\bar{c})$  &
\multicolumn{3}{l|}{5.17 pb} \\ \hline 
$\sigma(ggH_s)\text{BR}(H_s\rightarrow s\bar{s})$  &
\multicolumn{3}{l|}{7.24 pb} \\ \hline 
$\sigma(ggH_s)\text{BR}(H_s\rightarrow \gamma\gamma)$  &
\multicolumn{3}{l|}{7.95 fb} \\ \hline \hline
$\sigma(ggH)$  & \multicolumn{3}{l|}{38.72 fb} \\ \hline
$\sigma(ggH)\text{BR}(H\rightarrow t\bar{t})$  &
\multicolumn{3}{l|}{9.80 fb} \\ \hline 
$\sigma(ggH)\text{BR}(H\rightarrow \tilde{\chi}_1^0 \tilde{\chi}_1^0)$  &
\multicolumn{3}{l|}{5.73 fb} \\ \hline 
$\sigma(ggH)\text{BR}(H\rightarrow h H_s)$  &
\multicolumn{3}{l|}{8.08 fb} \\ \hline 
$\sigma(ggH)\text{BR}(H\rightarrow h H_s \to b\bar{b} + \tau\tau)$  &
\multicolumn{3}{l|}{4.26 fb} \\ 
$\sigma(ggH)\text{BR}(H\rightarrow h H_s \to \tau\tau + \tau\tau)$  &
\multicolumn{3}{l|}{0.45 fb} \\ \hline \hline
$\sigma(ggA_s)$  & \multicolumn{3}{l|}{9.31 fb} \\ \hline
$\sigma(ggA_s)\text{BR}(A_s\rightarrow b\bar{b})$  &
\multicolumn{3}{l|}{3.78 fb} \\ \hline 
$\sigma(ggA_s)\text{BR}(A_s\rightarrow \tau\tau)$  &
\multicolumn{3}{l|}{0.46 fb} \\ \hline \hline
$\sigma(ggA)$  & \multicolumn{3}{l|}{41.26 fb} \\ \hline
$\sigma(ggA)\text{BR}(A\rightarrow t\bar{t})$  &
\multicolumn{3}{l|}{11.24 fb} \\ \hline 
$\sigma(ggA)\text{BR}(A\rightarrow \tilde{\chi}_1^0 \tilde{\chi}_1^0)$  &
\multicolumn{3}{l|}{5.94 fb} \\ \hline
$\sigma(ggA)\text{BR}(A\rightarrow h A_s)$  &
\multicolumn{3}{l|}{4.95 fb} \\ \hline  
$\sigma(ggA)\text{BR}(A\rightarrow h A_s \to b\bar{b} + b\bar{b})$  &
\multicolumn{3}{l|}{1.15 fb} \\ 
$\sigma(ggA)\text{BR}(A\rightarrow h A_s \to b\bar{b} + \tau\tau)$  &
\multicolumn{3}{l|}{0.26 fb} \\ \hline 
$\sigma(ggA)\text{BR}(A\rightarrow Z H_s)$  &
\multicolumn{3}{l|}{7.78 fb} \\ \hline
$\sigma(ggA)\text{BR}(A\rightarrow Z H_s \to b\bar{b} + \tau\tau)$  &
\multicolumn{3}{l|}{1.08 fb} \\ \hline \hline
\end{tabular}}
\caption{The signal rates for D.1. \label{tab:ratesscenD1}} 
\end{center}
\end{table}
\noindent
\underline{\it D.1) SM-like Higgs decays into light CP-even
 singlet-like Higgs
 pairs:} The scenario D.1, Table~\ref{table:scenD.1}, 
 leads to a very light CP-even singlet Higgs $H_s$ with a mass of
9.6~GeV. The branching 
ratio of the SM-like $h$, that is copiously produced with a cross
section of 44~pb, into $H_sH_s$ makes up 10\%. The singlet $H_s$ is so
light that it cannot decay into bottom pairs and instead decays at
90\% into $\tau$ pairs. This leads then to cascade decay rates of
3.6~pb in the decay chain $h \to H_s H_s \to 4\tau$, {\it
  cf.}~Table~\ref{tab:ratesscenD1}. And even the
$(\tau\tau)(\mu\mu)$ final state reaches 32~fb. This channel should
therefore not only allow for $H_s$ discovery, but also for the
measurement of the triple Higgs coupling $\lambda_{h H_s H_s}$. Better
discovery prospects for $H_s$ are obtained in direct gluon fusion
production with subsequent decay into {\it e.g.}~$\tau$ pairs. As
$H_s$ is so light and not 100\% singlet-like its gluon fusion cross
section is large with 440~pb. The $\tau\tau$ final state rate is then
given by enormous 405~pb. And also the rare decays into charm quarks
amount to 5.2~pb, allowing for a measurement of the $H_s c\bar{c}$
coupling. The $2\gamma$ final state finally reaches 8~fb. The masses
of the doublets $H$ and $A$ are around 790~GeV, so that their gluon
fusion production cross sections only reach ${\cal O}(40\mbox{
  fb})$. Their masses are large enough so that they decay into top
quark pairs, also the lightest neutralino pair final state rates can reach $\sim
6$~fb. Furthermore, the branching ratio BR$(H\to hH_s)=0.21$ so that
the Higgs-to-Higgs cascade into the $(\tau\tau)(b\bar{b})$ final state
reaches 4.3~fb and can add to the $H_s$ search channel or
alternatively to the measurement of the $\lambda_{H H_s h}$
coupling. The pseudoscalar can decay via $h A_s$. However, the
$(\tau\tau)(b\bar{b})$ rate is small with 0.3~fb. The extraction of
the triple coupling $\lambda_{A A_s h}$ will be difficult here. But it
can add to the $A_s$ discovery, in particular as the gluon fusion cross section for
the very singlet-like $A_s$ is small. Finally, $A$ can decay into the
gauge-Higgs final state $ZH_s$ leading to 1~fb in the
$(\tau\tau)(b\bar{b})$ final state. \s

\begin{table}[!b]
\begin{center}
{\small \begin{tabular}{|l|l|l|l|}
\hline \hline
\textbf{D.2 (Point ID 110)} & \multicolumn{3}{l|}{\textbf{Scenario}} \\ \hline \hline
$M_{H_s},M_{h},M_{H}$ & 112.0 GeV &
126.3 GeV & 1288.2 GeV \\ \hline 
$M_{A_s}, M_{A}$ & 61.5 GeV &
\multicolumn{2}{l|}{1287.4 GeV} \\ \hline 
$|S_{H_1h_s}|^2, |P_{A_1 a_s}|^2$ & 0.63 & \multicolumn{2}{l|}{1} \\ \hline \hline
$\mu_{\tau\tau}$, $\mu_{bb}$ & 0.73 & \multicolumn{2}{l|}{0.62} \\ \hline
$\mu_{ZZ}$, $\mu_{WW}$, $\mu_{\gamma\gamma}$  & 0.90 & 1.03 & 1.06 \\ 
\hline \hline
$\tan\beta$, $\lambda$, $\kappa$ & 6.36 & 0.47 & 0.14 \\ \hline
$A_{\lambda}$, $A_ {\kappa}$, $\mu_{\text{eff}}$ & 1217.1 GeV & 19.6 GeV 
& 195.3 GeV \\ \hline
$A_ t$, $A_b$, $A_{\tau}$ & -1804.6 GeV & -1196.8 GeV & 1704.8 GeV \\ \hline
$M_1$, $M_2$, $M_3$ & 417.2 GeV & 237.5 GeV & 2362.2 GeV \\ \hline
$M_{Q_3}=M_{t_R}$, $M_{b_R}$ & 967.8 GeV  & \multicolumn{2}{l|}{3 TeV} \\ \hline
$M_{L_3}=M_{\tau_R}$, $M_{\text{SUSY}}$ & 2491.6 GeV & 
\multicolumn{2}{l|}{3 TeV} \\ \hline \hline
\end{tabular}}
\caption{The parameters defining scenario D.2, together with the Higgs
boson masses, singlet components and reduced signal rates of
$h$. \label{table:scenD.2}} 
\end{center}
\end{table}
\noindent
\underline{\it D.2) SM-like Higgs decays into light CP-odd
  singlet-like Higgs
  pairs:} In the scenario D.2, defined in
Table~\ref{table:scenD.2}, the SM-like Higgs can decay into a
pair of pseudoscalar singlets $A_s$. The latter is very light with a
mass around 62~GeV. The lightest scalar $H_s$ with a mass of 112~GeV
is close in mass to 
$h$, and both mix strongly, so that the gluon fusion production cross
section for $h$ only amounts to 27~pb, while the $H_s$ production is
rather large for a singlet-like boson and reaches 17~pb, as given in
Table~\ref{tab:ratesscenD2}. The $h$ 
reduced rates are still compatible with the LHC data, although the
final state rates for $\tau\tau$ and $b\bar{b}$ are somewhat on the
lower side. The $h$ cascade decay via an $A_s$ pair reaches in the
$(\tau\tau)(b\bar{b})$ final state a large cross section of
276~fb. This should make a measurement of the triple Higgs couplings
$\lambda_{h A_s A_s}$ possible. Also the $4\tau$ final state is
sizeable with 12~fb. The light $A_s$ can be directly produced, too, and
then searched for in the $\tau\tau$ final state with a rate of
92~fb. The copiously produced $H_s$ can be searched for in the
standard decay channels. Difficult, however, if not impossible is the
production of the doublet-like $H$ and $A$, as they have masses of
1.3~TeV. Since they are both rather down-component doublet-like, one might
consider associated production with a $b$-quark pair, in view of the
not so small $\tan\beta=6.4$. But also these cross sections remain
below 1~fb. \s

\begin{table}[!ht]
\begin{center}
{\small \begin{tabular}{|l|l|l|l|}
\hline \hline
\textbf{D.2 (Point ID 110)} & \multicolumn{3}{l|}{\textbf{Signal
    Rates}} \\ \hline \hline  
$\sigma(ggh)$  & \multicolumn{3}{l|}{27.37 pb} \\ \hline
$\sigma(ggh)\text{BR}(h\rightarrow A_sA_s)$  & 
\multicolumn{3}{l|}{1.85 pb} \\ \hline
$\sigma(ggh)\text{BR}(h\rightarrow A_sA_s\rightarrow bb + bb)$  & 
\multicolumn{3}{l|}{1.55 pb} \\ 
$\sigma(ggh)\text{BR}(h\rightarrow A_sA_s\rightarrow bb + \tau\tau)$  & 
\multicolumn{3}{l|}{276.30 fb} \\ 
$\sigma(ggh)\text{BR}(h\rightarrow A_sA_s\rightarrow \tau\tau + 
\tau\tau)$  & \multicolumn{3}{l|}{12.36 fb} \\ 
$\sigma(ggh)\text{BR}(h\rightarrow A_sA_s \rightarrow bb + 
\gamma\gamma)$  & \multicolumn{3}{l|}{0.34 fb} \\ \hline \hline
$\sigma(ggH_s)$  & \multicolumn{3}{l|}{17.25 pb} \\ \hline
$\sigma(ggH_s)\text{BR}(H_s\rightarrow b\bar{b})$  & 
\multicolumn{3}{l|}{14.64 pb} \\ \hline
$\sigma(ggH_s)\text{BR}(H_s\rightarrow \tau\tau)$  & 
\multicolumn{3}{l|}{1.50 pb} \\ \hline
$\sigma(ggH_s)\text{BR}(H_s\rightarrow \gam)$  & 
\multicolumn{3}{l|}{13.93 fb} \\ \hline
$\sigma(ggH_s)\text{BR}(H_s\rightarrow ZZ)$  & 
\multicolumn{3}{l|}{23.90 fb} \\ \hline
$\sigma(ggH_s)\text{BR}(H_s\rightarrow WW)$  & 
\multicolumn{3}{l|}{401.21 fb} \\ \hline
$\sigma(ggH_s)\text{BR}(H_s\rightarrow \mu\mu)$  & 
\multicolumn{3}{l|}{5.33 fb} \\ \hline
$\sigma(ggH_s)\text{BR}(H_s\rightarrow Z \gamma)$  & 
\multicolumn{3}{l|}{4.15 fb} \\ \hline \hline
$\sigma(ggA_s)$  & \multicolumn{3}{l|}{1.13 pb} \\ \hline
$\sigma(ggA_s)\text{BR}(A_s\rightarrow b\bar{b})$  & 
\multicolumn{3}{l|}{1.03 pb} \\ \hline
$\sigma(ggA_s)\text{BR}(A_s\rightarrow \tau\tau)$  & 
\multicolumn{3}{l|}{92.46 fb} \\ \hline \hline
$\sigma(ggH)$  & \multicolumn{3}{l|}{0.46 fb} \\ \hline 
$\sigma(bbH)$  & \multicolumn{3}{l|}{0.82 fb} \\ \hline\hline
$\sigma(ggA)$  & \multicolumn{3}{l|}{0.72 fb} \\ \hline 
$\sigma(bbA)$  & \multicolumn{3}{l|}{0.82 fb} \\ \hline\hline
\end{tabular}}
\caption{The signal rates for D.2. \label{tab:ratesscenD2}} 
\end{center}
\end{table}
Picking up the idea presented in \cite{Englert:2014uua}, sum rules for the
Higgs couplings can be exploited to give a hint to the missing states, 
in case the heavy Higgs bosons cannot be discovered. This allows then
to disentangle an NMSSM Higgs sector from a supersymmetric Higgs
sector with minimal particle content, if {\it
  e.g.}~only three Higgs bosons are discovered and not all of them are
scalar. From Eq.~(\ref{eq:gaugecoupl}) and the unitarity of the 
${\cal R}^S$ matrix, it can be derived that the NMSSM CP-even Higgs
couplings to vector bosons with respect to the SM coupling obey the sum rule
\beq
\sum_{i=1}^3 g_{H_i VV}^2 = 1 \; . \label{eq:vecsumrule}
\eeq
The $H_i$ couplings $G_{H_i tt/bb}$ to top and bottom quarks
(normalised to the SM) are given by
\beq
\frac{G_{H_i tt}}{g_{H^{\text SM} tt}}  &\equiv& g_{H_i tt} =
\frac{{\cal R}^S_{i2}}{\sin\beta} \\
\frac{G_{H_i bb}}{g_{H^{\text SM} bb}}  &\equiv& g_{H_i bb} =
\frac{{\cal R}^S_{i1}}{\cos\beta}  \;. \label{eq:couplings}
\eeq
Exploiting the unitarity of ${\cal R}^S$ leads to the sum rule
\beq
\frac{1}{\sum_{i=1}^3 g_{H_i tt}^2}  + \frac{1}{\sum_{i=1}^3 g_{H_i
    bb}^2} = 1 \;. \label{eq:yuksumrule}
\eeq
If three neutral Higgs bosons are discovered, but not all of them are CP-even,
then these rules are violated, while in the MSSM with only three
neutral Higgs bosons in total the sum rules would be
fulfilled. Applying this to our scenario, in case among the CP-even
Higgs bosons we only find $h$ and $H_s$ but not the heavy $H$, the
sums Eqs.~(\ref{eq:vecsumrule}) and (\ref{eq:yuksumrule}) result in 
\beq
\sum_{i=1}^2 g_{H_i VV}^2 \approx 1 
\eeq
for the gauge couplings and 
\beq
\frac{1}{\sum_{i=1}^2 g_{H_i tt}^2}  + \frac{1}{\sum_{i=1}^2 g_{H_i
    bb}^2} = 1.85
\eeq
for the Yukawa couplings. As the heavy doublet $H$ is dominated by the
$H_d$ component ${\cal R}^S_{31}$ and the coupling to down-type quarks
is additionally enhanced for large $\tan\beta$ values, {\it
  cf.}~Eq.~(\ref{eq:couplings}), the missing $H_3$ coupling to
down-type quarks has a large effect on the Yukawa coupling sum
rule. For the same reason the effect on the vector coupling sum rule
is negligible, {\it cf.}~Eq.(\ref{eq:gaugecoupl}). At a future
high-energy LHC the couplings to fermions will be
measurable at ${\cal O}(10-20\%)$ level accuracy \cite{CMS:2013xfa}. This is
largely sufficient to test for the deviation in the Yukawa coupling sum rule
in this scenario. 

\section{Conclusions \label{sec:concl}}
After the discovery of the Higgs boson at the LHC no direct sign of
new particles beyond the SM has shown up yet. Also the discovered Higgs
boson itself looks very SM-like. Still it could be the SM-like
resonance of an extended Higgs sector as it emerges in supersymmetric
theories. In particular, the NMSSM with five neutral Higgs bosons
gives rise to a very interesting phenomenology and a plethora of new
signatures. We have investigated in a large scan over the NMSSM
parameter range, what the prospects of pinning down the NMSSM
Higgs sector at the 13~TeV run of the LHC are. Hereby we have taken into
account the constraints that arise from the LEP, Tevatron and LHC Higgs boson
searches, from Dark Matter measurements and low-energy observables and
from exclusion bounds on SUSY particles. We find, that in the NMSSM both
the lightest or next-to-lightest Higgs boson can be the SM-like Higgs
boson. The LHC Higgs signal can also be built up by two Higgs
resonances that are almost degenerate and have masses close to 
125~GeV. Furthermore, viable scenarios can be found for low and high
$\tan \beta$ values. \s 

We have investigated the production rates of the
neutral Higgs bosons in the SM final states. A lot of scenarios should
be accessible at the LHC, some, however, will be challenging and
possibly not allow for the discovery of all NMSSM Higgs bosons. We
therefore subsequently focused on the subspace of the NMSSM, that we
call the Natural NMSSM, and which features heavy Higgs bosons with
masses below about 530~GeV and besides a light CP-odd singlet state a
light CP-even singlet-like Higgs boson with a mass between about
62 and 117~GeV. The study of the signal rates in the SM final
states reveals that the Natural NMSSM should give access to all neutral
NMSSM Higgs bosons in most of the scenarios. Where the discovery in
direct production is difficult, it can be complemented by searches in
Higgs-to-Higgs or Higgs-to-gauge boson-Higgs decays. It should
therefore be possible to discover the Natural NMSSM Higgs bosons at the next run
of the LHC or to strongly constrain this scenario. \s

Higgs decays into Higgs pairs offer an interesting alternative to the
discovery of Higgs states that may be difficult to be accessed in
direct production. At the same time they give access to the trilinear
Higgs self-coupling involved in the Higgs-to-Higgs decays. Taking the
results of our large parameter scan we have extracted benchmark
scenarios that highlight different properties of
Higgs-to-Higgs-decays. In analysing the various decay signatures, we
found multi-photon and multi-fermion signatures that not only lead to
very promising signal rates (far above the ones to be expected in 
SM Higgs pair production), but some of them are also unique to the
NMSSM and lead to 
spectacular final states with up to six photons. Another interesting
outcome is the possibility of CP-even singlet states that are below the bottom
pair threshold, so that the decays into other light fermions are
dominant and give access to the measurement of Higgs couplings to
taus, muons and even light quarks. Finally, we show in one
example, how coupling measurements could help to point to an
additional CP-even Higgs boson, in case not all CP-even states should
have been discovered. \s

In summary, our results show that the search for Higgs bosons needs to
be continued not only in the high- but also in the low-mass regions
below 125~GeV. Furthermore, we have to be prepared for new exotic signatures that
cannot appear in minimal supersymmetric extensions of the SM, but are
possible in the NMSSM due to significant signal rates arising from
Higgs pair production in Higgs decays. 

\subsubsection*{Acknowledgments}
We thank Alexander Nikitenko and Markus Schumacher for fruitful
discussions and advise. RN also thanks Peter Athron, Anthony Thomas and 
Anthony Williams for useful discussions. MM acknowledges discussions
with Conny Beskidt, Ramona Gr\"ober, Rui Santos and Michael Spira. 
MM is supported by the DFG SFB/TR9  ``Computational
Particle Physics''.
The work by RN was supported by the Australian 
Research Council through the ARC Center of Excellence in Particle
Physics at the Terascale. 
SFK is supported by the STFC Consolidated grant ST/J000396/1 and EU
ITN grant INVISIBLES 289442 .
KW gratefully acknowledges support of the Graduiertenkolleg GRK1694
``Elementarteilchen bei h\"ochster Energie und h\"ochster
Pr\"azision''. 

\setcounter{equation}{0}
\section*{Appendix}
\begin{appendix}
\section{Higgs Couplings in the Natural NMSSM}
In the following approximations for the Higgs couplings in the
parameter range of the Natural NMSSM will be given.  First of all, the
125~GeV Higgs boson has to be produced with SM-like rates. This means
that it must have large enough couplings to top quarks and hence a
large $H_u$ component. At the same time, it is advantageous not to
have a too large $H_d$ component, as then the coupling to $b$ quarks
is suppressed, leading to enhanced branching ratios that account for a
possible slight suppression in the production. Due to the unitarity of
the mixing matrix rotating the interaction to the Higgs mass
eigenstates, this means, that the other doublet-like heavier Higgs boson must be
$H_d$-like. The light CP-even and CP-odd Higgs bosons, $H_s$ and $A_s$
respectively, must be singlet-like in order to avoid the exclusion bounds. In
summary, for the natural NMSSM the approximate
compositions of the CP-even Higgs mass eigenstates expressed in terms of the
mixing matrix ${\cal R}_{ij}$ ($i=H_s,h,H$, $j=h_d,h_u,h_s$), in the
notation of Section 
\ref{sec:naturalnmssm}, are given by
\beq
(H_s,h,H)^T = {\cal R}^S (h_d,h_u,h_s)^T = 
\left(\begin{array}{ccc}
0 & 0 & 1 \\ 0 & 1 & 0 \\ 1 & 0 & 0 
\end{array}\right) (h_d,h_u,h_s)^T \;.
\eeq
For the composition of the pseudoscalars we have,
\beq
(A_s,A)^T = {\cal R}^P (a,a_s)^T = 
\left(\begin{array}{cc}
0 & 1 \\ 1 & 0 
\end{array}\right) (a,a_s)^T \;,
\eeq
with $a= \sin\beta a_d + \cos\beta a_u$.
Comparing this with the actual composition as result of our scan,
the actual composition is approximated rather well, with at most 30\%
deviations. With these assumptions we get for the trilinear
Higgs couplings involved in the Higgs-to-Higgs decays, in units of the
SM coupling, 
\beq
g_{H h h} &=& \frac{\cos\beta}{M_Z^2} (\lambda^2 v^2- M_Z^2) \\
g_{H h H_s} &=& \frac{-v}{\sqrt{2} M_Z^2} ( \lambda A_\lambda+2\kappa
\mu) \\
g_{H H_s H_s} &=& \frac{\lambda v^2}{M_Z^2} (\lambda\cos\beta-\kappa\sin\beta) 
\\
g_{hH_sH_s}&=&\frac{\lambda  v^2}{M_Z^2} (\lambda  \sin \beta -\kappa  \cos 
\beta ) \\
g_{h A A_s} &=& \frac{\lambda v \sin\beta}{2 M_Z^2} (\sqrt{2} A_\lambda
-2 \kappa v_s) \\
g_{hA_sA_s}&=&\frac{\lambda  v^2}{M_Z^2} (\kappa  \cos \beta +\lambda  \sin 
\beta ) \\
g_{H_s A A_s} &=& \frac{-\lambda v^2 \kappa}{M_Z^2} \\
g_{HA_sA_s}&=&\frac{\lambda  v^2}{M_Z^2} (\kappa  \sin \beta +\lambda  \cos 
\beta ) \;.
\eeq
\end{appendix}

\vspace*{0.5cm}

\end{document}